\documentclass{article}

\usepackage{PRIMEarxiv}
\usepackage[utf8]{inputenc} 
\usepackage[T1]{fontenc} 
\usepackage{amssymb}
\usepackage{lipsum}
\usepackage{parskip}
\usepackage{stmaryrd}

\usepackage{mathrsfs}
\usepackage[colorlinks]{hyperref}
\usepackage{graphicx}
\usepackage{amsmath}
\usepackage{amssymb}
\usepackage[dvipsnames]{xcolor}
\usepackage{subfig}
\usepackage{comment}

\usepackage{appendix}
\usepackage{here}
\usepackage{gensymb}
\usepackage{mathrsfs}
\usepackage{ctable}
\usepackage{array}
\newcolumntype{C}[1]{>{\centering\arraybackslash}p{#1}} 
\usepackage{bm}
\usepackage{multirow}
\usepackage{textcomp}
\usepackage{algpseudocode}
\usepackage{siunitx}[=v2]
\DeclareSIUnit{\wtpercent}{wt\%}
\newcommand{\murm}{%
  \ifmmode
    \mathchoice
        {\hbox{\normalsize\textmu}}
        {\hbox{\normalsize\textmu}}
        {\hbox{\scriptsize\textmu}}
        {\hbox{\tiny\textmu}}%
  \else
    \textmu
  \fi
}

\usepackage{booktabs}
\usepackage{empheq, etoolbox}
\title{High-fidelity level-set modeling of diffusive solid-state phase transformations for polycrystalline materials}

\author{
  Nitish Chandrappa \\
  Mines Paris, PSL University\\
  Centre for material forming (CEMEF), UMR CNRS\\
  06904 Sophia Antipolis, France\\
  \texttt{nitish.chandrappa@minesparis.psl.eu} \\
  \AND
  Marc Bernacki \\
  Mines Paris, PSL University\\
  Centre for material forming (CEMEF), UMR CNRS\\
  06904 Sophia Antipolis, France\\
  \texttt{marc.bernacki@minesparis.psl.eu} \\
}

\begin{document}
\maketitle
\begin{abstract}
The formation of microstructures in metallic alloys during hot metal forming involves simultaneous metallurgical complex phenomena. Traditional high-fidelity numerical frameworks used on the polycrystalline scale tend to focus on single-phase microstructures or isolate phase transformations from grain boundary migration mechanisms. The level-set method is highlighted as effective in proposing a global framework for modeling multiphase polycrystalline materials and diffusive solid-state phase transformations. This framework includes novel techniques for efficient large-scale microstructural representation, strong coupling with ThermoCalc software for real-time thermodynamic data, application for ternary alloys and beyond by taking solute drag aspects, and the use of advanced nucleation models. Numerous applications are then illustrated. 

\end{abstract}


\section{Introduction}
\label{intro}

The microstructural evolution in metallic alloys in the context of hot metal forming typically involves a complex interplay of multiple phenomena that occur simultaneously. Industrial metallurgical products are often subjected to complex thermomechanical treatments or processing conditions such as high plastic deformation at elevated temperatures. Under these conditions, solid-state phase transformations within alloyed materials become inevitable, along with other concurrent phenomena. Currently, most state-of-the-art numerical predictions primarily focus on single-phase microstructural evolution or exclusively model phase transformation, often neglecting the consideration of other coexisting phenomena. Although delving into individual phenomena provides valuable insights, focusing solely on singular aspects might inadvertently lead to oversights and limitations in comprehending the holistic behavior of these materials. Such a narrow focus could lead to overlooking crucial interactions that may influence the alloy's mechanical strength, thermal stability, or its susceptibility. Neglecting the concomitant nature of these transformation aspects could limit our ability to predict and optimize material performance under diverse operating conditions.

Advances in computational resources have paved the way for more intricate models (such as atomistic \cite{jin2006atomic, mishin2010atomistic, biglari2013simulation} and full-field mesoscopic models \cite{janssens2010computational}) based on a complete and explicit representation of the microstructure. These models enable to closely follow the topological evolution of the microstructure during a transformation and have the potential to capture complex evolution aspects, and hence their predictive capabilities are much wider. Full-field models (FFM) provide a good equilibrium between the high-level description of the microstructure and the computational demands. Among the various approaches within the scope of FFM, each numerical method has its unique scope of application, with some better suited to specific scenarios than others. For modeling recrystallization and grain growth under high plastic deformation, the level-set method \cite{bernacki2011level, FAUSTY2018578, alvarado2021, MairePhD}  stands out as a robust choice and has been effectively utilized. At the mesoscopic scale, Phase-field methods \cite{hillert2004definitions, Yeon2001, Pariser2001, Huang2006, Mecozzi2007PhaseFM, Militzer2006} have been well established as the popular choice of method to simulate diffusive solid-state phase transformation (DSSPT) due to its thermodynamic consistency. However, the potential of level-set methods remains largely unexplored in the realm of DSSPT. The developments of Bzowski et al. \cite{bzowski2018application} is one of the only few works based on level-set method for DSSPT. In these works, one level-set function per grain is considered, and to model solute diffusion, the diffusion equation is resolved in the parent phase while that in the product phase his considered as negligible. 

The authors thus proposed a global level-set framework in \cite{chandrappa2023level} aimed at thoroughly exploring the capabilities of using the level-set method to simulate DSSPT in a finite element (FE) context. Within this framework, the grain coloration technique, proposed in \cite{Scholtes2015, scholtes2016development} was used to limit the number of level-set functions required to describe a large-scale microstructure. The diffusion aspects are modeled in both the parent and the product phase. Illustrations for phase transformation in simple binary alloyed microstructures were demonstrated. The objective of this work is to generalize the previously proposed numerical framework for ternary alloys and beyond by taking solute drag aspects \cite{hutchinson2006growth, purdy2011alemi} into account. This work also aims to integrate nucleation models into the framework, to enrich boundary conditions to broaden the scope of application, and to develop an efficient numerical coupling with ThermoCalc software \cite{thermocalc} to seamlessly extract the necessary thermodynamic data. 

An overview of the numerical formulation proposed in \cite{chandrappa2023level} is provided in the next section, followed by the improvements and the generalizations incorporated. A benchmark analysis of the numerical model, followed by illustrations in large scale microstructures, are presented in the third section. The model's capabilities in simulating other diffusive solid-state phenomena are also demonstrated. The concluding remarks and the prospective works are then discussed in the last part.

\section{Numerical formulation}
\label{numform}

\begin{figure}[!htb]
	\centering
	\captionsetup{justification=centering,margin=1cm}
	\subfloat[Classification of different grains]{\includegraphics[width=6cm]{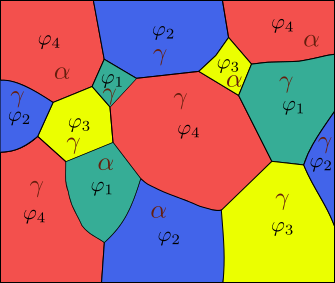}\label{primebasisGLS}}
    \hspace{1em}
	\subfloat[Classification of regions of different phases]{\includegraphics[width=6cm]{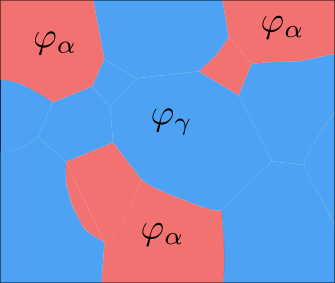}\label{secondbasisLS}}
	\caption{Biphasic microstructure indicating the underlying the level-set representation}
	\label{LSbasis}
\end{figure}

The growth regime of a microstructural evolution involving diffusive phase transformation is influenced by two coupled processes: (i) solute partitioning between the phases involved in transformation, and (ii) subsequent migration of the interface network along with other contributions to the interface kinetics. As outlined in the previous publication \cite{chandrappa2023level}, level-set (LS) functions are employed to represent and govern the interface migration aspects. In the context of a polycrystalline microstructure, a set of global level set (GLS) functions, $\varphi_i$, are used to efficiently represent and classify large number of grains based on the grain coloration theorem \cite{Scholtes2015, scholtes2016development}. However, in the context of multiphase microstructure, secondary LS functions are necessary to distinguish the regions composed of different phases. The $\varphi_i$ GLS functions constitute the primary basis of LS representation, while the level-set functions used to classify the phases forms the secondary basis. Fig.\ref{primebasisGLS} gives an illustration of a biphasic microstructure with the color code indicating the family of GLS functions used to depict different grains. Meanwhile, Fig.\ref{secondbasisLS} indicates the regions composed of $\alpha$ and $\gamma$ phases, and the secondary basis $\varphi_\alpha$ LS function employed to characterize them. 

The presence of different phases introduce material discontinuities. Consequently, within the classical level-set framework, resolving solute redistribution may require separate treatment in the domains of different phases by explicitly taking into account the necessary jump conditions at the interphase boundaries. Incorporating these jump conditions necessitates localizing the evolving interphase boundaries at each resolution step, a task facilitated by the secondary level-set function ($\varphi_\alpha$). As iterated in \cite{chandrappa2023level}, rather than relying on the sharp interface characteristics (in theory) provided by the LS representation, we adopt a diffuse interface hypothesis across the interphase boundaries while addressing solute diffusion within the domain. This approach aims to smooth out discontinuities and integrate a unified equation for modeling solute partitioning throughout the domain. It's worth noting that the secondary basis level-set function $\varphi_\alpha$ is initially constructed from the input phase characteristic function ($\chi_\alpha$) \cite{chandrappa2023level}. Then, as the evolution proceeds, $\varphi_\alpha$ is updated accordingly. The transition from $\varphi_\alpha$ to a diffuse interface representation ($\phi$) is achieved using a hyperbolic tangent relation between them, which is extensively explained in our previous publication \cite{chandrappa2023level}:

\begin{equation}
\phi = \frac{1}{2}tanh\left(\frac{3\varphi_\alpha}{\eta}\right)+\frac{1}{2}.
\label{tanhfrel}
\end{equation}The $\eta$ in the above relation is a diffuse interface thickness parameter.

After resolving the solute diffusion equation, the resulting solute distribution influences the kinetics of interface migration, alongside other inherent effects depending on the various considered phenomena. The resulting motion of the multi-phase grain interface network are then governed by the transport of primary basis GLS functions ($\varphi_i$), based on their classical level-set description. 

\subsection{Solute redistribution}
Upon establishing a diffuse interface description, $\forall i\in \{C,X\}$, a total concentration field variable ($x_i$) can be represented as a continuous field between the concentrations in the individual phases ($x_i^\alpha, x_i^\gamma$) for each solute $i$:
\begin{equation}
x_i = \phi x_i^\alpha + (1-\phi)x_i^\gamma.
\label{cmix}
\end{equation}
As detailed in \cite{chandrappa2023level}, through the imposition of solute flux continuity for each phase and assuming a constant concentration ratio between the phases at the interface throughout, the following unified diffusion equation is derived, adhering to Fick's laws of diffusion:

\begin{equation}
\frac{\partial x_i}{\partial t} = \bm{\nabla}\cdot\left\{D^*(\phi)\left[\bm{\nabla} x_i -\frac{x_i(k_p^i-1)}{1+\phi(k_p^i-1)}\bm{\nabla}\phi \right]\right\}.
\label{diffmixform}
\end{equation}

With further simplifications, the above Eq.\eqref{diffmixform} could be reformulated into the following flux form:
\begin{equation}
\frac{\partial x_i}{\partial t} = \bm{\nabla}\cdot\left[D^*(\phi)\bm{\nabla}x_i - x_i\bm{\mathcal{A}}(\phi)\right].
\label{CDRFluxform}
\end{equation}
A Convective-Diffusive-Reactive (CDR) formulation is then obtained by expanding the above equation:
\begin{equation}
\frac{\partial x_i}{\partial t} +  \left(\bm{\mathcal{A}}-\bm{\nabla}D^*\right)\cdot\bm{\nabla}x_i - D^*\Delta x_i + \mathcal{R}x_i = 0.
\label{CDReqn} 
\end{equation} 

In the above CDR formulation:
\begin{itemize}
    \item $D^*$ is called the mixed diffusivity, expressed as:
    \begin{equation}
    D^*(\phi) = \frac{D_\gamma^i + \phi(k_p^iD^i_\alpha - D^i_\gamma)}{1+\phi(k_p^i-1)},
    \label{mixeddiffus}
    \end{equation} where $D^i_\alpha$ and $D^i_\gamma$ represent the diffusivity of the $i$th solute element in $\alpha$ and $\gamma$ phases, respectively. $k_p^i$ is the constant concentration ratio respected by the redistribution of the solute atoms between the parent and the product phase at the interphase. This ratio is assumed to be equal to the equilibrium partitioning ratio ($k_{p_{eq}}^i$), corresponding to the conditions at the instant $t$:
    \begin{equation}
    k_p^i\left(t\right) = \frac{x_i^\alpha}{x_i^\gamma} \approx k_{p_{eq}}^i\left(t\right) = \frac{\left.x_i^\alpha\right|_{eq}}{\left.x_i^\gamma\right|_{eq}}
    \label{kpdef},
    \end{equation}where $\left.x^\alpha_i\right|_{eq}$ and $\left.x^\gamma_i\right|_{eq}$ are the equilibrium concentrations of $\alpha$ and $\gamma$ phases respectively at temperature $T$ corresponding to an instant $t$.

    \item $\bm{\mathcal{A}}$ is a part of the advective coefficient and is expressed as:
    \begin{equation}
    \bm{\mathcal{A}}(\phi)= \frac{D^*(\phi)(k_p^i-1)}{1+\phi(k_p^i-1)}\bm{\nabla}\phi.
    \label{AdefinCDReqn}
    \end{equation}

    \item $\mathcal{R}$ represents the reactive term and is defined as:
    \begin{equation}
    \mathcal{R}=\bm{\nabla}\cdot\bm{\mathcal{A}}.
    \label{RdefinCDReqn}
    \end{equation}
    \label{DAR_def}
\end{itemize}

For a test function, $\psi \in H^1(\Omega)$, the generic P1 finite element weak formulation of Eq.\eqref{CDReqn} could be simplified to the following form:
\begin{equation}
\begin{split}
    \int_\Omega \frac{\partial x_i}{\partial t}\psi \ d\Omega  +\int_\Omega\bm{\mathcal{A}}\cdot\bm{\nabla}x_i\psi \ d\Omega + \int_\Omega D^*\bm{\nabla}\psi\cdot \bm{\nabla}x_i \ d\Omega \\ + \int_\Omega \mathcal{R}x_i\psi \ d\Omega - \int_{\partial \Omega} \psi D^*\bm{\nabla}x_i\cdot\bm{n}_b \ dS = 0,
\end{split}
\label{diffWF}
\end{equation} where $\bm{n}_b$ is the outward unit normal to the domain boundary and the last boundary integral term is subject to the imposed boundary conditions. 

In our previous works \cite{chandrappa2023level}, we employed homogeneous Neumann conditions on $x_i$ for solute mass conservation, i.e., $\left.\bm{\nabla}x_i\cdot \bm{n}_b\right|{\partial \Omega} = 0$. At first glance, these assumptions may seem reasonable, focusing on pure solute diffusion within the computational domain without any influx or outflux of solute atoms. However, these conditions are incomplete and are only valid in a specific scenario. The mathematical formulation in Eq.\eqref{CDRFluxform} suggests that applying homogeneous Neumann-type boundary conditions on $x_i$ conserves solely the diffusive flux ($-D^*\bm{\nabla}x_i$), while the advective flux ($x_i\bm{\mathcal{A}}$) remains non-conserved.

If we apply the Reynolds's transport theorem on $x_i$ for a fixed domain boundary, for mass conservation:
\begin{equation}
    \frac{d}{dt}\int_\Omega x_i d\Omega = \int_\Omega \frac{\partial x_i}{\partial t}d\Omega=0.
    \label{reytranstheo}
\end{equation}

Plugging Eq.\eqref{CDRFluxform}, and applying the divergence theorem, we obtain:
\begin{equation}
    \frac{d}{dt}\int_\Omega x_i d\Omega = \int_\Omega \frac{\partial x_i}{\partial t}d\Omega= \int_{\partial \Omega} \left[D^*(\phi)\bm{\nabla}x_i - x_i\bm{\mathcal{A}}(\phi)\right]\cdot \bm{n}_b dS = 0.
    \label{reytransexpand}
\end{equation}
Thus, the above equation holds true for:
\begin{equation}
    \left.\left[-D^*(\phi)\bm{\nabla}x_i + x_i\bm{\mathcal{A}}(\phi)\right]\cdot \bm{n}_b\right|_{\partial \Omega} = 0.
    \label{SRBC}
\end{equation}
The aforementioned Robin-type condition constitutes the correct and complete boundary condition for Eq.\ref{CDReqn}.

Equation \eqref{AdefinCDReqn} implies that $\bm{\mathcal{A}}\rightarrow 0$ far from a phase interface where $\bm{\nabla}\phi\rightarrow 0$. In this scenario, Eq.\eqref{SRBC} simplifies to a pure Neumann-type boundary condition, i.e., $\left.-D^*\bm{\nabla}x_i\cdot \bm{n}_b\right|_{\partial \Omega} = 0$. Therefore, a pure Neumann-type boundary condition is a particular case, applicable only when the local boundary region does not interact with any phase interface. In cases where phase interfaces approach certain sections of the boundary region, the local advective flux should also be conserved through a Robin type boundary condition as in Eq.\eqref{SRBC}.

Thus, for the resolution of the solute redistribution, there's a continuous local switch between homogeneous Neumann-type and Robin-type boundary conditions. This switch is based on the interaction of phase interfaces with the domain boundaries during the phase transformation.

Applying the generic Robin-type boundary condition to the weak form in Eq.\eqref{diffWF}, we eventually obtain:
\begin{equation}
\int_\Omega \frac{\partial x_i}{\partial t}\psi \ d\Omega  +\int_\Omega\bm{\mathcal{A}}\cdot\bm{\nabla}x_i\psi \ d\Omega + \int_\Omega D^*\bm{\nabla}\psi\cdot \bm{\nabla}x_i \ d\Omega + \int_\Omega \mathcal{R}x_i\psi \ d\Omega - \int_{\partial \Omega} x_i\psi \bm{\mathcal{A}}\cdot\bm{n}_b \ dS = 0.
\label{diffWFwithBC} 
\end{equation}

\subsection{Interface migration}
For a polycrystalline microstructure with $N_g$ grains whose interface network is represented by a set of $N_{LS}$ GLS functions ($\varphi_i$), the following convective equations need to be resolved for modeling the interface migration:
\begin{equation}
\begin{cases}
\frac{\partial \varphi_i}{\partial t} + \bm{v}\cdot\bm{\nabla}\varphi_i=0 \\
\varphi_i(\bm{x}, t=0) = \varphi_i^0(\bm{x})
\end{cases}	\forall i \in \{1,2,...,N_{LS}\}.
\label{GLStranseqn}
\end{equation} The velocity field ($\bm{v}$) in this application is based as a function of various driving pressures affecting the microstructural evolution.

In \cite{chandrappa2023level}, the following Convective-Diffusive level-set transport equations were formulated for grain/phase interface network migration for each of the $ith$ GLS function ($\varphi_i$) during a $\gamma\rightarrow\alpha$ phase transformation:
\begin{equation}
\begin{gathered}
\frac{\partial \varphi_i}{\partial t} + \left[\bm{v_{\Delta G_{\gamma\alpha}}}+\bm{v_{\llbracket \mathcal{E} \rrbracket}}\right]_i\cdot\bm{\nabla}\varphi_i - \left[\sum_{l\in \mathscr{S}}\chi_{l}M_{l}\sigma_{l}\right]\Delta\varphi_i = 0 \qquad \forall i \in \{1,2,...,N_{LS}\},
\label{lsCDeqn_old}
\end{gathered}
\end{equation}where $\mathscr{S}=\{\gamma\gamma, \gamma\alpha, \alpha\alpha\}$ with $\chi_{\gamma\gamma}, \chi_{\gamma\alpha},\text{ and } \chi_{\alpha \alpha}$ being the interface characteristic functions, $M$ is the interface mobility, and $\sigma$ is the interfacial energy. The convective velocity components are defined as:
\begin{equation}
    \begin{gathered}
        \bm{v_{\Delta G_{\gamma\alpha}}}=\chi_{\gamma\alpha}M_{\gamma\alpha}\Delta G_{\gamma\rightarrow\alpha} \bm{n} \\ 
        \bm{v_{\llbracket \mathcal{E} \rrbracket}}=\left[\sum_{l\in \mathscr{S}}\chi_{l}M_{l}\llbracket \mathcal{E} \rrbracket_{l} \right]\bm{n}.
    \end{gathered},
    \label{vdGandvdEdesc_basic}
\end{equation} with $\bm{n}$ being the outward unit normal to a migrating interface.

$\Delta G_{\gamma\rightarrow\alpha}$ corresponds to the driving pressure for phase transformation of $\gamma$ phase into $\alpha$ phase, and $\llbracket \mathcal{E} \rrbracket$ constitutes a driving pressure linked to the jump in stored energy ($\mathcal{E}$) due to plastic deformation.

The formulation in Eq.\eqref{lsCDeqn_old} corresponds to a scenario where the net driving pressure ($P$) is assumed to be composed of the phase transformation contribution, the stored energy component and the capillarity effects ($-\kappa \sigma$): 
\begin{equation}
    P=\Delta G_{\gamma\rightarrow\alpha} + \llbracket \mathcal{E} \rrbracket - \kappa \sigma.
    \label{drivpress_old}
\end{equation}

In a biphasic polycrystalline context, accounting for different kinds of interfaces, this net driving pressure is rewritten in a generalized form as:
\begin{equation}
\begin{gathered}
P = \chi_{\gamma\alpha}\Delta G_{\gamma\rightarrow\alpha} + \sum_{l\in \mathscr{S}}\chi_{l}\left(\llbracket \mathcal{E} \rrbracket_{l} - \kappa \sigma_{l}\right).
\end{gathered}
\label{totpresspolycryst_old}
\end{equation}
As a reminder, the interface migration velocity ($\bm{v}$) concerning a microstructural evolution at the mesoscopic scale is assumed to be a product of the interface mobility and the net local driving pressures describing the various involved phenomena at the interface \cite{christian2002theory}: $\bm{v}=MP\bm{n}$. The interface mobility is also expressed in a generalized form accounting for different types of interfaces such that, $M = \sum_{l\in \mathscr{S}}\chi_{l} M_l$.

In the context of Ternary alloys and beyond, comprising of slow diffusing substitutional elements, an additional driving pressure contribution due to the solute drag effects linked to these slow diffusing elements could be taken into account. Accounting for this contribution, Eq.\eqref{totpresspolycryst_old} could be modified as:
\begin{equation}
\begin{gathered}
P = \chi_{\gamma\alpha}\Delta G_{\gamma\rightarrow\alpha} + \sum_{l\in \mathscr{S}}\chi_{l}\left(\llbracket \mathcal{E} \rrbracket_{l} - \kappa \sigma_{l} + \left[P_{SD}\right]_l\right),
\end{gathered}
\label{totpresspolycryst}
\end{equation} where $P_{SD}$ is the solute drag pressure associated to the drag effects.

Upon prescribing the above modified interface kinetics into the transport equations (Eqs.\eqref{GLStranseqn}), and with further simplifications using the metric properties of a LS function, we obtain the following form $\forall i \in \{1,2,...,N_{LS}\}$:
\begin{equation}
\begin{gathered}
\frac{\partial \varphi_i}{\partial t} + \left[\bm{v_{\Delta G_{\gamma\alpha}}}+\bm{v_{\llbracket \mathcal{E} \rrbracket}}\right]_i\cdot\bm{\nabla}\varphi_i -  \sum_{l\in \mathscr{S}}\chi_{l}M_{l}\left[P_{SD}\right]_l - \left[\sum_{l\in \mathscr{S}}\chi_{l}M_{l}\sigma_{l}\right]\Delta\varphi_i = 0,
\label{lsCDReqn}
\end{gathered}
\end{equation}where,
\begin{equation}
\bm{n}_i=-\frac{\bm{\nabla}\varphi_i}{\Vert\bm{\nabla}\varphi_i\Vert}=-\bm{\nabla}\varphi_i \implies \kappa_i = \bm{\nabla}\cdot\bm{n}_i=-\Delta \varphi_i
\label{GLSmetprop}.
\end{equation}

The driving pressure due to solute drag contributes passively by opposing the actively available interface kinetics (linked to the principally active driving pressures), as the solutes tend to bind to the migrating interface due to segregation. Thus, the magnitude of the solute drag pressure ($\Delta G_{SD} = \lvert P_{SD} \rvert$) is generally assumed to be a function of the interface migration velocity itself, $\Delta G_{SD} = f(\bm{v})$. This has the effect of rendering the overall interface kinetics description non-linear. The dependence on the velocity is such that, in the vicinity of zero velocity, or at sufficiently high interfacial velocities, the drag pressure should theoretically vanish. The latter is due to the fact that at high velocities, there is not enough scope or time for the solute atoms to interact with the rapidly migrating interface, thus reducing the potential to impart any retardation effects of significance. For modeling the solute drag driving pressure ($P_{SD}$), Cahn's solute drag model \cite{cahn1962impurity} provides a simplified description for the magnitude of the drag pressure ($\Delta G_{SD}$):

\begin{equation}
    \Delta G_{SD}\left(\bm{v}\right) = \frac{\alpha_C x_X^0 \left\lVert \bm{v}\right\rVert}{1+\beta_C^2\left\lVert \bm{v}\right\rVert^2},
    \label{DeltaG_SD_def_num}
\end{equation} where $\alpha_C$ and $\beta_C$ are parameters defined as functions of temperature, interfacial solute diffusivity, interface width and the binding energy. Cahn proposed the following definitions for these parameters: 

\begin{equation}
    \alpha_C=4N_Vk_BT\int_{-\infty}^{+\infty}\frac{\sinh^2\left(\frac{E(z)}{2k_BT}\right)}{D^X_\Gamma (z)} \ dz,
    \label{alphaCahndef}
\end{equation}

\begin{equation}
    \frac{\alpha_C}{\beta_C^2}=\frac{N_V}{k_BT}\int_{-\infty}^{+\infty}\left(\frac{\partial E}{\partial z}\right)^2 D^X_\Gamma (z) \ dz.
    \label{betaCahndef}
\end{equation}

Even though Cahn's model lacks physical sense across interphase boundaries and is more apt for grain boundaries, the simplicity in its description is attractive for numerical implementation, especially in a FE framework, considering the non-linearity of a solute drag pressure on the velocity field. Cahn's simplified drag pressure is capable of capturing the general trends expected for solute drag effects whether it is for grain interfaces or phase interfaces as illustrated in \cite{nitishphd}. In first approximation, we shall thus consider Cahn's simplified description in this work to model solute drag pressure. 

The solute drag driving pressure component can be written as:
\begin{equation}
    P_{SD}\left(\bm{v}, \bm{n}\right) = -\Delta G_{SD}\frac{(\bm{v}\cdot\bm{n})}{\lVert \bm{v} \rVert} = -\frac{\alpha_C x_X^0 (\bm{v}\cdot\bm{n})}{1+\beta_C^2\left\lVert \bm{v}\right\rVert^2} 
    \label{P_SD_def_num}.
\end{equation}
With the application of Cahn's solute drag model even for phase interfaces, for the parameters, $\left.\left(\alpha_C, \beta_C\right)\right|_{l=\gamma\alpha}$, instead of using the analytical definitions provided by Cahn in Eq.\eqref{alphaCahndef} and Eq.\eqref{betaCahndef}, the idea is to consider them as a set of temperature dependent model parameters to be fitted in accordance with relevant experimental results. This avoids the complexity of precisely quantifying some of the physical parameters required by their analytical definitions, especially across phase interfaces. So, by selectively controlling the two parameters ($\alpha_C, \beta_C$), we could attempt to converge to expected magnitudes for the drag pressure across phase interfaces.

Since the solute drag pressure renders the velocity field non-linear. if we consider a fully implicit time discretization scheme for the resolution of Eq.\eqref{lsCDReqn}, we will obtain a non-linear formulation. Considering Euler implicit time discretization, Eq.\eqref{lsCDReqn} could be discretized as:
\begin{equation}
\frac{\partial \varphi_i}{\partial t} \approx \frac{\varphi_i^{k+1}-\varphi_i^{k}}{\Delta t} = \mathcal{F}(\varphi_i^{k+1}), \qquad \forall i \in \{1,2,...,N_{LS}\}
\label{lsEqnTimeDdemo},
\end{equation} where
\begin{equation}
    \mathcal{F}(\varphi_i^{k+1}) = \left[\sum_{l\in \mathscr{S}}\chi_{l}M_{l}\sigma_{l}\right]\Delta\varphi_i^{k+1} + \sum_{l\in \mathscr{S}}\chi_{l}M_{l}\left[P_{SD}(\bm{v}^{k+1})\right]_l - \left[\bm{v_{\Delta G_{\gamma\alpha}}}+\bm{v_{\llbracket \mathcal{E} \rrbracket}}\right]_i\cdot\bm{\nabla}\varphi_i^{k+1},
    \label{FvarphiDefAtTimestepk}
\end{equation}is a time dependent operator, with $k$ being the index of time stepping.

If one considers $P_{SD}^{k+1}$ in the second term in Eq.\eqref{FvarphiDefAtTimestepk}, one obtains from Eq.\eqref{P_SD_def_num}, 
\begin{equation}
    P_{SD}(\bm{v}^{k+1}) = -\frac{\alpha_C x_X^0 (\bm{v}\cdot\bm{n})^{k+1}}{1+\beta_C^2\left\lVert \bm{v}^{k+1}\right\rVert^2}.
    \label{P_SD_def_tdiscret}
\end{equation} Since the interface migration velocity is normal to the interface, at any position $\bm{x}$, for the \textbf{signed distance} $i$th GLS function, such that $\varphi_i^{k+1}(\bm{x})\geq 0$, it is possible to express: 

\begin{equation}
    (\bm{v}\cdot\bm{n})^{k+1} = \frac{\varphi_i^{k+1}-\varphi_i^{k}}{\Delta t}  \approx \frac{\partial \varphi_i}{\partial t},
    \label{v.ndefwithLSF}
\end{equation} and,
\begin{equation}
    \left\lVert \bm{v}^{k+1}\right\rVert = \frac{\left|\varphi_i^{k+1}-\varphi_i^{k}\right|}{\Delta t}.
    \label{vnormdefwithLSF}
\end{equation}

So, $P_{SD}^{k+1}$ would then become:
\begin{equation}
    P_{SD}^{k+1} = -\frac{\alpha_C x_X^0 \left(\frac{\varphi_i^{k+1}-\varphi_i^{k}}{\Delta t}\right)}{1+\beta_C^2\left( \frac{\varphi_i^{k+1}-\varphi_i^{k}}{\Delta t} \right)^2}.
    \label{P_SD_def_tdiscret_nonlin}
\end{equation}
Clearly the term in the denominator due to $\left\lVert \bm{v}^{k+1}\right\rVert$ manifests the operator $\mathcal{F}$ in Eq.\eqref{FvarphiDefAtTimestepk} and hence the time discretized formulation in Eq.\eqref{lsEqnTimeDdemo} non-linear. To avoid this non-linearity and simplify the resolution procedure, as seen in the works of \cite{furstoss2021role}, the following explicit scheme based assumption is made:
\begin{equation}
    \left\lVert \bm{v}^{k+1}(\bm{x})\right\rVert^2 \approx \left\lVert \bm{v}^{k}(\bm{x})\right\rVert^2 = \left(\frac{\varphi_i^{k}(\bm{x})-\varphi_i^{k-1}(\bm{x})}{\Delta t}\right)^2 = \left\lVert \bm{v}^{old}(\bm{x})\right\rVert^2  \quad \text{ for } i \mid \varphi_i^{k}(\bm{x}) \geq 0,
\end{equation} where $\bm{v}^{k}$ or $\bm{v}^{old}$ is the net migration velocity field prescribed to compute the GLS solution $\varphi_i^k$ at the current state. Even though this formulation deviates slightly from the fully implicit time scheme, for the small ranges of time steps expected to be considered in the GLS resolution, the errors accumulated are expected to be of the small order while considerably simplifying the resolution procedure, thanks to the linearization. Taking this into account, Eq.\eqref{P_SD_def_tdiscret_nonlin} could be rewritten as:
\begin{equation}
    P_{SD}^{k+1} \approx -\left(\frac{\alpha_C x_X^0}{1+\beta_C^2\left\lVert \bm{v}^{old}\right\rVert^2}\right)\left(\frac{\varphi_i^{k+1}-\varphi_i^{k}}{\Delta t}\right).
    \label{P_SD_def_tdiscret_lin}
\end{equation}
This would then linearize the operator $\mathcal{F}$ in Eq.\eqref{FvarphiDefAtTimestepk} and hence the time discretized formulation in Eq.\eqref{lsEqnTimeDdemo}. Although not exactly, but such a strategy of splitting the operator $\mathcal{F}$ into implicit and explicit parts weakly mimics the ideologies of IMplicit-EXplicit (IMEX) time integration methods \cite{crouzeix1980methode, ascher1997implicit}.  
Substituting Eq.\eqref{P_SD_def_tdiscret_lin} into the time discretized formulation and going back to the continuum description in Eq.\eqref{lsCDReqn}, we obtain:
\begin{equation}
\begin{gathered}
\frac{\partial \varphi_i}{\partial t} + \left[\bm{v_{\Delta G_{\gamma\alpha}}}+\bm{v_{\llbracket \mathcal{E} \rrbracket}}\right]_i\cdot\bm{\nabla}\varphi_i +  \sum_{l\in \mathscr{S}}\chi_{l}M_{l}\left(\frac{\alpha_C x_X^0}{1+\beta_C^2\left\lVert \bm{v}^{old}\right\rVert^2}\right)_l\frac{\partial \varphi_i}{\partial t}= \left[\sum_{l\in \mathscr{S}}\chi_{l}M_{l}\sigma_{l}\right]\Delta\varphi_i.
\label{lsCDeqnlin1}
\end{gathered}
\end{equation}
By imposing, 
\begin{equation}
    \mathscr{M}_{SD} = 1+\sum_{l\in \mathscr{S}}\chi_{l}M_{l}\left(\frac{\alpha_C x_X^0}{1+\beta_C^2\left\lVert \bm{v}^{old}\right\rVert^2}\right)_l
    \label{MSDdef},
\end{equation}we obtain:
\begin{equation}
\begin{gathered}
\mathscr{M}_{SD}\frac{\partial \varphi_i}{\partial t} + \left[\bm{v_{\Delta G_{\gamma\alpha}}}+\bm{v_{\llbracket \mathcal{E} \rrbracket}}\right]_i\cdot\bm{\nabla}\varphi_i = \left[\sum_{l\in \mathscr{S}}\chi_{l}M_{l}\sigma_{l}\right]\Delta\varphi_i.
\label{lsCDeqnlin2}
\end{gathered}
\end{equation} Eventually, we can express the above equation into the following linearized convective-diffusive (CD) formulation $\forall i \in \{1,2,...,N_{LS}\}$:
\begin{equation}
\begin{gathered}
\frac{\partial \varphi_i}{\partial t} + \frac{1}{\mathscr{M}_{SD}}\left[\bm{v_{\Delta G_{\gamma\alpha}}}+\bm{v_{\llbracket \mathcal{E} \rrbracket}}\right]_i\cdot\bm{\nabla}\varphi_i - \frac{1}{\mathscr{M}_{SD}}\left[\sum_{l\in \mathscr{S}}\chi_{l}M_{l}\sigma_{l}\right]\Delta\varphi_i = 0.
\label{lsCDeqnlinfinal}
\end{gathered}
\end{equation} In the following, we shall refer $\mathscr{M}_{SD}$ as the solute drag pressure coefficient. When the solute drag effects are absent or negligible, $\mathscr{M}_{SD} \rightarrow 1$, while when they are non-negligible, $\mathscr{M}_{SD} > 1$. In contrast to the previous formulation (Eq.\eqref{lsCDeqn_old}) without the solute drag aspects, the new formulation only seems to involve a factor, $\mathscr{M}_{SD}$, which has the global effect of heterogeneously lowering the interface mobility (or increasing the drag resistance) for migration as a function of the local interface kinetics.

In \cite{chandrappa2023level}, it has been highlighted that a better description for $\bm{v_{\Delta G_{\gamma\alpha}}}$ is essential in the context of polycrystals comprising of multi-junctions, and thus the following description was proposed:
\begin{equation}
    \bm{v_{\Delta G_{\gamma\alpha}}}(\bm{x},t)=\sum_{i=1}^{N_{LS}}\sum_{\substack{j=1 \\ j \neq i}}^{N_{LS}}\chi_{G_i}M_{ij}\exp{\left(-\beta_e|\varphi_j|\right)}\chi_{\gamma\alpha}\Delta G_{\gamma\rightarrow\alpha}\mathscr{F}_s(-\bm{n}_j),
    \label{vdeltaG}
\end{equation}

However, the use of all the $N_{LS}$ level-set functions is not necessary unlike in the case of stored energy component, since $\Delta G_{\gamma\rightarrow\alpha}$ is valid only on the phase interfaces and not on the entire grain/phase interface network. So, an alternative and simple approach could be proposed that takes into account only the secondary basis level-set function representing all the $\alpha$ phase grains, $\varphi_\alpha$:
\begin{equation}
    \bm{v_{\Delta G_{\gamma\alpha}}}(\bm{x},t)= M_{\gamma\alpha}\chi_{\gamma \alpha}\Delta G_{\gamma\rightarrow\alpha}(\bm{x},t)(-\bm{\nabla}\varphi_\alpha)
    \label{vdeltaG_v2}.
\end{equation}

\subsubsection*{Computation of the solute drag pressure coefficient, \texorpdfstring{$\mathscr{M}_{SD}$}{}:}
\label{C2_Comp_MSD}
To compute the solute drag pressure coefficient in Eq.\eqref{MSDdef}:
\begin{equation*}
    \mathscr{M}_{SD} = 1+\sum_{l\in \mathscr{S}}\chi_{l}M_{l}\left(\frac{\alpha_C x_X^0}{1+\beta_C^2\left\lVert \bm{v}^{old}\right\rVert^2}\right)_l \quad \forall l \in\{\gamma\gamma, \gamma\alpha, \alpha\alpha\}, 
\end{equation*} we need the respective laws/ expressions governing $\left\{\alpha_C (T),\beta_C (T)\right\}_{l}$ and $M_l(\bm{x},t,T)$ provided, and $\chi_l$ and $\lVert\bm{v}^{old}\rVert$ need to be computed. The computation of $\chi_l$ functions is already discussed in \cite{chandrappa2023level}. In order to compute $\lVert\bm{v}^{old}\rVert$, let us consider its discrete definition:
\begin{equation}
    \left\lVert \bm{v}^{old}(\bm{x})\right\rVert =\frac{ \left\lvert\varphi_i^{k}(\bm{x})-\varphi_i^{k-1}(\bm{x})\right\rvert}{\Delta t} \quad \text{ for } i \mid \varphi_i^{k}(\bm{x}) \geq 0,
    \label{voldformaldef}
\end{equation}where $\varphi_i^k$ and $\varphi_i^{k-1}$ are the GLS solutions available at the current and the previous states of time, respectively. In practice, the above definition is reformulated as:
\begin{equation}
    \left\lVert \bm{v}^{old}(\bm{x})\right\rVert =\max\limits_{i\in\{1,...,N_{LS}\}} \left(\mathcal{I}_i^k(\bm{x})\frac{\left\lvert\varphi_i^{k}(\bm{x})-\varphi_i^{k-1}(\bm{x})\right\rvert}{\Delta t}\right),
    \label{voldnumcompdef}
\end{equation}where $\mathcal{I}_i^k(\bm{x})$ is used as an indicator function associated to the $i$th GLS function such that $\varphi_i^k(\bm{x})\geq 0$, i.e.,
\begin{equation}
    \mathcal{I}_i^k(\bm{x})=
    \begin{cases}
        1 & \text{if} \quad \varphi_i^k(\bm{x}) \geq 0 \\
        0 & \text{otherwise}
    \end{cases}
    \label{indicatorfuncdef}.
\end{equation}

At the initial state of time ($k=0$), $\lVert v^{old} \rVert=0$, as $\varphi_i^0 = \varphi_i^{-1}$ The solute drag pressure coefficient needs to be computed before the next resolution step of Eq.\eqref{lsCDeqnlinfinal}.

\subsection{ThermoCalc Coupling}

As described in \cite{chandrappa2023level}, the driving pressure related to the diffusive phase transformation ($\Delta G_{\gamma\rightarrow\alpha}$) is prescribed through a linearization of the phase diagram using a thermodynamic database such as ThermoCalc \cite{thermocalc}. The driving pressure is subsequently expressed as a function of the current temperature, and the phase interface concentrations as follows:
\begin{equation}
    \begin{gathered}
        \Delta G_{\gamma \rightarrow \alpha} (T, x^\alpha_C, x^\gamma_C) = \Delta S^{\gamma\alpha}\left[(T^R - T) + 0.5m^{\alpha/(\alpha+\gamma)}_{A-C}\left(x^\alpha_C - \left.x^\alpha_C\right|_{R}\right) \right. \\ + \left. 0.5m^{\gamma/(\alpha+\gamma)}_{A-C}\left(x^\gamma_C - \left.x^\gamma_C\right|_{R}\right) \right],
    \end{gathered}
    \label{dGdpformCperP}
\end{equation} where $C$ is any interstitial element (Carbon for instance). $T^R$ is the reference temperature at which the phase diagram is locally linearized, with $\left(\left.x^\alpha_C\right|_{R}, \left.x^\gamma_C\right|_{R}\right)$ being the corresponding equilibrium concentrations. $m^{\alpha/(\alpha+\gamma)}_{A-C}$, and $m^{\gamma/(\alpha+\gamma)}_{A-C}$ are the local slopes of the linearized phase boundaries, and $\Delta S^{\gamma\alpha}$ is the entropy difference between the two phase at the temperature $T$.

From Eqs.\eqref{cmix} and \eqref{kpdef}, the above description is rewritten in the form of the total concentration variable:
\begin{equation}
\begin{gathered}
\Delta G_{\gamma \rightarrow \alpha} (T, x_C) = \Delta S^{\gamma\alpha}\left[T^R - T + 0.5m^{\alpha/(\alpha+\gamma)}_{A-C}\left(\frac{k_p^Cx_C}{1 + \phi(k_p^C-1)} - \left.x^\alpha_C\right|_{R}\right) \right. \\ + \left. 0.5m^{\gamma/(\alpha+\gamma)}_{A-C}\left(\frac{x_C}{1 + \phi(k_p^C-1)} - \left.x^\gamma_C\right|_{R}\right)\right]
\end{gathered}
\label{dGdpformTC}
\end{equation}

In the context of ternary alloys ($A-C-X$), if we assume ortho-equilibrium (complete local-equilibrium) between the phase for both the solutes $C$ and $X$, the driving pressure could be expressed using the same principles of phase diagram linearization \cite{Mecozzi2007PhaseFM, nitishphd} as follows: 
\begin{equation}
\begin{gathered}
\Delta G_{\gamma \rightarrow \alpha} (T, x_C, x_X) = \Delta S^{\gamma\alpha}\left(T^R - T\right) + 0.5\Delta S^{\gamma\alpha}\sum_{q\in\{C,X\}} \left[m^{\alpha/(\alpha+\gamma)}_{A-q}\left(\frac{k_p^qx_q}{1 + \phi(k_p^q-1)} - \left.x^\alpha_q\right|_{R}\right) \right. \\ +  \left. m^{\gamma/(\alpha+\gamma)}_{A-q}\left(\frac{x_q}{1 + \phi(k_p^q-1)} - \left.x^\gamma_q\right|_{R}\right)\right].
\end{gathered}
\label{dGdpformTCTernaryOE}
\end{equation}
Ortho-equilibrium hypothesis is feasible if both $C$ and $X$ are interstitial solutes or if the rates of diffusion of $C$ and $X$ are fairly consistent.

However, if the ternary alloy is composed of a slow-diffusing substitutional element ($X$) along with a faster diffusing interstitial element ($C$), ortho-equilibrium may not be feasible as it could be difficult to simultaneously satisfy the mass balances for both $C$, and $X$. In such a scenario, a constrained phase equilibria is assumed. Within this research framework, the para-equilibrium (PE) hypothesis \cite{hillert2004definitions} is selected as the constrained phase equilibrium model. According to PE conditions, the slow-diffusing element $X$ is presumed to be completely immobile, allowing only the fast-diffusing element $C$ to redistribute between the involved phases. Thus, local equilibrium is maintained solely for the $C$ element. This hypothesis is generally applicable when the diffusion of $X$ element is not possible within the experimental timeframe or the selected heat treatment. The driving pressure is then derived by extracting the hypothetical phase diagrams obtained under the PE conditions and subsequently linearizing them as previously:
\begin{equation}
\begin{gathered}
\Delta G_{\gamma \rightarrow \alpha}^{PE} (T, x_C) = \Delta S^{\gamma\alpha}_{PE}\left[T^R - T + 0.5m^{\alpha/(\alpha+\gamma)}_{[A-C]_{PE}}\left(\frac{k_p^Cx_C}{1 + \phi(k_p^C-1)} - \left.x^\alpha_C\right|_{R}^{PE}\right) \right. \\ + \left. 0.5m^{\gamma/(\alpha+\gamma)}_{[A-C]_{PE}}\left(\frac{x_C}{1 + \phi(k_p^C-1)} - \left.x^\gamma_C\right|_{R}^{PE}\right)\right],
\end{gathered}
\label{dGdpformTCTernaryPE}
\end{equation}where $m^{\alpha/(\alpha+\gamma)}_{[A-C]_{PE}}$ and $m^{\gamma/(\alpha+\gamma)}_{[A-C]_{PE}}$ are the slopes of the linearized imaginary solvus lines constructed under PE assumptions, while $\left.x^\alpha_C\right|_{R}^{PE}$ and $\left.x^\gamma_C\right|_{R}^{PE}$ are the corresponding constrained equilibrium concentrations at $T^R$.

\begin{figure}[!htb]
	\centering
	\includegraphics[height=16cm]{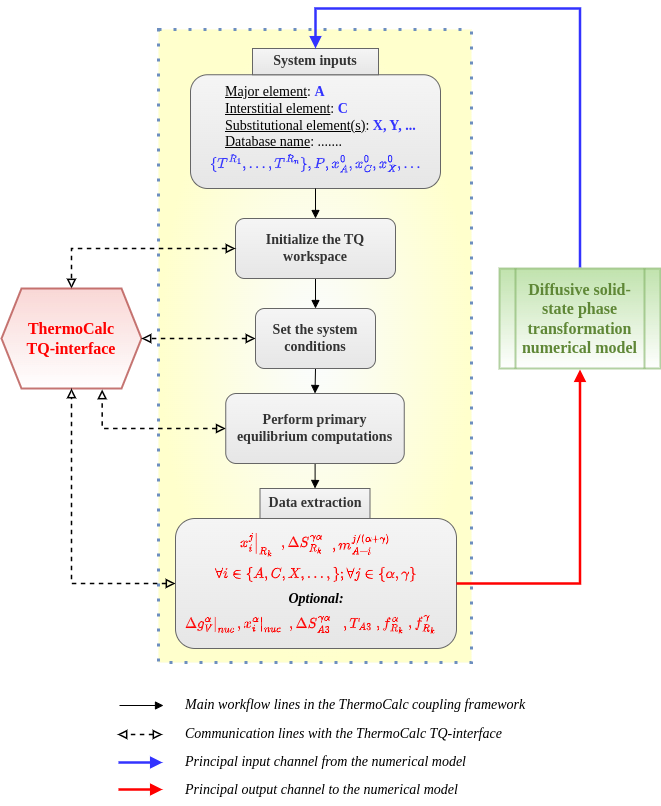}
	\captionsetup{justification=centering,margin=2cm}
	\caption{General overview of the ThermoCalc coupling with the numerical model}
	\label{TCcouplingframework}
\end{figure}

In previous works \cite{chandrappa2023level}, the interaction with ThermoCalc was manual, limiting the number of reference temperatures for data extraction. Furthermore, the ThermoCalc console or its graphical user interface does not facilitate the seamless calculation of certain thermodynamic data under para-equilibrium conditions. These difficulties were overcome by establishing a coupling between the numerical model and the TQ-interface (SDK)
of ThermoCalc to enable automatic computation of relevant thermodynamic data over a large set of reference points under both ortho-equilibrium as well as para-equilibrium conditions. Fig.\ref{TCcouplingframework} gives a general overview of the ThermoCalc coupling framework using the TQ-interface. For a given alloy system, the desired thermodynamic conditions for data extraction are passed as inputs from the numerical model to the ThermoCalc coupling interface. The coupling model then provides a matrix of desired thermodynamic data for a given range of temperatures over a large set of points. The ThermoCalc data extraction is typically performed as a pre-processing step in the numerical framework. The discrete and rich thermodynamic dataset generated is then utilized in the computational loop with the assistance of an interpolation model, enabling the interpolation of necessary thermodynamic data for any temperature at each time step. This approach implies local piecewise linearization of the phase diagram at the current temperature, $T^R=T$. This is crucial for accurately capturing strong topological evolution of the solvus surfaces (or lines) in a phase diagram, particularly for non-isothermal or continuous cooling transformations involving a long thermal path ($\left|T_{initial}-T_{final}\right|$). This helps to enrich the driving pressure description that is based on strong assumptions, and potentially minimize errors. Interested readers are encouraged to refer \cite{nitishphd} for further specifications on the coupling.

\subsection{Nucleation model}

Precisely modeling nucleation events at the mesoscopic scale is complex and often more suitable for much lower scales of modeling, such as atomistic simulations. The literature offers a wide array of hypotheses and parameters aimed at characterizing nucleation events in phase transformation. These parameters encompass various aspects such as the nucleation start temperature ($T_{N_s}$), nucleation temperature range ($\delta T_N$), shield distance ($\delta l_N$), shield time ($\delta t_N$), parameters related to nucleus shape, and composition-based factors influenced by local parent phase features, among others. Additionally, adhering to the Continuous Nucleation Theory (CNT) introduces supplementary parameters such as incubation time and frequency factor. However, due to experimental constraints in observing nucleation events, quantifying most of these parameters independently presents significant challenges. Therefore, in this study, we will employ a simplified approach, aiming for potential refinements in subsequent research. The nucleus shape will be assumed to be spherical, and only a select few parameters from the aforementioned list will be taken into account.

\begin{figure}[!htb]
	\centering
	\includegraphics[height=6cm]{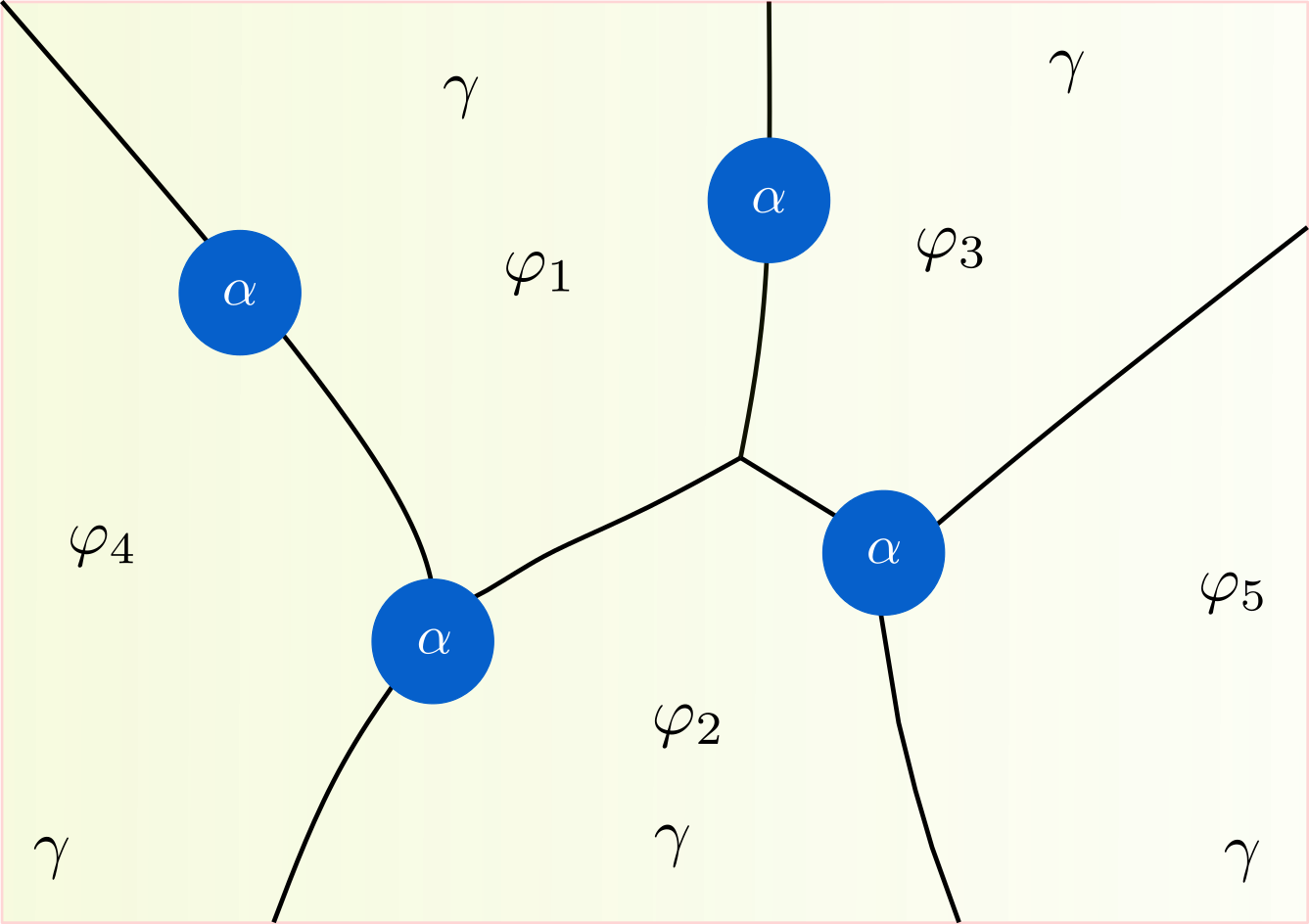}
	\captionsetup{justification=centering,margin=2cm}
	\caption{Illustration of spherical nuclei of $\alpha$ phase (blue) nucleating in a 2D parent $\gamma$ phase}
	\label{NucGBGC}
\end{figure}

The most favored sites for nucleating the new phase are the grain corners (triple junctions in 2D, quadruple junctions in 3D), followed by grain edges and surfaces. At lower cooling rates, grain corners are preferred. However, at higher cooling rates, the nucleation density is expected to increase, raising the likelihood of saturating the grain corners for nucleation. This situation consequently opens up opportunities for nucleation on grain edges and surfaces \cite{offerman2002grain, mecozzi2008role}. Fig.\ref{NucGBGC} gives a 2D representation of nucleation of $\alpha$ phase on grain corners and grain boundaries in the parent $\gamma$ phase.

The nucleation start temperature ($T_{N_s}$) is one of the parameters used to qualify the onset of nucleation of the new phase. As the temperature decreases below the transformation temperature ($T_{A3}$), the potential driving pressure for initiating the stable phase's nucleation typically increases. Considering capillary effects, the likelihood of a critical nucleus forming near the $T_{A3}$ temperature is relatively low. Conversely, as the temperature decreases further and more driving pressure becomes available, the likelihood of nucleation increases. Therefore, it's logical to assume that $T_{N_s}$ is generally set slightly below the transformation temperature. This difference, known as the nucleation undercooling ($\Delta T_N = T_{A3} - T_{N_s}$), is often determined practically by observing the temperature at which a certain initial percentage of the product phase fraction is experimentally detected, as mentioned in literature such as \cite{mecozzi2008role}. 

Nucleation can occur in a site-saturated or continuous manner. In site-saturated nucleation, all nuclei form simultaneously below the nucleation start temperature, while continuous nucleation involves nuclei forming over a certain period. In the context of continuous cooling transformation, the latter mode of nucleation translates to nucleation occurring over a particular temperature span, referred as nucleation temperature range ($\delta T_N$). In a study by Offerman et al. \cite{offerman2002grain} investigating nucleation, the existence of a distinct temperature range is suggested, implying a preference for continuous mode of nucleation over site-saturated scenarios. They also noted a correlation between the cooling rate and this temperature range. Specifically, they observed that $\delta T_N$ becomes more pronounced with higher cooling rates while being relatively smaller at lower cooling rates. $\delta T_N$ is generally assumed to be a modeling parameter \cite{mecozzi2008role} adjusted to align with relevant experimental results.

\begin{figure}[!htb]
	\centering
	\includegraphics[height=6cm]{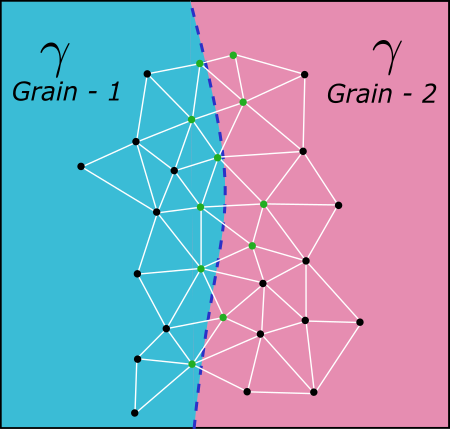}
	\captionsetup{justification=centering,margin=2cm}
	\caption{Characterization of computational nodes for grain boundary nucleation in a 2D context}
	\label{NucGB_sites}
\end{figure}

From a numerical modeling perspective, a site characteristic function ($\chi_{sn}$) is adopted to indicate the sites (and their computational nodes) available for nucleation:
\begin{equation}
    \chi_{sn}(\bm{x},t)=
    \begin{cases}
        1 & \text{ if } \bm{x} \in \text{Choice of nucleation site} \\
        0 & \text{otherwise}
    \end{cases}, \quad \forall t \mid T_{N_s}-\delta T_N < T(t) \leq T_{N_s}
    \label{charfuncNS}.
\end{equation}
If the chosen nucleation site is on the grain boundaries, then $\chi_{sn}(\bm{x},t)=\chi_{\gamma\gamma}$ qualifies all the eligible nodes on the $\gamma$-phase grain boundaries as illustrated in green in Fig.\ref{NucGB_sites}. 

\begin{figure}[!htb]
	\centering
	\includegraphics[height=6cm]{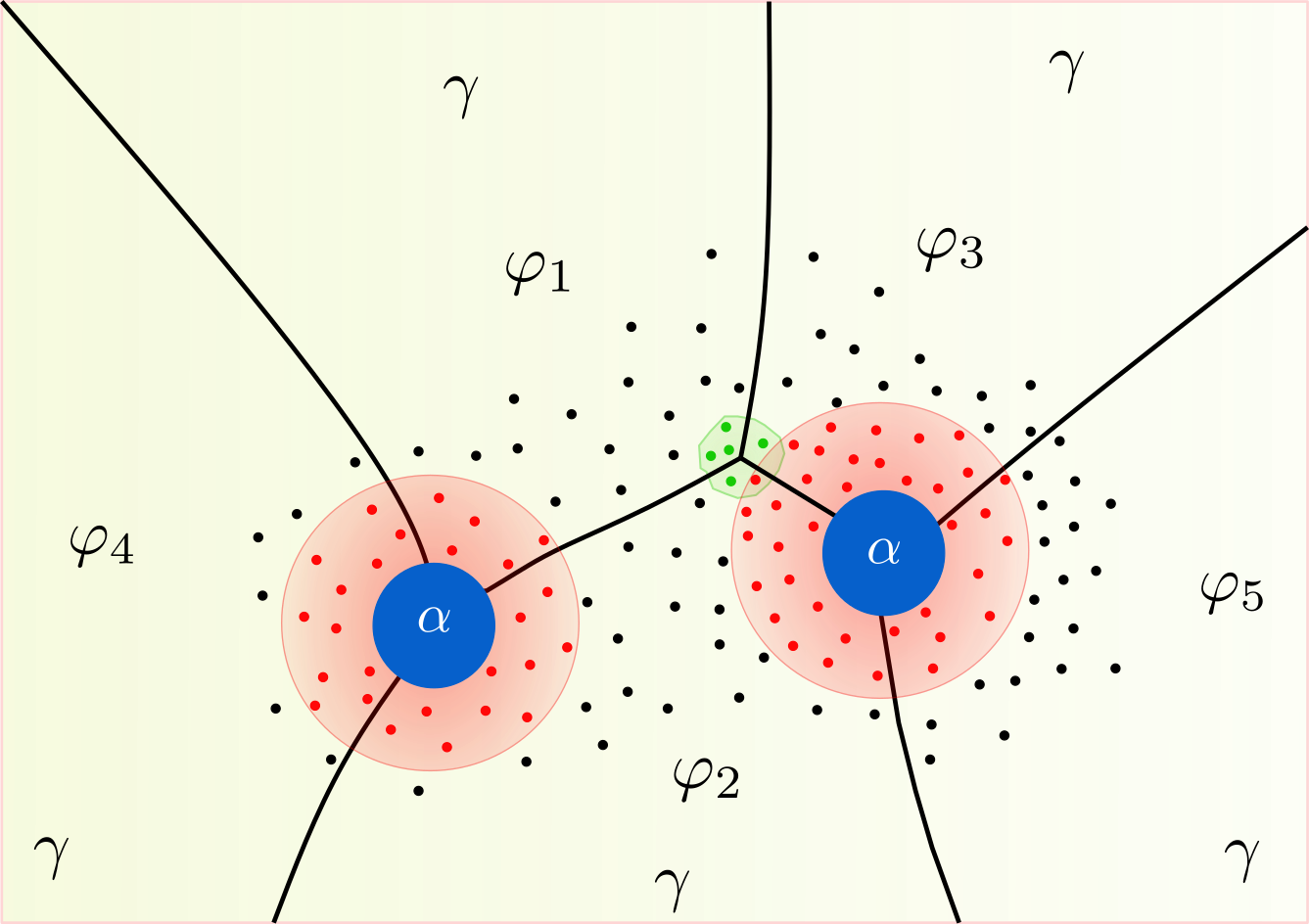}
	\captionsetup{justification=centering,margin=2cm}
	\caption{Characterization of nucleation sites for grain corner nucleation: eligible nodes (green), shielded ineligible nodes (red)}
	\label{MJnucsitedemo}
\end{figure}

Alternatively, when opting for grain corners as the choice of nucleation site, the nodes are qualified as follows:
\begin{algorithmic}
    \Require $\epsilon_{MJ}$, $\varphi_i$
    \State  $\text{count }\gets 0$
    \State  $i \gets 0$
    \For {$\bm{x}$\text{ in }$\Omega$} 
        \While{$i < N_{LS}$}
            \If{$\lvert \varphi_i(\bm{x}) \rvert \leq \epsilon_{MJ}$}
                \State $\text{count }\gets \text{count }+1$
            \EndIf
        \EndWhile
        \If{$\text{count }\geq d_i+1$}\Comment{$d_i$ is the dimensionality}
            \State $\chi_{sn}(\bm{x}) \gets 1$
        \Else
            \State $\chi_{sn}(\bm{x}) \gets 0$
        \EndIf 
    \EndFor
\end{algorithmic}
Here, $\epsilon_{MJ}$ characterizes a small distance threshold around the multiple junction. In 3D, if a node has at least $4$ GLS functions that meet the condition, it qualifies as a multi-junction node available for nucleation. In 2D, the equivalent criterion is fulfilled with $3$ GLS functions. A similar criterion can be applied to qualify grain edge nodes in 3D, where the condition is met by $3$ GLS functions. Fig.\ref{MJnucsitedemo} gives a 2D representation of the small localized region in green around the triple junction satisfying the above condition and hence the nodes within this zone are provisionally qualified as eligible sites. 

The site characteristic function is however utilized in conjunction with a shield distance parameter, $\delta l_N$. As discussed in \cite{mecozzi2008role}, shield distance is an inter-nucleation distance parameter, shielding pre-existing nuclei from the formation of additional ones in their proximity. From a physical perspective, this parameter reflects that upon nucleus formation, the surrounding local region becomes less conducive to further nucleation due to subsequent changes in local characteristics, such as solute concentration. This parameter primarily serves as a modeling parameter due to the challenges in consistently quantifying it, stemming from the lack of substantial physical underpinning. In the current model, at any instant $t$, $\delta l_N$ is taken as, $\delta l_N(t)=k_{sh}r^*_{num}(t)$, where $k_{sh}$ is a shield distance factor, and $r^*_{num}$ is the numerically imposed critical radius of the nucleus. 

In many scenarios, the critical radii of nuclei are significantly smaller compared to the size of the considered domain. Consequently, accurately capturing these small nuclei poses substantial computational demands, leading to them being rarely precisely prescribed at their theoretical critical radii. To address this, a minimum radius is set based on the underlying mesh resolution, while a maximum size is imposed to prevent uncharacteristically large nuclei. Thus, the numerical radius of a nucleus ($r^*_{num}$) in this model is determined by the following expression:
\begin{equation}
    r^*_{num} = \min \left(\max \left(r^*_{th}, k_1h_{min}\right), k_2h_{min} \right)
    \label{Numradnuc},
\end{equation} where $r^*_{th}$ is the theoretically estimated critical radius, $h_{min}$ is the minimum size of the underlying mesh resolution, $k_1$ is a constant governing the minimum number of mesh elements to be included along the radius, and $k_2 (> k_1)$ represents another constant limiting the maximum size permissible for a nucleus. The theoretical critical radius technically depends on the site of nucleation, however in this context, it is roughly estimated as follows:
\begin{equation}
    r^*_{th} = \frac{(d_i-1) \sigma_{\gamma\alpha}}{\Delta g^\alpha_V},
    \label{Theoradnuc}
\end{equation} where $d_i$ denotes the dimensionality, and $\Delta g^\alpha_V(T)$ is the driving pressure for nucleation at the current temperature. This driving pressure can be extracted during the pre-processing stage, thanks to the ThermoCalc coupling, and subsequently interpolated at any temperature as previously discussed.

Therefore, the shield distance parameter, combined with the site characteristic function, further filters out nodes that are ineligible for nucleation (i.e., $\varphi_\alpha (\bm{x}) \geq -\delta l_N$). The shielded zone and the nodes within it are highlighted in red in Fig.\ref{MJnucsitedemo}. So, eventually one of the nodes within the green zone is shielded and hence marked in red as ineligible for nucleation. 

\begin{figure}[!htb]
	\centering
	\includegraphics[height=13cm]{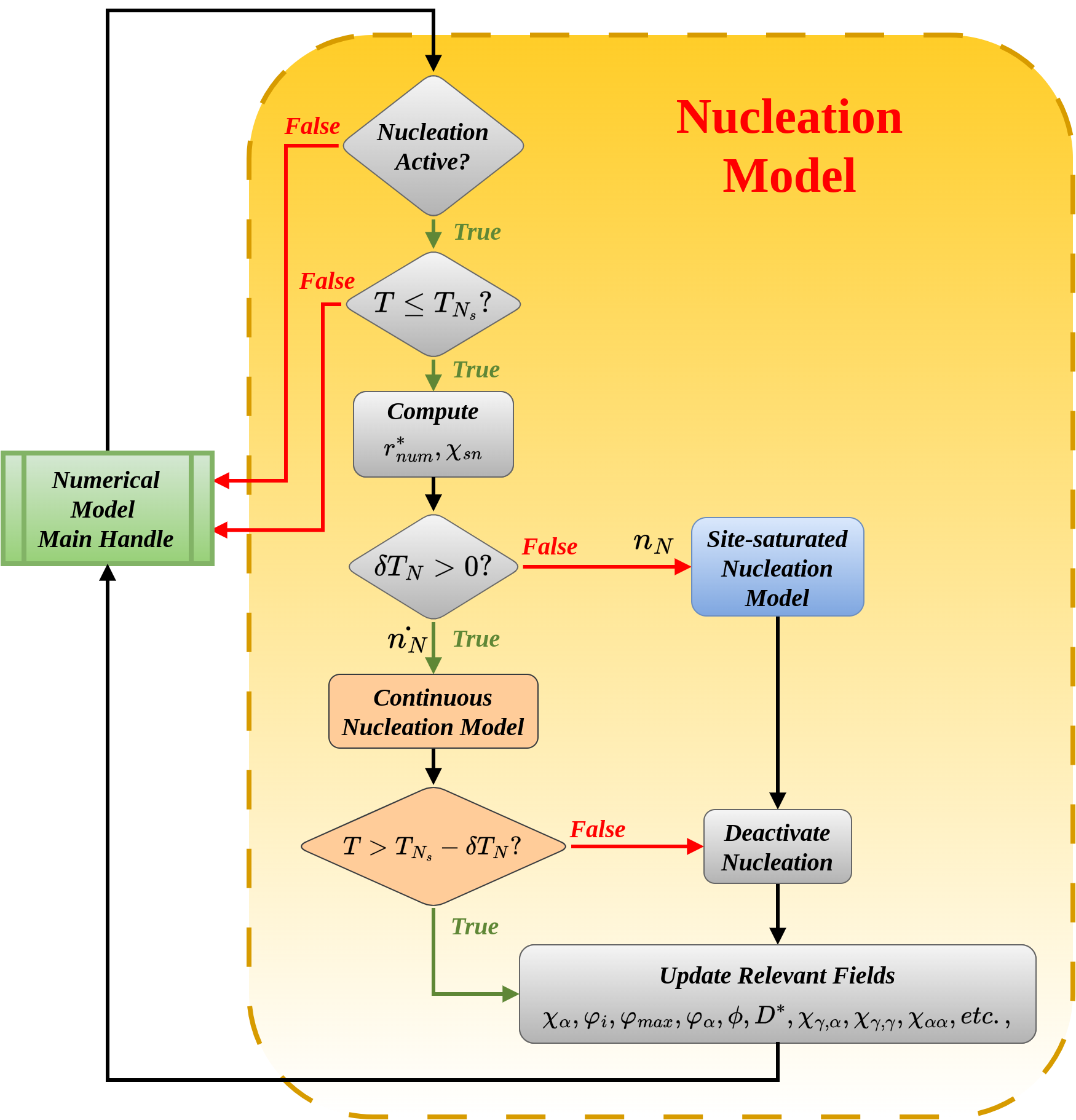}
	\captionsetup{justification=centering,margin=2cm}
	\caption{General workflow of the nucleation model}
	\label{nucmodelflowchart}
\end{figure}

The functioning of the nucleation model adopted within this framework is summarized in Fig.\ref{nucmodelflowchart}.
Currently, the model adopts a constant nucleation rate for the continuous nucleation scenario. The implementation of Continuous Nucleation Theory (CNT) has not been incorporated into this work. The nucleation rate ($\dot{n_N}$) is estimated using the cooling rate $\dot{T}$, the imposed nucleation density ($\rho_N$) and the nucleation temperature range as follows:
\begin{equation}
    \dot{n_N} = \frac{\rho_N V_0\dot{T}}{\delta T_N},
    \label{nucrate}
\end{equation} where $V_0$ represents the total volume of the domain (or area $A_0$ in 2D). The nucleation density could be approximated using the expected final fraction of the product phase (e.g., $f^\alpha_f$) and its anticipated average grain size ($r_\alpha^f$) as outlined in \cite{mecozzi2008role}:
\begin{equation}
    \rho_N = \frac{3f^\alpha_f}{4\pi (r_\alpha^f)^3}.
    \label{nucdensity}
\end{equation}
In 2D, the above relation is modified to $\rho_N = \frac{f^\alpha_f}{\pi (r_\alpha^f)^2}$.

Upon the creation of a nucleus, its introduction triggers a local change in the phase status, subsequently altering the local diffusivity field. This alteration allows the solute redistribution equation to naturally adjust the local composition. Thus, the local composition isn't manually altered upon nucleus introduction to avoid tampering with the solute mass conservation. At the moment of creation, the newly formed nucleus is dynamically integrated into one of the existing GLS functions, $\varphi_i$, or a new function is created if needed. As the local phase features are updated, the associated dependent fields, including the secondary basis function $\varphi_\alpha$, and the diffuse interface function ($\phi$), etc., are correspondingly modified.

\subsection{Overview of the numerical framework}

\begin{figure}[!htb]
	\centering
	\includegraphics[height=16cm]{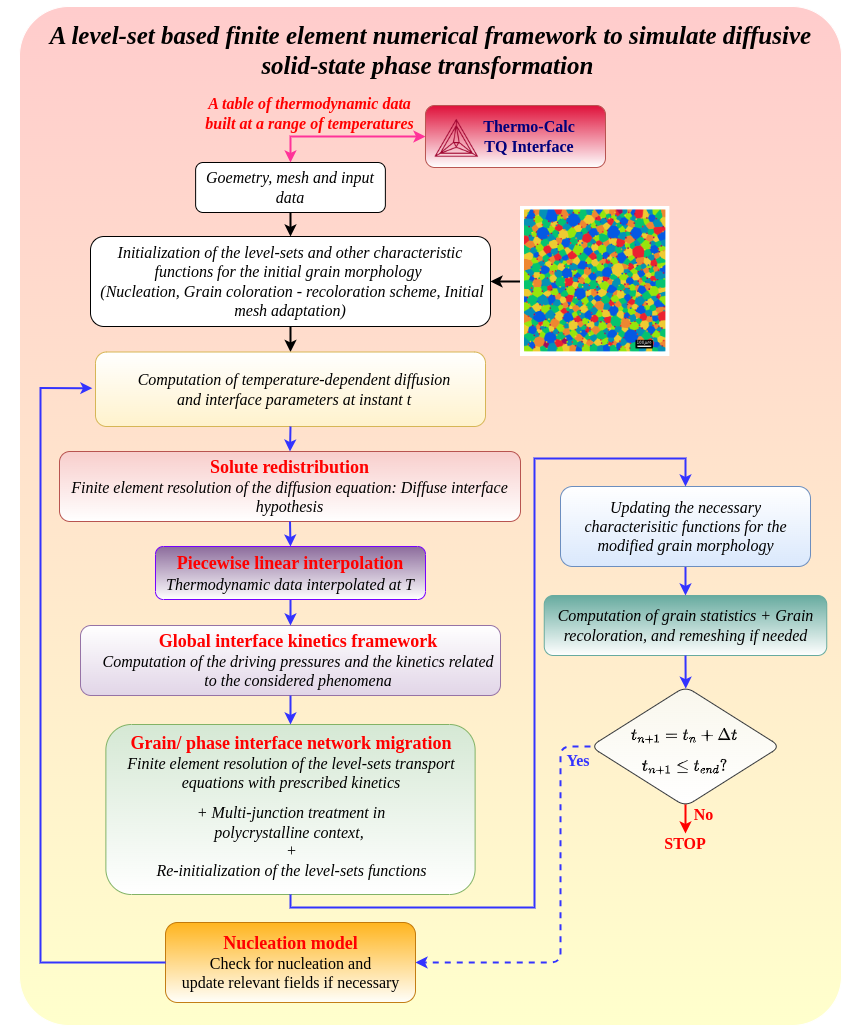}
	\captionsetup{justification=centering,margin=2cm}
	\caption{General overview of the numerical framework}
	\label{GenNumframework}
\end{figure}

Fig.\ref{GenNumframework} presents a general outline of the numerical framework highlighting the key stages and tasks involved. The numerical simulations are carried out in a FE context using unstructured triangular meshes with a P1 interpolation, and employing an implicit backward Euler time scheme for the time discretization. Each system associated with Eq.\eqref{diffWF} and the weak formulation of Eqs.\eqref{lsCDeqnlinfinal} is assembled using typical P1 finite elements with a Streamline Upwind Petrov-Galerkin (SUPG) stabilization for the convective terms \cite{Brooks1982}. A particular adaptive meshing/ remeshing strategy is adopted along the grain and interphase boundaries, outlined in \cite{chandrappa2023level}. The boundary conditions applied to the GLS functions ($\varphi_i$) are classical null Neumann boundary conditions, while the total solute concentration field is imposed with Robin-type conditions. It should be highlighted that, post the resolution of the LS transport equations, a particular numerical treatment \cite{Merriman1994MotionOM} is performed on the $\varphi_i$ functions to avoid any kinematic incompatibilities at the multiple junctions:

\begin{equation}
    \hat{\varphi_i}=\frac{1}{2}\left(\varphi_i-\max_{j\neq i}\varphi_j\right) \quad \forall i\in \{1,...,N_{LS}\}.
\label{mjtreat}
\end{equation}

It is then followed by a reinitialization procedure of the GLS functions using a recent efficient strategy \cite{shakoor2015efficient} to restore the metric properties of a signed distance function. 

\section{Results and discussion}
\label{resdisc}

\subsection{Benchmarking}
To validate this numerical framework, the semi-analytical mixed-mode (SAM) sharp interface model proposed by Bos et al. \cite{bos2007mixed} for diffusive phase transformation is used. Furthermore, the results are compared with those obtained from a similar analysis conducted by Mecozzi et al. \cite{mecozzi2011quantitative} using their phase-field numerical-model (PF-NM) framework.

\begin{figure}[!htb]
	\centering
	\includegraphics[width=7cm]{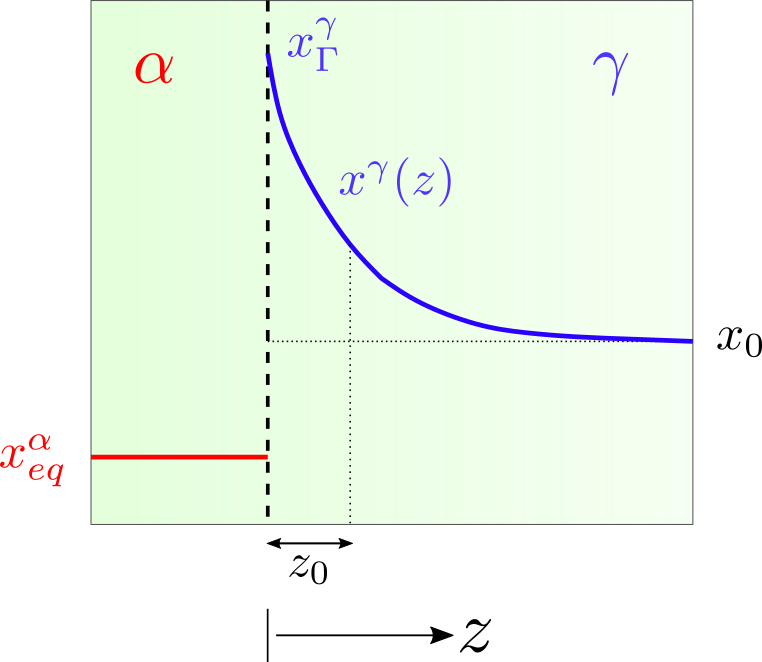}
	\captionsetup{justification=centering,margin=2cm}
	\caption{Solute concentration profile assumed in the semi-analytical model by Bos et al. \cite{bos2007mixed}}
	\label{bosSAcprofdemo}
\end{figure}

In the context of $\gamma \rightarrow \alpha$ phase transformation, the semi-analytical formulation assumes an analytical function of position in a semi-infinite domain to model the solute concentration profile in front of the $\alpha / \gamma$ interface. The assumed concentration profile is then constrained by flux and mass balance relations. The diffusion in the product $\alpha$ phase is assumed to be instantaneous, and hence the concentration profile within this phase is set to be homogeneous at its equilibrium concentration. Fig.\ref{bosSAcprofdemo} depicts the representation  of the concentration profile in a 1D domain within the sharp interface based semi-analytical model. A single $\alpha$ phase nucleus is assumed to form in the infinite $\gamma$ phase matrix.

The driving pressure for phase transformation is assumed to be proportional to the deviation of $\gamma$ phase interface concentration ($x^\gamma_\Gamma$) from its equilibrium concentration ($x^{\gamma}_{eq}$):
    \begin{equation}
    \Delta G_{\gamma \rightarrow \alpha}=\Upsilon\left(x^{\gamma}_{eq}-x^{\gamma}_\Gamma\right),
    \label{deltaGdbosSAdescr}
\end{equation}where $\Upsilon$ is a temperature dependent proportionality factor derived from thermodynamic databases. Interested readers are referred to \cite{bos2007mixed, nitishphd} for further details on the formulation and the resolution procedure.

\subsubsection*{Simulation conditions:}
An isothermal austenite decomposition ($\gamma \rightarrow \alpha$) in a steel alloy (Fe - \SI{0.1}{\wtpercent}C - \SI{0.5}{\wtpercent}Mn) at \SI{1000}{\kelvin} is considered for benchmarking. This scenario mirrors the case studied by Mecozzi et al. \cite{mecozzi2011quantitative} in their comparison of a phase-field numerical model with the aforementioned semi-analytical method. The phase transformation is assumed to occur under the para-equilibrium (PE) hypothesis. Notably, this study does not consider the solute drag effects potentially induced by the Mn element, and the capillarity effects at the phase interface are neglected for this analysis. For this particular case study, following \cite{mecozzi2011quantitative}, we impose an additional hypothesis of equal undercoolings in both the phases when deriving the driving pressure via phase diagram linearization. This added assumption enables the use of a coherent thermodynamic description between the sharp interface model and the numerical model. So, the driving pressure description in Eq.\eqref{dGdpformCperP} for the numerical model could be simplified to the following form:
\begin{equation}
    \Delta G_{\gamma \rightarrow \alpha} = \Delta S^{\gamma\alpha}m^{\gamma/(\alpha+\gamma)}_{Fe-C} \left(x^\gamma_C - \left.x^\gamma_C\right|_{eq}\right),
    \label{SAMhypo_Num_DG_Descr}
\end{equation} where $x^\gamma_C$ is the interface concentration in the $\gamma$ phase. By comparing the above equation with the driving pressure formulation of the semi-analytical sharp interface model as described in Eq.\eqref{deltaGdbosSAdescr}, we can deduce that, $\Upsilon = -m^{\gamma/(\alpha+\gamma)}_{Fe-C}\Delta S^{\gamma\alpha}$. The above derivation under the equal undercooling assumption is thoroughly outlined in \cite{mecozzi2011quantitative, nitishphd}.

The required mobility, carbon diffusivity, and thermodynamic data have been adopted from \cite{mecozzi2011quantitative} to maintain consistency with the comparison. Therefore, the temperature dependency of phase interface mobility and carbon diffusivity in the austenite phase has been considered based on an Arrhenius-type law as follows:
\begin{equation}
    \begin{gathered}
        M_{\gamma\alpha}=3.5\times10^{17}\exp\left(\frac{-Q_m}{RT}\right) \quad in \quad \SI{}{\micro\meter\tothe{4}\per\joule\per\second}\\
        D_\gamma^C=1.5\times10^7\exp\left(\frac{-Q^C_\gamma}{RT}\right) \quad in \quad \SI{}{\micro\meter\tothe{2}\per\second}
    \end{gathered},
    \label{BosSAMobDiffDef}
\end{equation} where $Q_m=\SI{140}{\kilo\joule\per\mole}$, and $Q^C_\gamma=\SI{142.1}{\kilo\joule\per\mole}$.

The thermodynamic data extracted according to \cite{mecozzi2011quantitative}, with a reference temperature of $T^R = \SI{1073}{\kelvin}$, have been tabulated in Tab.\ref{SAMbosTDdata}. These data have been derived by linearizing the pseudo-binary Fe-C phase diagrams under the para-equilibrium hypothesis.

\begin{table}[!htb]
\centering
\begin{tabular}{@{} l *{5}{C{6em}} @{}}
\specialrule{.2em}{.1em}{.1em}
\toprule
 $T^R$ (\SI{}{\kelvin}) & $\left.x_C^{\alpha}\right|_R$ (\SI{}{\wtpercent})& $\left.x_C^{\gamma}\right|_R$ (\SI{}{\wtpercent}) & $\Delta S^{\gamma\alpha}$ (\SI{}{\joule\per\kelvin\per\micro\meter\cubed})& $m^{\alpha/(\alpha+\gamma)}_{Fe-C}$ (\SI{}{\kelvin\per\wtpercent}) & $m^{\gamma/(\alpha+\gamma)}_{Fe-C}$ (\SI{}{\kelvin\per\wtpercent})\\ 
 [0.5ex]
 \hline
 $1073$ & $0.009$ & $0.279$ & $3.46\times10^{-13}$ & $-10250$ & $-186.2$ \\ [1ex] 
 \bottomrule
\specialrule{.2em}{.1em}{.1em}
 \end{tabular}
 \captionsetup{justification=centering, margin=2.5cm}
 \caption{ThermoCalc data extracted at $T^R=\SI{1073}{\kelvin}$ from \cite{mecozzi2011quantitative} for Fe - \SI{0.1}{\wtpercent}C - \SI{0.5}{\wtpercent}Mn}
 \label{SAMbosTDdata}
\end{table}

In this comparison, two scenarios are examined: (i) A pseudo-1D domain measuring \SI{100}{\micro\meter} in length (with a relatively small and insignificant breadth of \SI{0.15}{\micro\meter}), and (ii) A square 2D domain of size $100\times100$ \SI{}{\micro\meter\squared}. Initially, the domain is entirely austenitic. Upon instantaneous cooling below the $T_{A3}$ temperature to \SI{1000}{\kelvin}, a ferrite nucleus is allowed to grow at the center of the domain for a duration of \SI{200}{\second}. In the LS numerical-model (LS-NM), the initial ferrite nucleus radius ($r_\alpha^i$) is set to \SI{0.25}{\micro\meter} due to numerical restrictions associated with the FE mesh size (whereas it starts from \SI{0}{\micro\meter} in the SAM). For the LS-NM, the initial ferrite nucleus composition is established at \SI{0.001832}{\wtpercent}C, which is the critical nucleus composition at $T=\SI{1000}{\kelvin}$ under the para-equilibrium (PE) constraints, extracted from ThermoCalc. However, in the SAM, the nucleus composition is directly set to the equilibrium composition for the ferrite phase (\SI{0.016013}{\wtpercent}C) under PE, assuming instantaneous diffusion in this phase. Throughout the numerical simulation, a uniform time step of $\Delta t=\SI{0.01}{\second}$ is utilized. An adaptive meshing/remeshing strategy is employed with a diffuse interface thickness of $\eta = \SI{0.3}{\micro\meter}$ for the 1D case and $\eta = \SI{1}{\micro\meter}$ for the 2D case. The mesh resolution within the diffuse interface ($h_{min}$) is specified as \SI{5}{\nano\meter} for the 1D case and \SI{70}{\nano\meter} for the 2D case.

\subsubsection*{Results:}

\begin{figure}[!htbp]
	\centering
	\captionsetup{justification=centering,margin=2cm}
	\subfloat[$t=\SI{0}{\second}$]{\includegraphics[width=15cm]{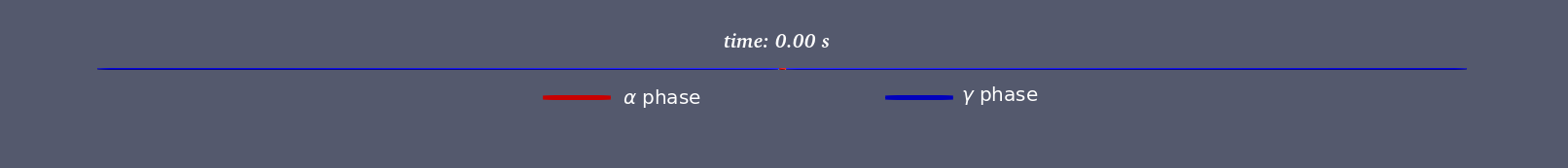}\label{PhaseDistt0LSNMMec1D}} \\
	\subfloat[$t=\SI{200}{\second}$]{\includegraphics[width=15cm]{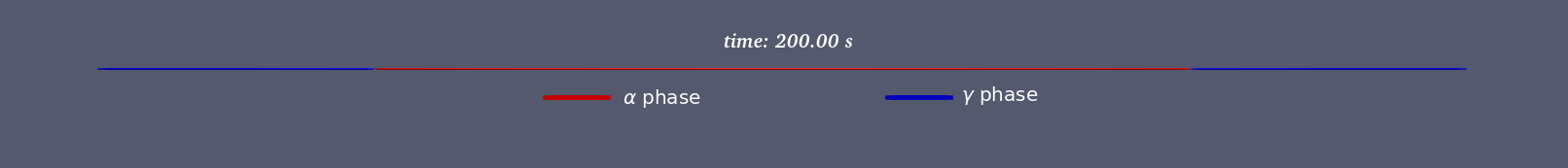}\label{PhaseDistt200LSNMMec1D}}
	\caption{Phase distribution obtained by the level-set based numerical model - 1D case}
	\label{PhaseDistLSNMMec1D}
\end{figure}

\begin{figure}[!htbp]
	\centering
	\captionsetup{justification=centering,margin=2cm}
	\subfloat[$t=\SI{0}{\second}$]{\includegraphics[width=6cm]{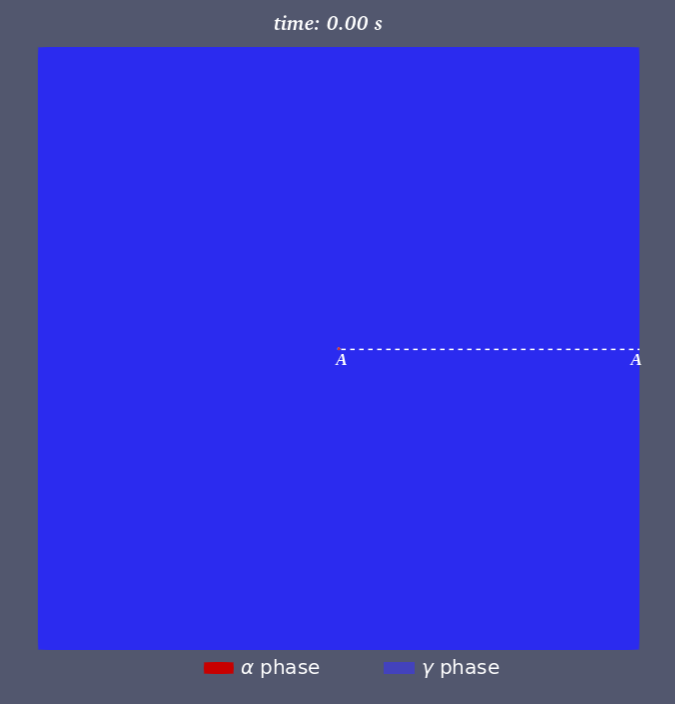}\label{PhaseDistt0LSNMMec2D}}
        \hspace{1em}
	\subfloat[$t=\SI{200}{\second}$]{\includegraphics[width=6cm]{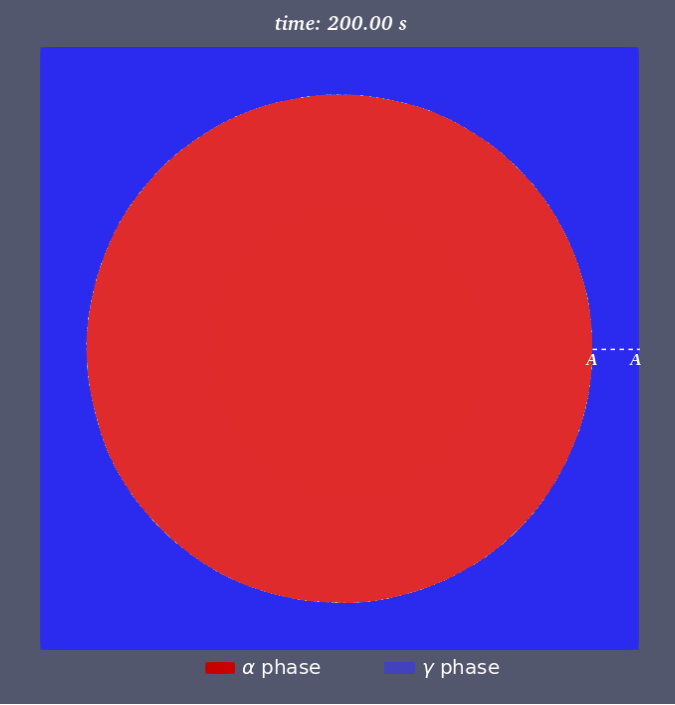}\label{PhaseDistt200LSNMMec2D}}
	\caption{Phase distribution obtained by the level-set based numerical model - 2D case}
	\label{PhaseDistLSNMMec2D}
\end{figure}

Figs.\ref{PhaseDistLSNMMec1D} and Figs.\ref{PhaseDistLSNMMec2D} illustrate the phase distribution at the initial state and at the end of $t=\SI{200}{s}$, obtained by the LS-NM for the 1D and the 2D cases, respectively. 

\begin{figure}[htbp]
	\centering
	\includegraphics[width=12cm]{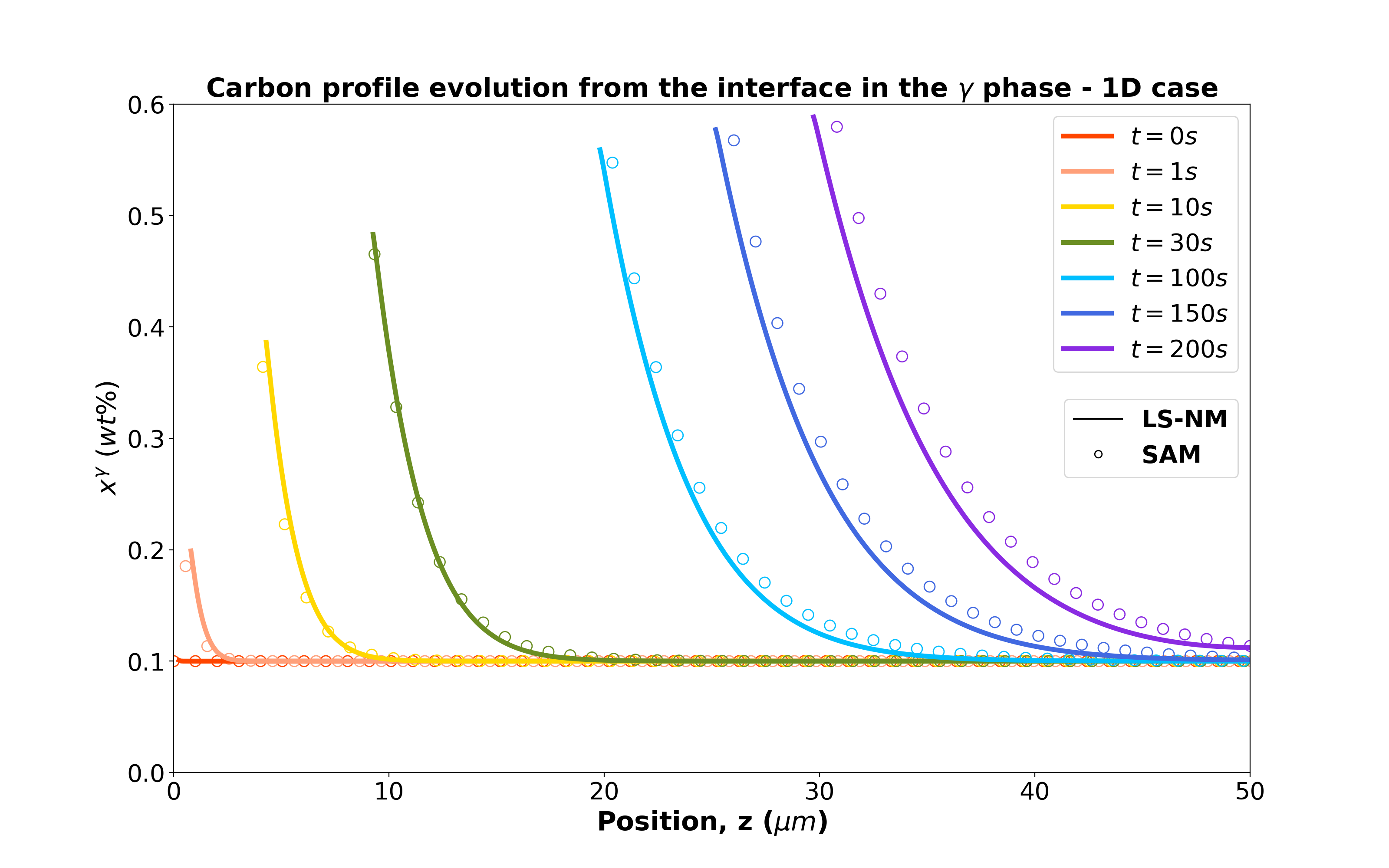}
	\captionsetup{justification=centering,margin=2cm}
	\caption{Evolution of the carbon concentration profiles in the austenite phase - 1D case}
	\label{CgammaProfCompMec1D}
\end{figure}

\begin{figure}[htbp]
	\centering
	\includegraphics[width=12cm]{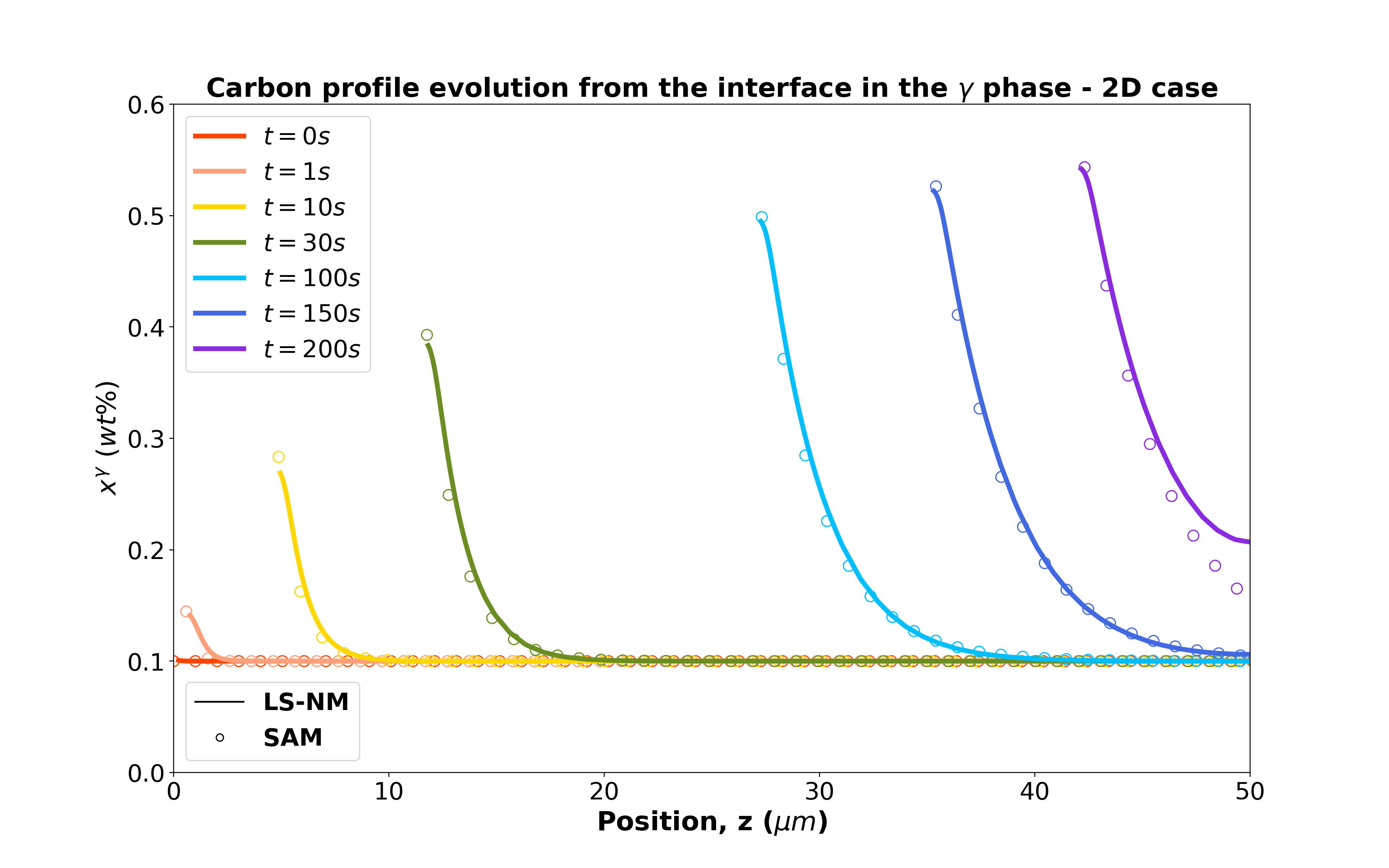}
	\captionsetup{justification=centering,margin=2cm}
	\caption{Evolution of the carbon concentration profiles in the austenite phase - 2D case analyzed along the A-A line described in Fig.\ref{PhaseDistLSNMMec2D}}
	\label{CgammaProfCompMec2D}
\end{figure}

Fig.\ref{CgammaProfCompMec1D} depicts a comparison between the LS-NM and the SAM concerning the evolution of carbon concentration profiles ahead of the $\alpha / \gamma$ phase interface within the austenite side at different time intervals for the 1D scenario. The solid lines denote the numerical solution, whereas the open circular markers denote the semi-analytical solution. Similarly, Fig.\ref{CgammaProfCompMec2D} presents this comparison for the 2D case.

The curves in Fig.\ref{CgammaProfCompMec1D} and Fig.\ref{CgammaProfCompMec2D} demonstrate good agreement at different instants between the LS-NM and the SAM in predicting the solute concentration profiles in the austenite phase for both the 1D and the 2D cases (analyzed along the A-A line described in Fig.\ref{PhaseDistLSNMMec2D}). In the 2D case, a slight deviation between the two models at $t=\SI{200}{\second}$ is noticeable. This discrepancy likely stems from the effects of boundary conditions in the numerical model as the phase interface approaches the boundaries of the 2D domain. In contrast, in the 1D case, the phase interface predicted by the numerical model at $t=\SI{200}{\second}$ is further away from the boundaries, thus unaffected by boundary conditions. It's important to note that while the semi-analytical model operates within a semi-infinite domain, the numerical model operates within a bounded domain with imposed boundary conditions. Analyzing these concentration profiles reveals the mixed-mode nature of the transformation kinetics. It's clear that the interfacial concentration doesn't immediately attain the local equilibrium concentration, leading to finite interface migration. At the same time, diffusion in the bulk of the austenite phase doesn't occur instantaneously, resulting in a gradient between the interfacial concentration and the bulk concentration.

\begin{figure}[htbp]
	\centering
	\captionsetup{justification=centering,margin=2cm}
	\subfloat[1D case]{\includegraphics[width=8cm]{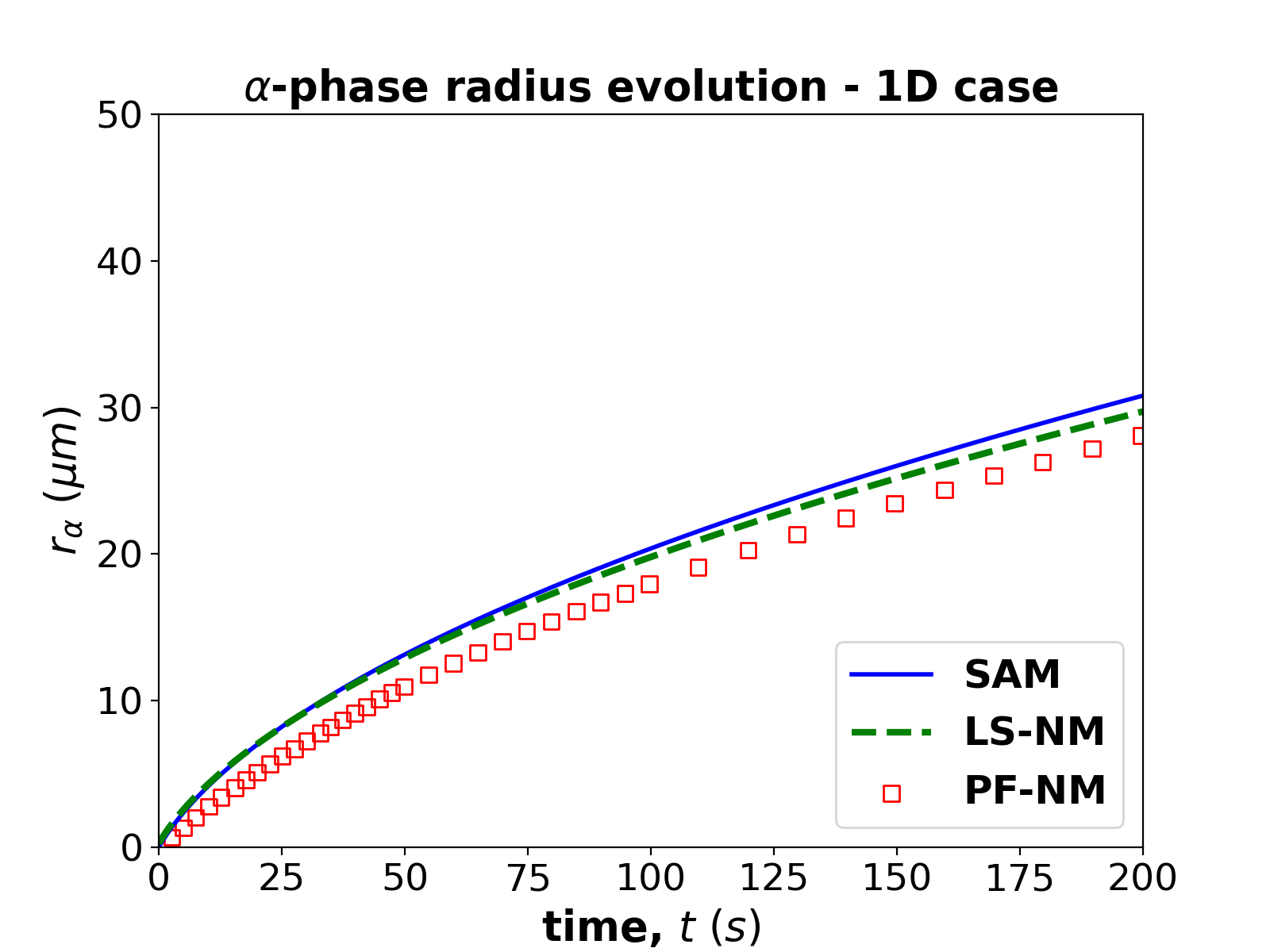}\label{RadEvolMec1D}}
	\subfloat[2D case]{\includegraphics[width=8cm]{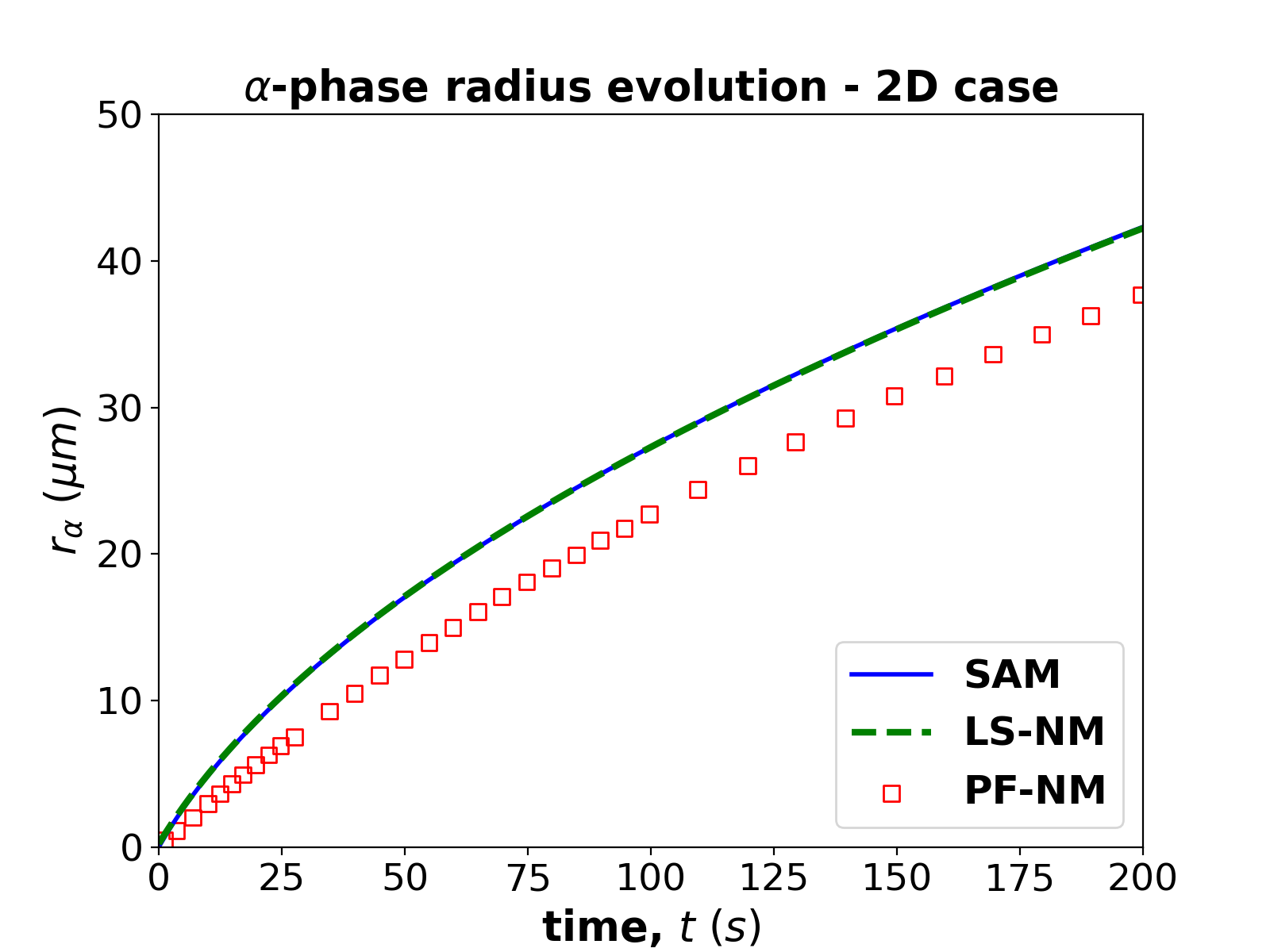}\label{RadEvolMec2D}}
	\caption{Comparison of the evolution of phase interface predicted by different models}
	\label{RadEvolMec2011}
\end{figure}

\begin{figure}[htbp]
	\centering
	\captionsetup{justification=centering,margin=2cm}
	\subfloat[1D case]{\includegraphics[width=8cm]{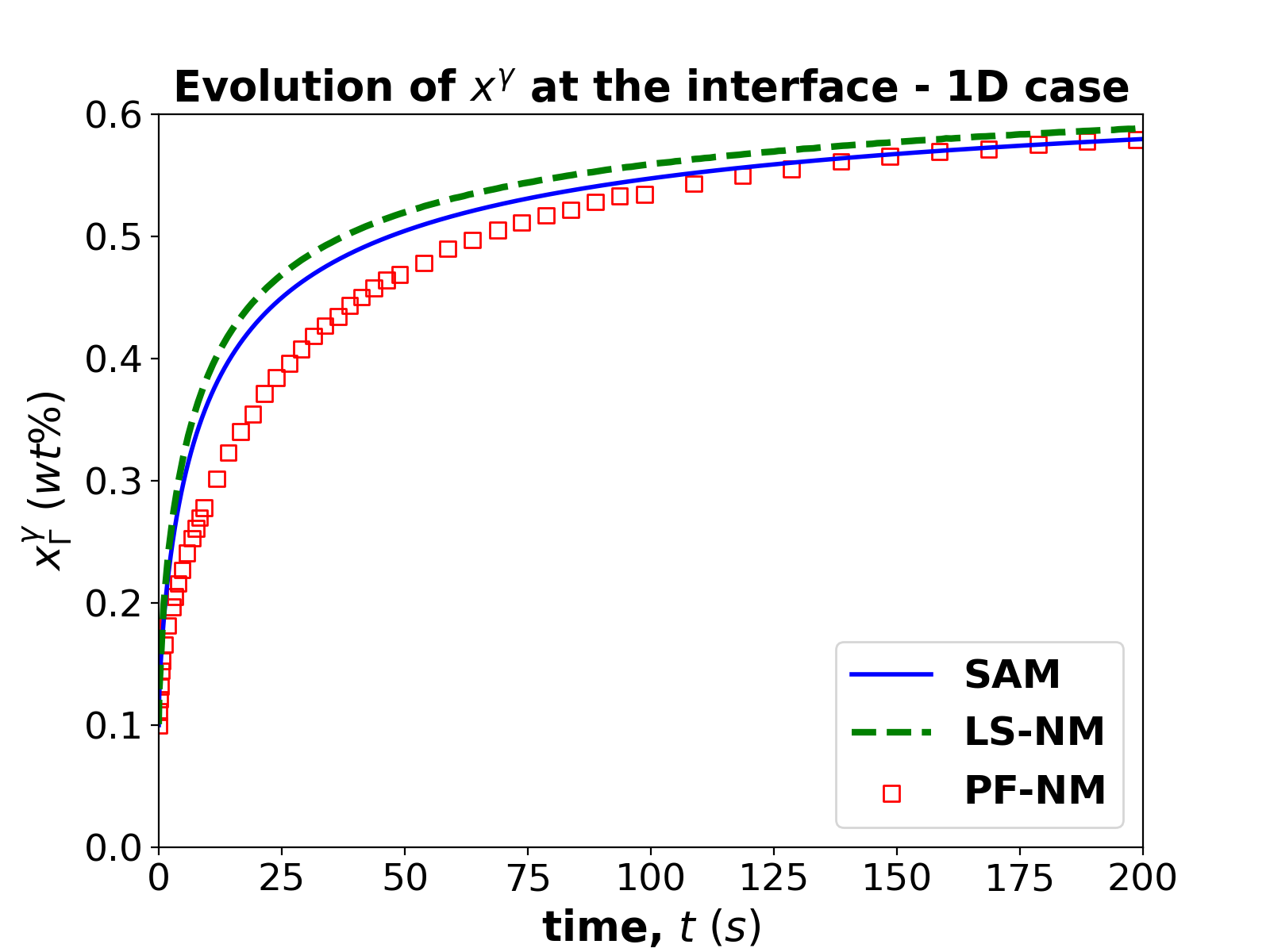}\label{CgintEvolMec1D}}
	\subfloat[2D case]{\includegraphics[width=8cm]{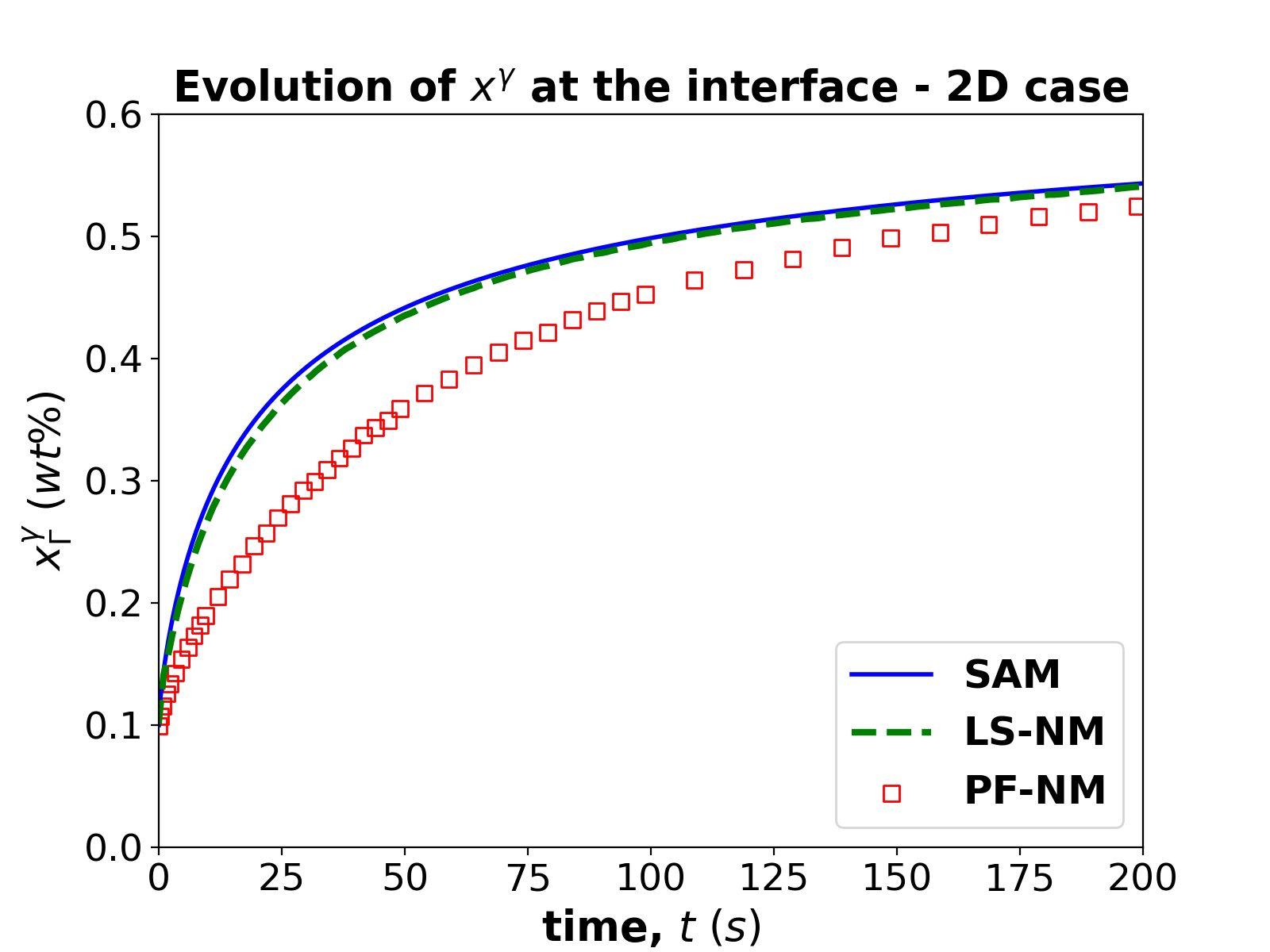}\label{CgintEvolMec2D}}
	\caption{Comparison of the evolution of interfacial carbon concentration in the austenite side as predicted by different models}
	\label{CgintEvolMec2011}
\end{figure}

In Figs.\ref{RadEvolMec2011}, the evolution of the radius of the ferrite phase nucleus between the SAM, LS-NM, and the PF-NM is compared. On the other hand, the evolution of the average interfacial carbon concentration in the austenite phase across these different models is compared in Figs.\ref{CgintEvolMec2011}. The PF-NM results used in these plots were sourced from \cite{mecozzi2011quantitative}, where an identical analysis was performed. These curves clearly show that the predictions from the LS-NM closely match those of the sharp interface SAM, in contrast to the PF-NM, which predicts slower transformation kinetics. Although the PF-NM predicts relatively lower interfacial carbon concentrations, which theoretically should translate to more driving pressure during the transformation, this does not lead to faster interface kinetics. In the phase-field method, the migration of interface kinetics is formulated using an effective or numerical interface mobility, which depends on the diffuse interface thickness parameter and the physical interface mobility. As highlighted in \cite{mecozzi2011quantitative}, during the phase-field simulation, the effective interface thickness is not constant and is lower than the initially imposed value. In contrast to the phase-field method, the level-set numerical method relies solely on the physical interface mobility to dictate interface migration kinetics. Since the migration of level-sets is theoretically synonymous to that of the sharp interface method, only the velocity field over the iso-zero contour of the level-set ($\varphi_\alpha=0$) is of prime importance. This differs from PF-NM, where kinetics are considered across the entire diffuse interface. This sharp interface characteristic of LS-NM in interface migration could perhaps reflect the closer resemblance observed in its predictions compared to the fully diffuse approach of PF-NM.

However, Mecozzi et al. \cite{mecozzi2011quantitative} demonstrate in the context of PF-NM that enhancing mesh resolution and ensuring an adequate number of discretization points within the diffuse interface indeed leads to better convergence in their model's predictions.
\subsection{2D biphasic polycrystal cases}

\subsubsection*{Binary alloy: A step towards a realistic large scale case}

\begin{figure}[!htbp]
	\centering
	\includegraphics[width=6cm]{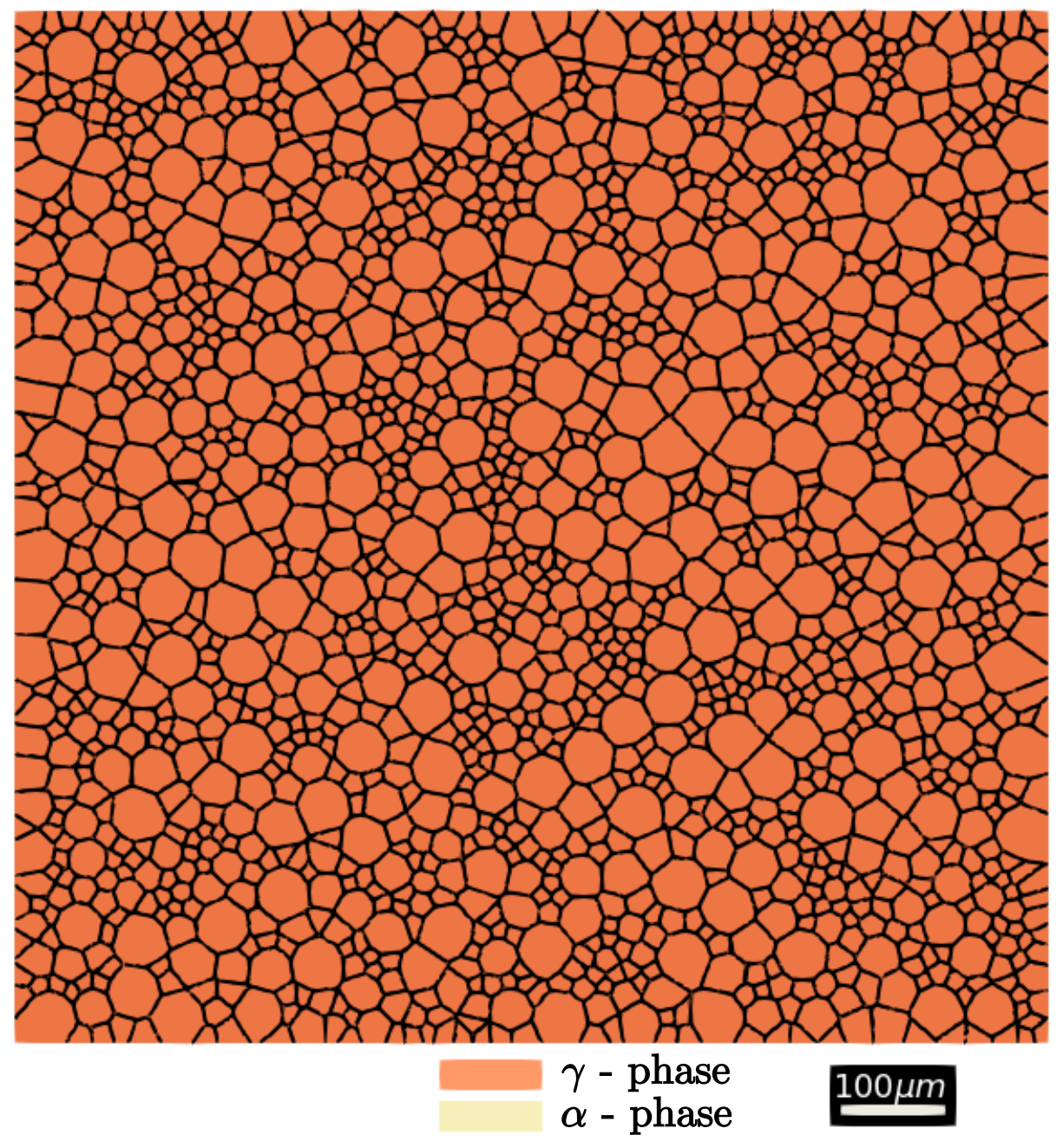}
	\captionsetup{justification=centering,margin=2cm}
	\caption{Illustration of the initial austenitic microstructure with $1592$ grains (the grain interfaces are highlighted in black)}
	\label{LSC_micro_init}
\end{figure}

\begin{figure}[htbp]
	\centering
	\includegraphics[width=8cm]{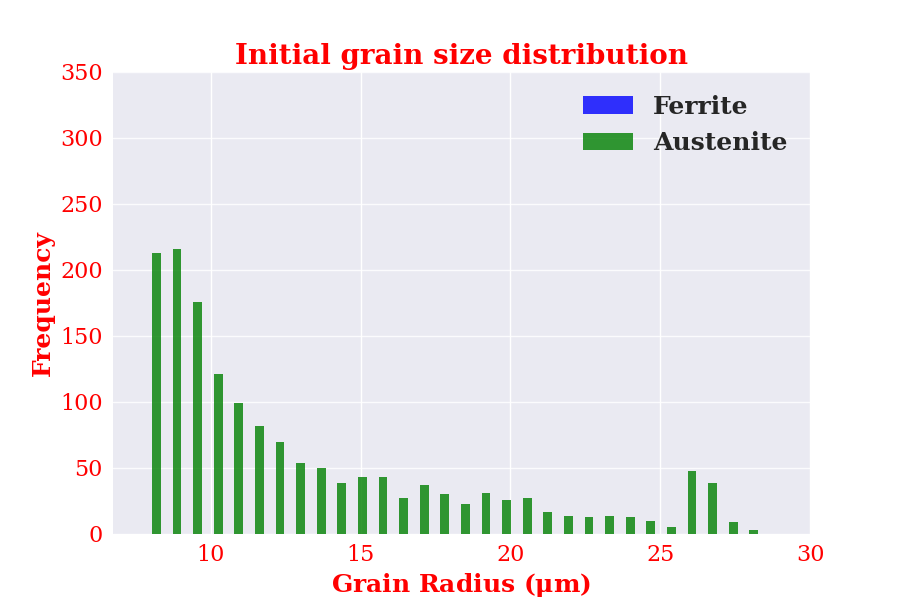}
	\captionsetup{justification=centering,margin=2cm}
	\caption{Grain size distribution of the initial large-scale microstructure}
	\label{LSC_GSD_Initi}
\end{figure}

We now examine austenite decomposition in a binary steel alloy (Fe - \SI{0.02}{\wtpercent}C) within a large-scale microstructure, emulating the complexity often encountered in industrial settings. At atmospheric pressure, the austenitization temperature for this alloy is approximately $T_{A3}=\SI{1175}{\kelvin}$. We consider an initial polycrystalline microstructure in a square domain of size \SI{1}{\milli\meter}, situated slightly above the $T_{A3}$ temperature with $T^i=\SI{1176}{\kelvin}$. At this stage, the microstructure is composed entirely of austenite grains. Fig.\ref{LSC_micro_init} illustrates the morphology of the initial microstructure, featuring $1592$ austenite grains. The initial grain morphology has been obtained by respecting the grain size distribution (evaluated for each grain as the radius of a circle of equivalent area) depicted in Fig.\ref{LSC_GSD_Initi} with an average grain radius of about \SI{14}{\micro\meter}. 

\begin{figure}[htbp]
	\centering
	\includegraphics[width=8cm]{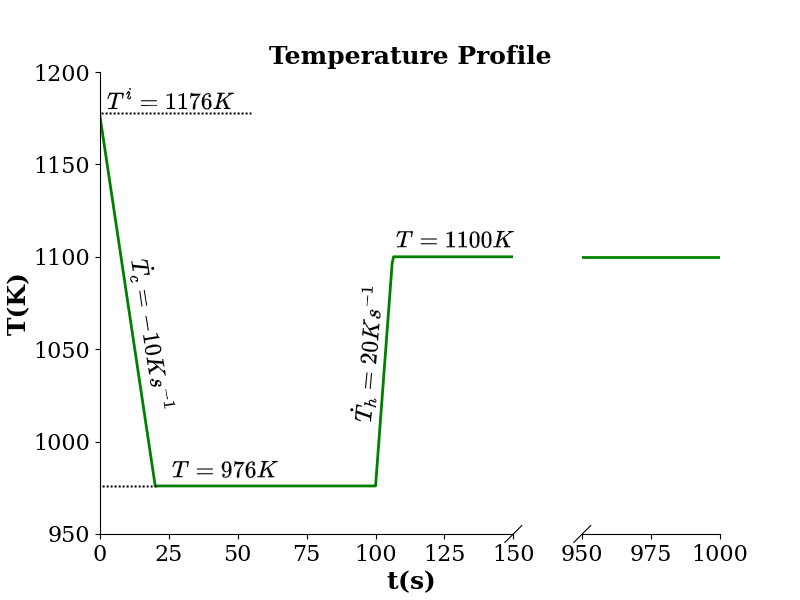}
	\captionsetup{justification=centering,margin=2cm}
	\caption{Thermal path imposed during the transformation: the broken axis representation indicates that the same temperature is maintained during that period}
	\label{LSC_tempprof}
\end{figure}

The microstructure is cooled at a constant rate of $\dot{T}=\SI{-10}{\kelvin\per\second}$ to $T^f=\SI{976}{\kelvin}$. Following the cooling phase, the microstructure is maintained at this temperature for an additional $80$ seconds to ensure complete transformation. It is then rapidly reheated to \SI{1100}{\kelvin} at a rate of \SI{20}{\kelvin\per\second} to enhance grain boundary mobility and thus promote grain growth effects. After the reheating, the microstructure is held at \SI{1100}{\kelvin} for the remaining duration of the simulation until $t=\SI{1000}{\second}$. The thermal path corresponding to this scenario is illustrated in Fig.\ref{LSC_tempprof}.

The interface mobility and carbon diffusivity data are adopted from \cite{krielaart1998kinetics, Mecozzi2007PhaseFM, lide2004crc} with Arrhenius type law for temperature dependence:
\begin{equation}
    \begin{gathered}
    M = 6\times10^{17}\exp{\left(\frac{-140000}{RT}\right)}, \quad \text{in} \quad \SI{}{\micro\meter\tothe{4}\per\joule\per\second}\\
    D_\alpha^C=2.2\times10^8 \exp\left(\frac{-122500}{RT}\right), \quad \text{in} \quad \SI{}{\micro\meter\tothe{2}\per\second}\\
    D_\gamma^C=1.5\times10^7 \exp\left(\frac{-142100}{RT}\right), \quad \text{in} \quad \SI{}{\micro\meter\tothe{2}\per\second}
    \end{gathered}.
    \label{MobDiffdata}
\end{equation}
The value for the interface energy is taken following \cite{loginova2003phase, Huang2006}, i.e. $\sigma_{\gamma\alpha}=\SI{1.0e-6}{\joule\per\milli\meter\squared}$. In this study, mobility and interfacial energy are assumed to be homogeneous, isotropic, and identical across phase and grain interfaces of both phases. It is however crucial to emphasize that the generalized kinetic description employed in this numerical model enables the seamless integration of any heterogeneity or anisotropy aspects if required.

Taking into account the available driving pressures for ferrite nucleation at different temperatures below the austenitization temperature, and factoring in the capillarity effects, the nucleation start temperature is estimated to be $T_{N_s}=\SI{1166}{\kelvin}$. This temperature is chosen to increase the probability of ferrite nuclei entering the growth regime of the transformation. This delay in nucleation could also loosely represent the notion of an incubation time. In this scenario, nucleation is configured to occur solely at the triple junctions (grain corners), continuously, until these sites are saturated, and spanning over a temperature range of $\delta T_N=\SI{40}{\kelvin}$. Under these conditions, a total of $2207$ ferrite nuclei are generated.

The resolution time step is fixed at $\Delta t = \SI{0.02}{\second}$. An adaptive mesh with periodic remeshing is utilized. The local mesh size within the diffuse phase interface is set as $h_{min}=\SI{0.5}{\micro\meter}$, and the diffuse phase interface thickness is set as $\eta=\SI{8}{\micro\meter}$ ($\approx 20h_{min}$).

\begin{figure}[htbp]
	\centering
	\includegraphics[width=8cm]{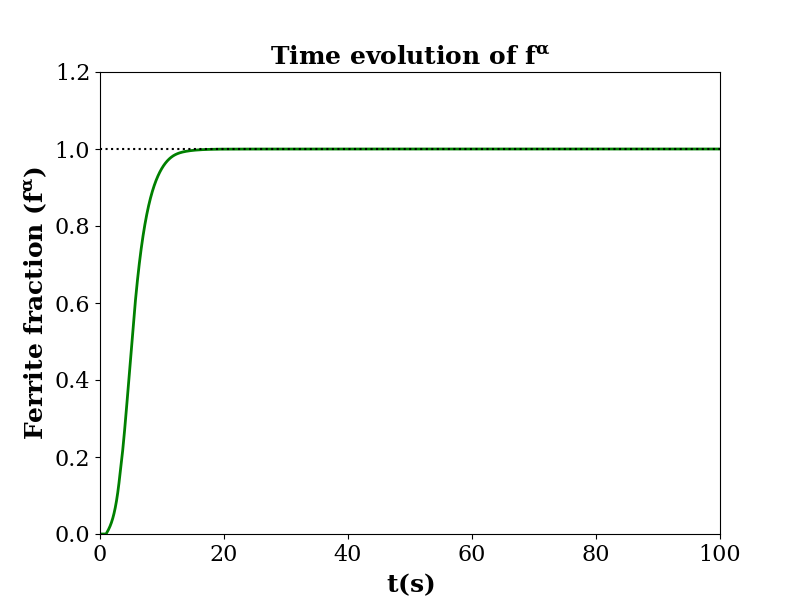}
	\captionsetup{justification=centering,margin=2cm}
	\caption{Time evolution of the ferrite fraction during the transformation}
	\label{LSC_fevol}
\end{figure}

In Fig.\ref{LSC_fevol}, the time evolution of the ferrite ($\alpha$) fraction is illustrated. It is evident that the ferrite fraction converges to unity, consistent with the ThermoCalc prediction indicating complete transformation to a ferritic microstructure below \SI{990}{\kelvin}.

\begin{figure}[htbp]
	\centering
	\captionsetup{justification=centering,margin=1cm}
        \subfloat[$t=\SI{1}{\second}, T=\SI{1166}{\kelvin}$]{\includegraphics[width=6cm]{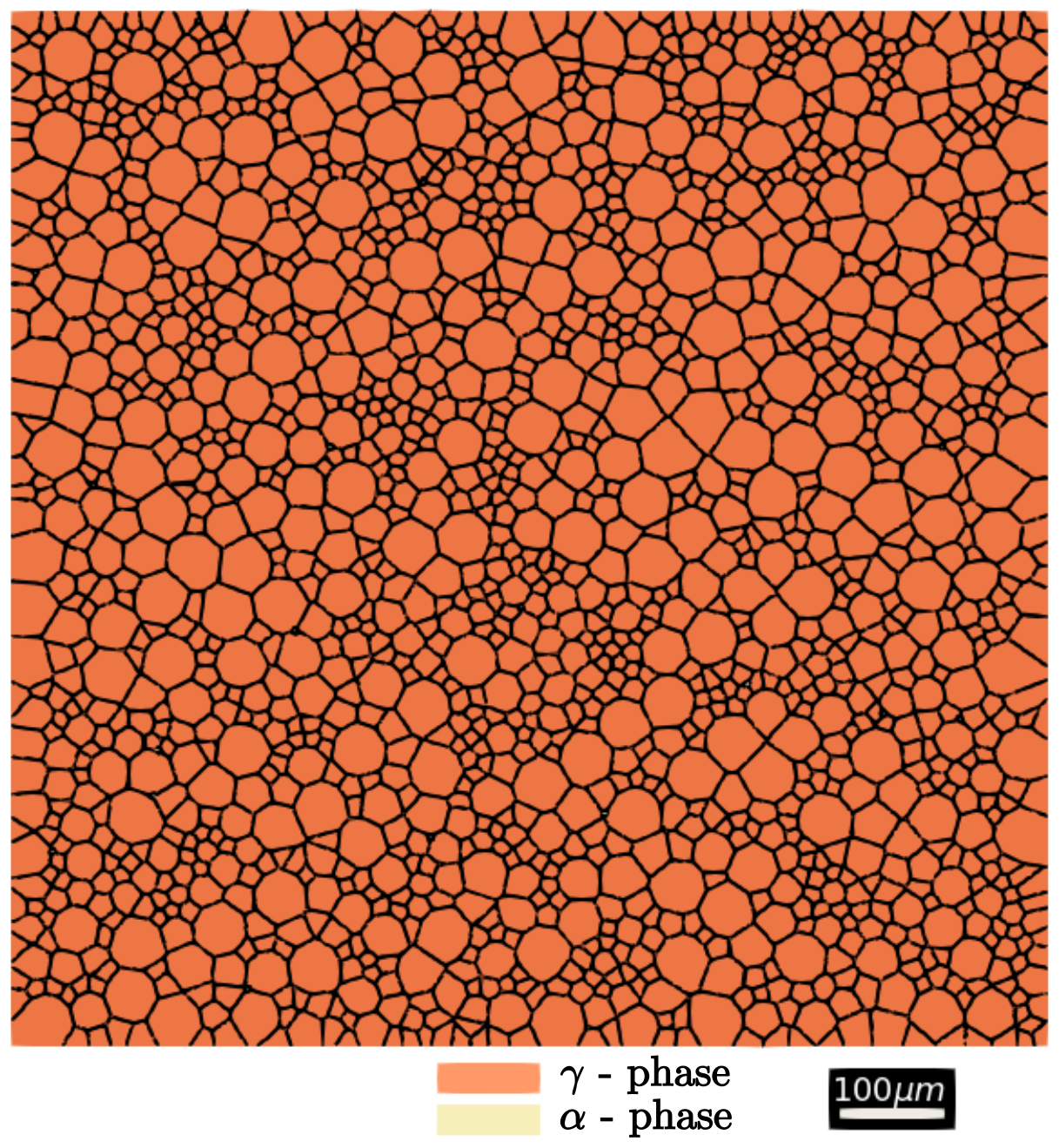}\label{PolyDSSPT_LSC_PD_0}}
        \hspace{0.5em}
 \subfloat[$t=\SI{3}{\second}, T=\SI{1146}{\kelvin}$]{\includegraphics[width=6cm]{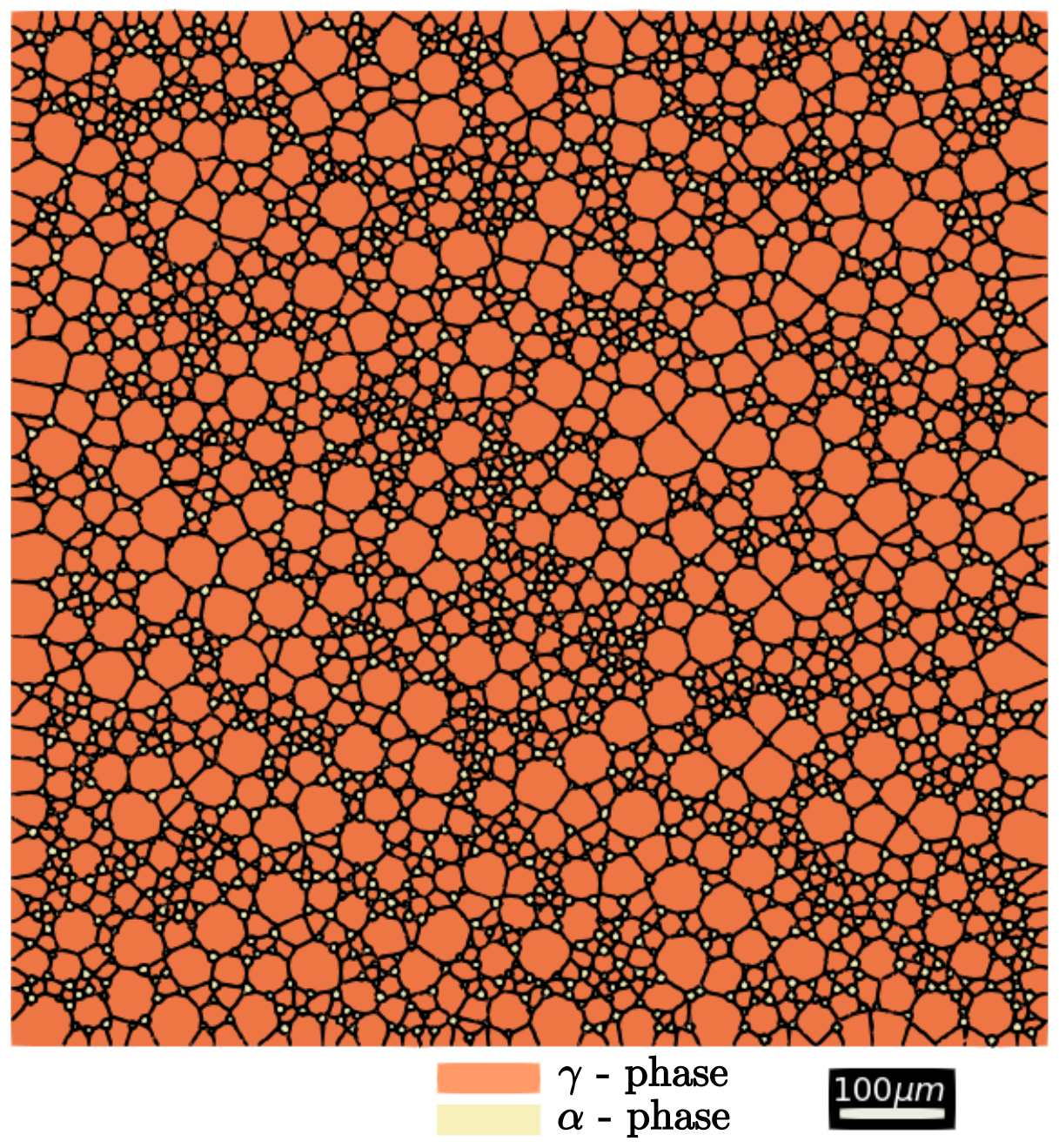}\label{PolyDSSPT_LSC_PD_2}}\\
	\subfloat[$t=\SI{5}{\second}, T=\SI{1126}{\kelvin}$]{\includegraphics[width=6cm]{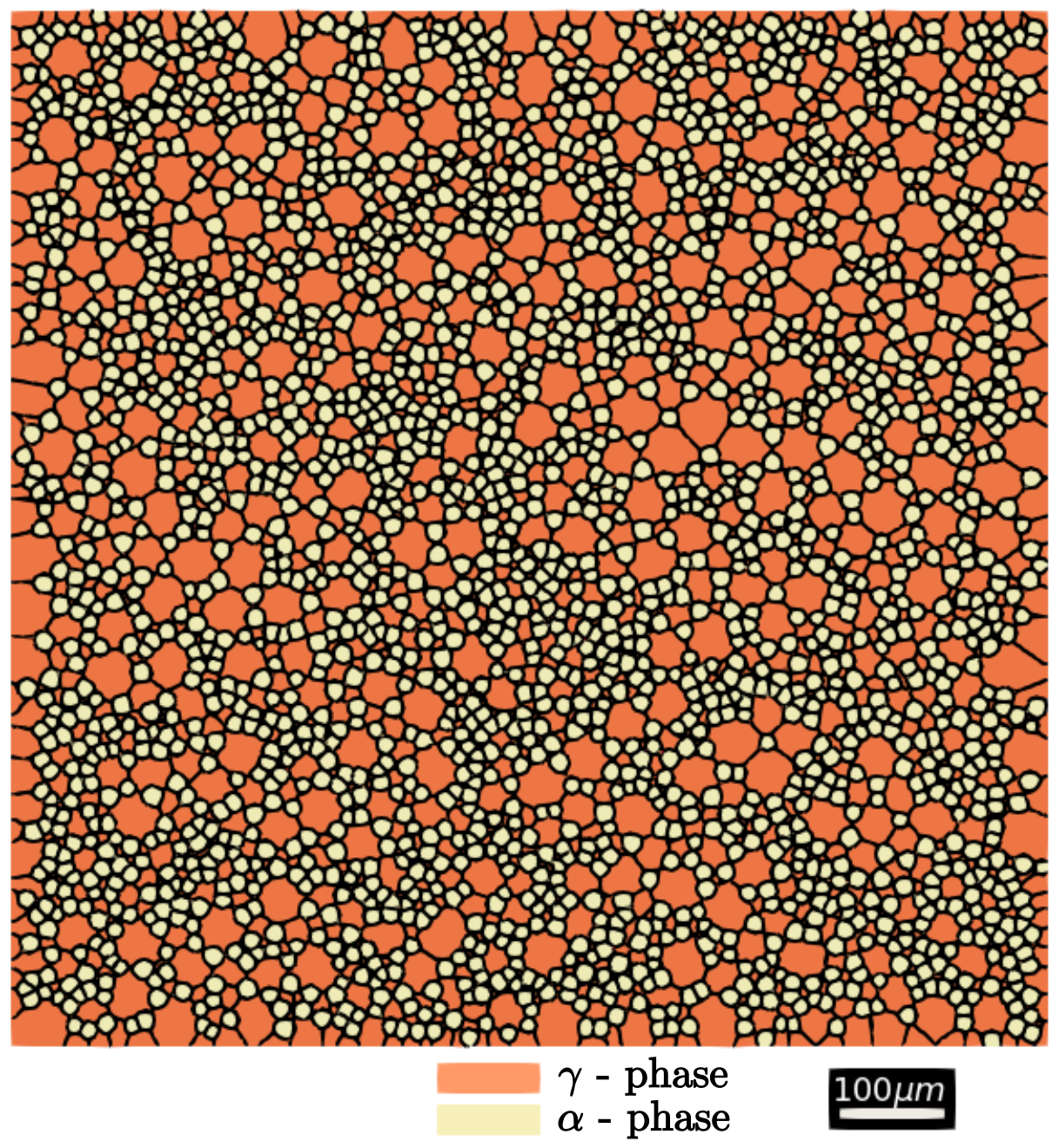}\label{PolyDSSPT_LSC_PD_3}}
 \hspace{0.5em}
	\subfloat[$t=\SI{20}{\second}, T=\SI{976}{\kelvin}$]{\includegraphics[width=6cm]{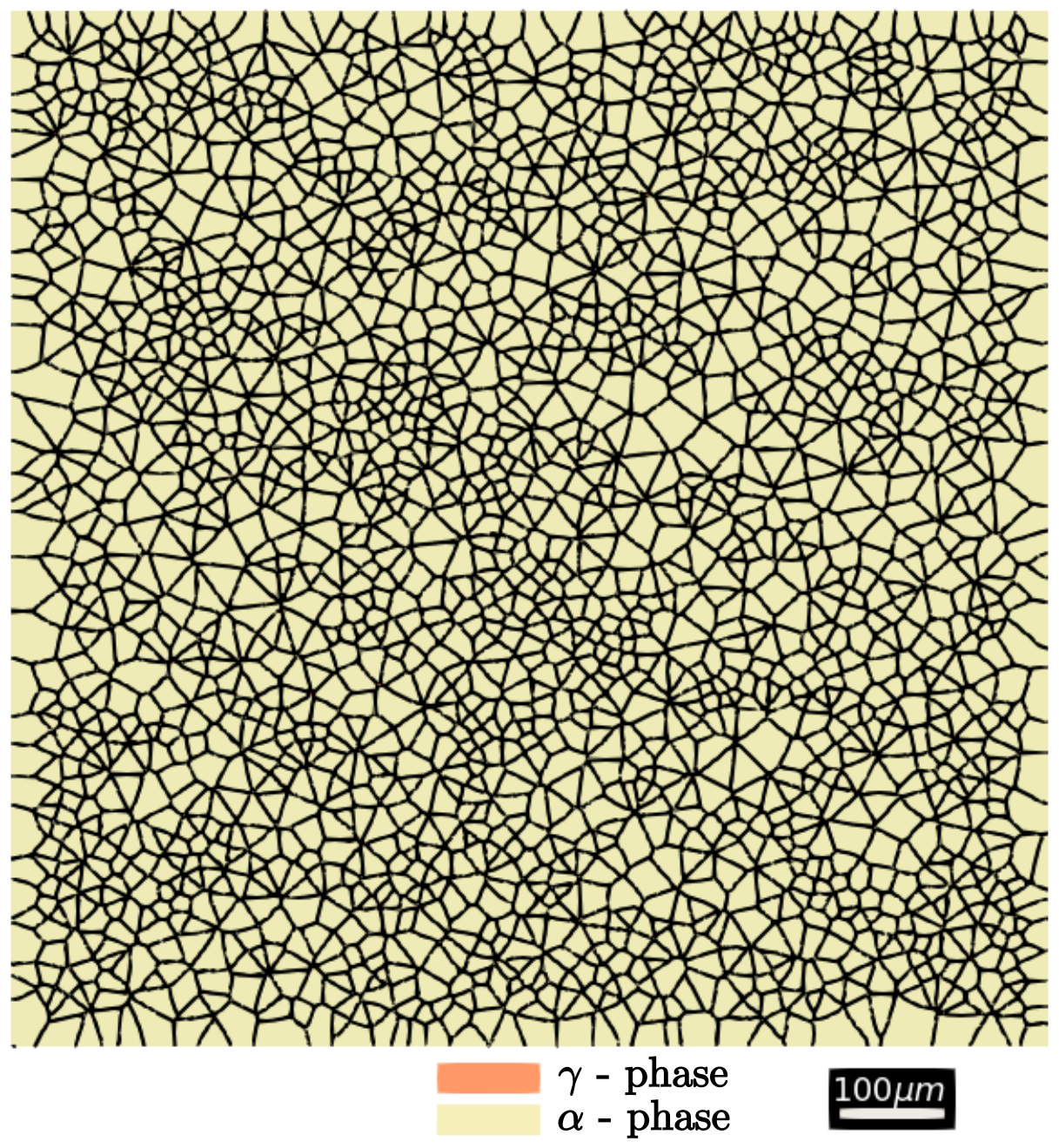}\label{PolyDSSPT_LSC_PD_5}}
	\caption{Snapshots of austenite decomposition into ferrite in a large-scale microstructure, at different instants till the end of cooling}
	\label{LSC_PD}
\end{figure}

\begin{figure}[!htbp]
	\centering
	\captionsetup{justification=centering,margin=1cm}
        \subfloat[$t=\SI{1}{\second}, T=\SI{1166}{\kelvin}$]{\includegraphics[width=7cm]{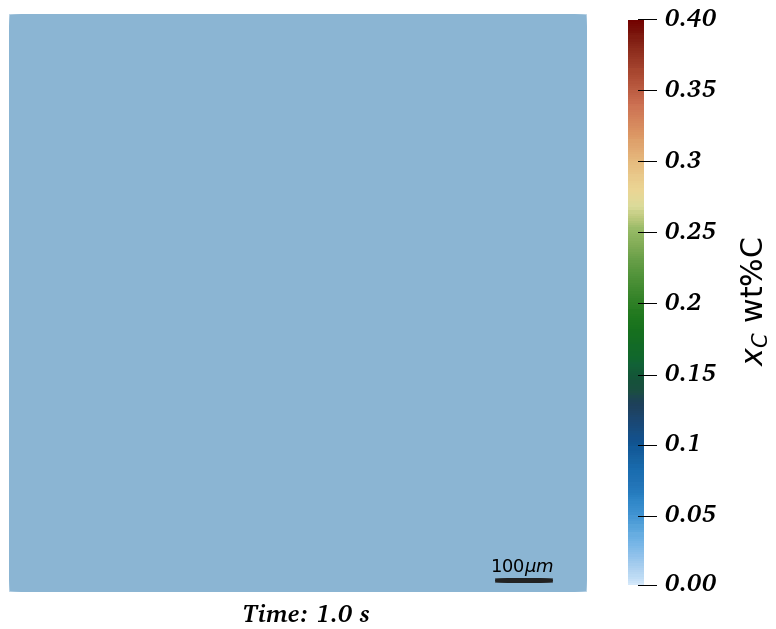}\label{PolyDSSPT_LSC_SD_0}}
        \hspace{0.5em}
 \subfloat[$t=\SI{3}{\second}, T=\SI{1146}{\kelvin}$]{\includegraphics[width=7cm]{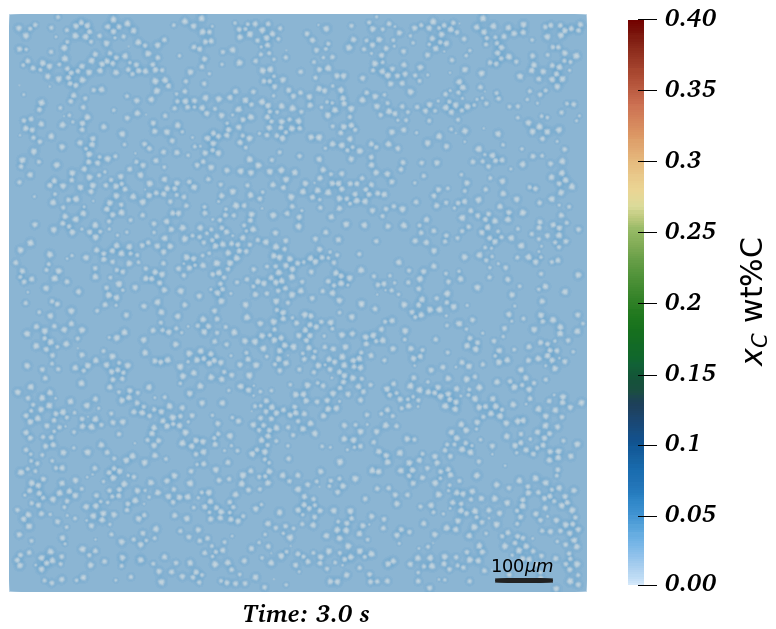}\label{PolyDSSPT_LSC_SD_2}}\\
	\subfloat[$t=\SI{5}{\second}, T=\SI{1126}{\kelvin}$]{\includegraphics[width=7cm]{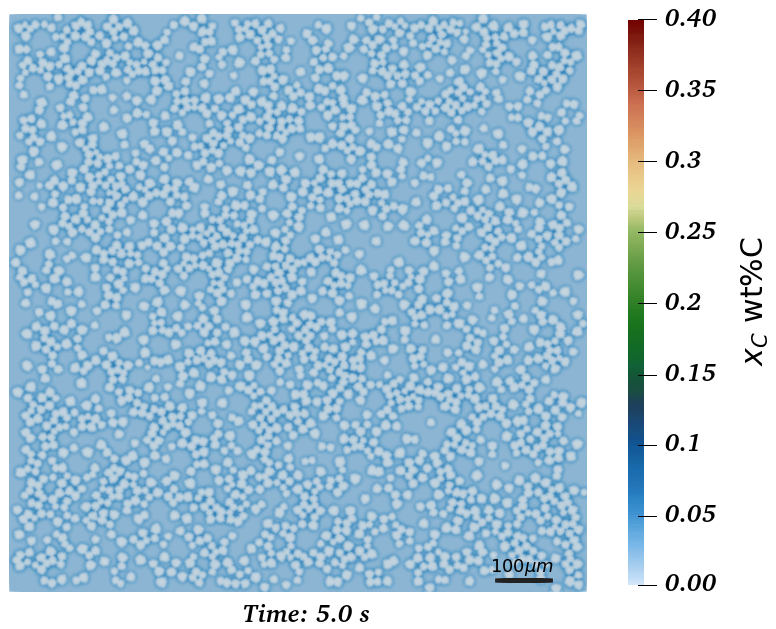}\label{PolyDSSPT_LSC_SD_3}}
 \hspace{0.5em}
	\subfloat[$t=\SI{20}{\second}, T=\SI{976}{\kelvin}$]{\includegraphics[width=7cm]{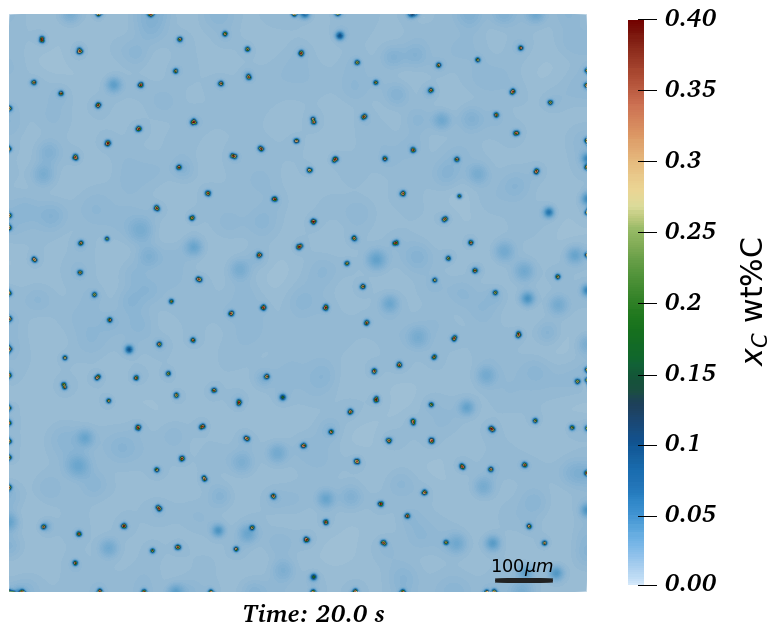}\label{PolyDSSPT_LSC_SD_5}}
	\caption{Snapshots of carbon evolution between the phases at different instants in a large-scale microstructure till the conclusion of cooling}
	\label{LSC_SD}
\end{figure}

Figs.\ref{LSC_PD} present the snapshots of the transforming microstructure at different intervals during the cooling stage. The associated carbon redistribution maps are illustrated in Figs.\ref{LSC_SD}. The continuous mode of nucleation results in a dispersion in the eventual ferrite grain size distribution. Early-formed nuclei gain an initial advantage in growth over subsequently appearing ones. Moreover, the first few nuclei experience minimal hard impingement until new neighboring nuclei gradually emerge. The presence of local nuclei clusters also contribute to the dispersion in ferrite grain size due to the enhanced hard impingement among the ferrite grains. Nucleation limited to grain corners and the relatively high nucleation density due to rapid cooling tend to yield globally finer equiaxed ferrite grains as evident from the Figs.\ref{LSC_PD}. The carbon enrichment in the austenite phase is noticeable as the transformation proceeds. The tiny dark spots observed in the Fig.\ref{PolyDSSPT_LSC_SD_5} is characterized by the formation of minuscule trapped austenite islands or pockets. Due to the high cooling rate, there is a rapid evolution resulting in a significant influx of carbon atoms into those austenite grains, which are thoroughly surrounded by a large number of ferrite grains. This substantial carbon enrichment consequently reduces the driving pressure for phase transformation in these areas, as local equilibrium is rapidly achieved. This results in trapped austenite, characterized by visible solute gradients. Given sufficient time, these solute gradients may diffuse into the bulk of the grains, enabling the additional growth of ferrite grains that extend into these confined austenite islands.

\begin{figure}[htbp]
	\centering
	\includegraphics[width=16cm]{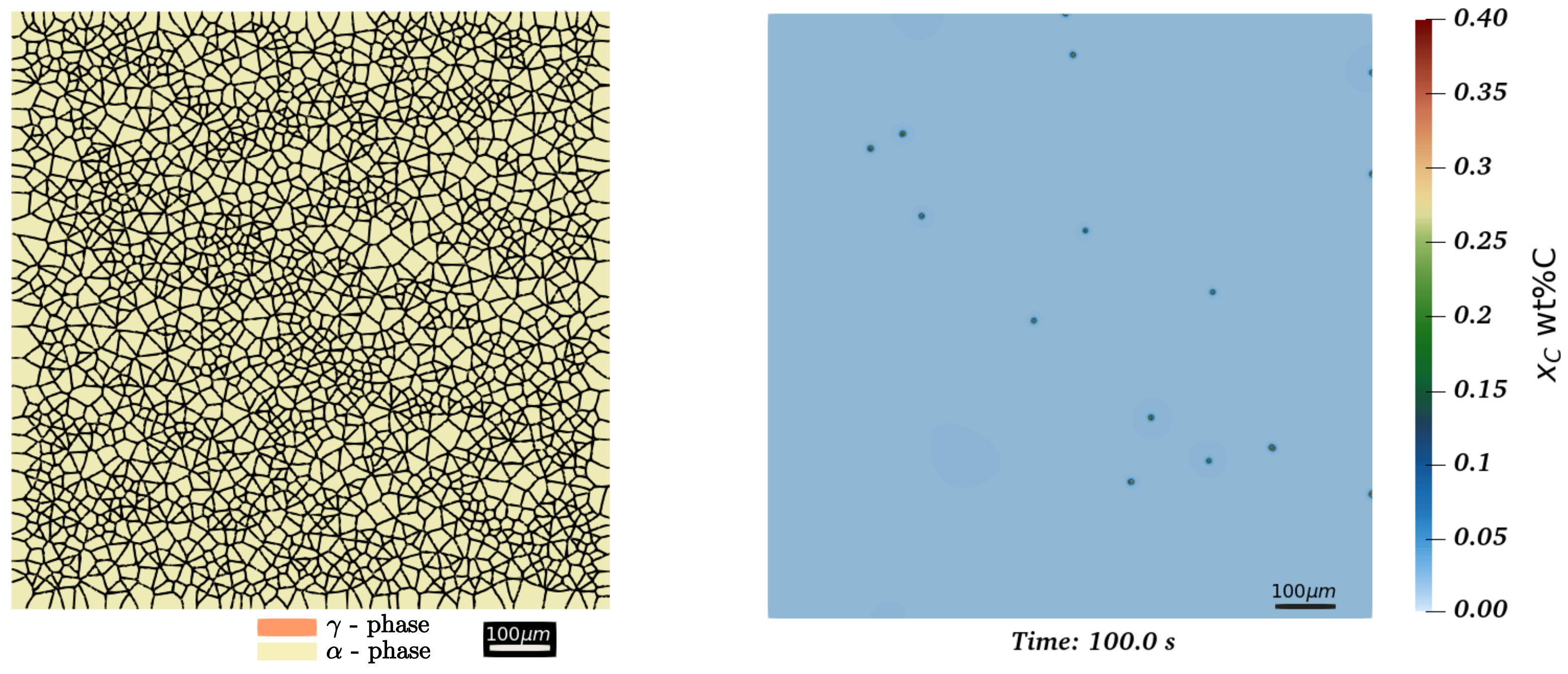}
	\captionsetup{justification=centering,margin=2cm}
	\caption{Configuration obtained post complete transformation to ferrite at the end of $t=\SI{100}{\second}$: Phase distribution (left), Carbon concentration field (right)}
	\label{PolyDSSPT_PDSD_PostPT}
\end{figure}
Figs.\ref{PolyDSSPT_PDSD_PostPT} represent the state of the ferritic microstructure obtained after the transformation stage (at $t=\SI{100}{\second}$). The triple junctions formed by the ferrite grains in Fig.\ref{PolyDSSPT_PDSD_PostPT} are clearly far from the expected equilibrium configuration (\SI{120}{\degree} for isotropic interfacial properties). Given the right activation, the system could further reduce the excess free energy, primarily existing as surface energy, through the grain growth (GG) phenomena. However, at temperatures as low as \SI{976}{\kelvin}, the interface mobility remains low, and the effects of GG are negligible within the timescales of interest. Therefore, to initiate the necessary activation, the microstructure is reheated to a higher temperature (\SI{1100}{\kelvin}) and maintained there to enhance the GG effects. It should be remarked that the GG stage of the simulation is resolved at a higher time step (\SI{0.5}{\second} in this case).
\begin{figure}[htbp]
	\centering
	\captionsetup{justification=centering,margin=1cm}
        \subfloat[$t=\SI{200}{\second}, T=\SI{1100}{\kelvin}$]{\includegraphics[width=6cm]{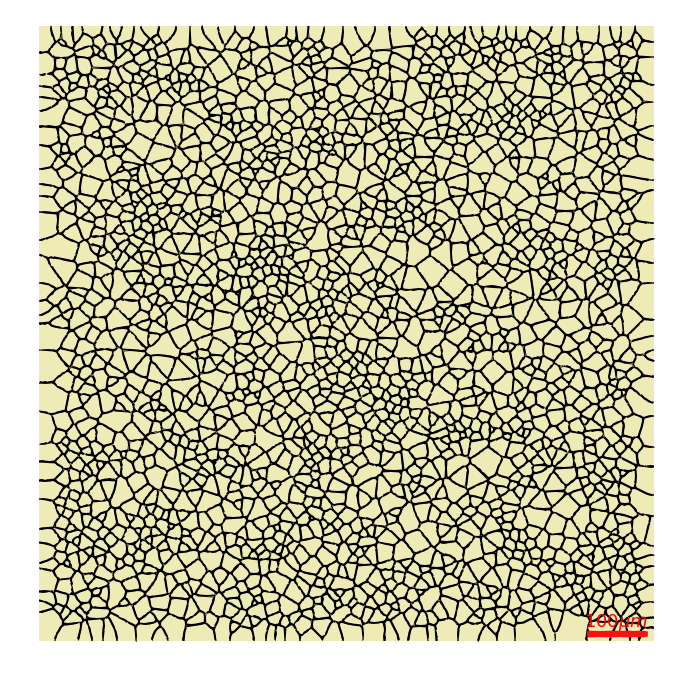}\label{PolyDSSPT_LSC_GG_0}}
	\subfloat[$t=\SI{400}{\second}, T=\SI{1100}{\kelvin}$]{\includegraphics[width=6cm]{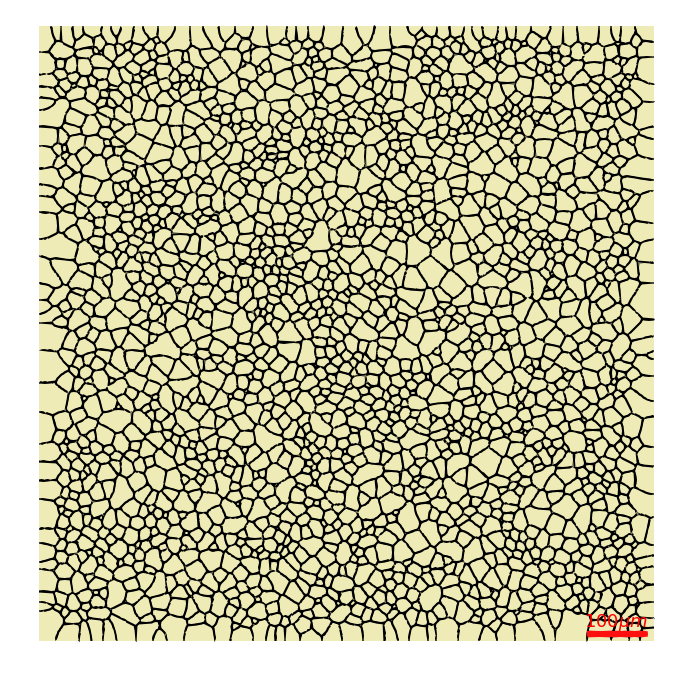}\label{PolyDSSPT_LSC_GG_1}}\\
 \subfloat[$t=\SI{700}{\second}, T=\SI{1100}{\kelvin}$]{\includegraphics[width=6cm]{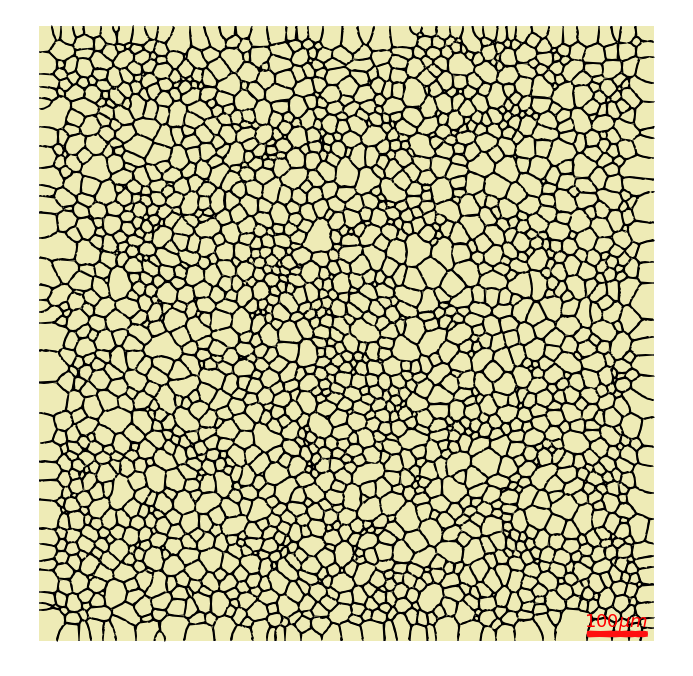}\label{PolyDSSPT_LSC_GG_2}}
	\subfloat[$t=\SI{1000}{\second}, T=\SI{1100}{\kelvin}$]{\includegraphics[width=6cm]{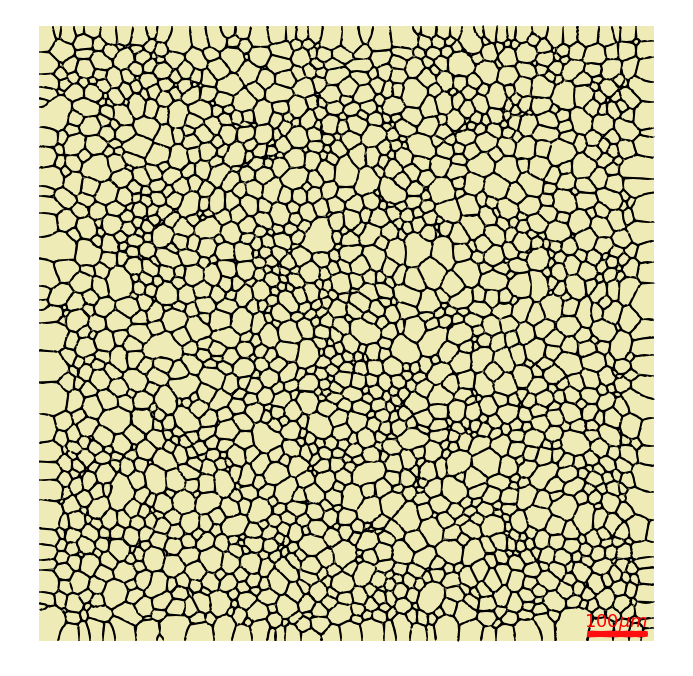}\label{PolyDSSPT_LSC_GG_3}}\\
	\caption{Snapshots of grain growth effects observed in ferrite grains at different instants after reheating}
	\label{LSC_GG}
\end{figure}

\begin{figure}[htbp]
	\subfloat[At the beginning of GG stage ($t=\SI{100}{\second}$)]{\includegraphics[width=8cm]{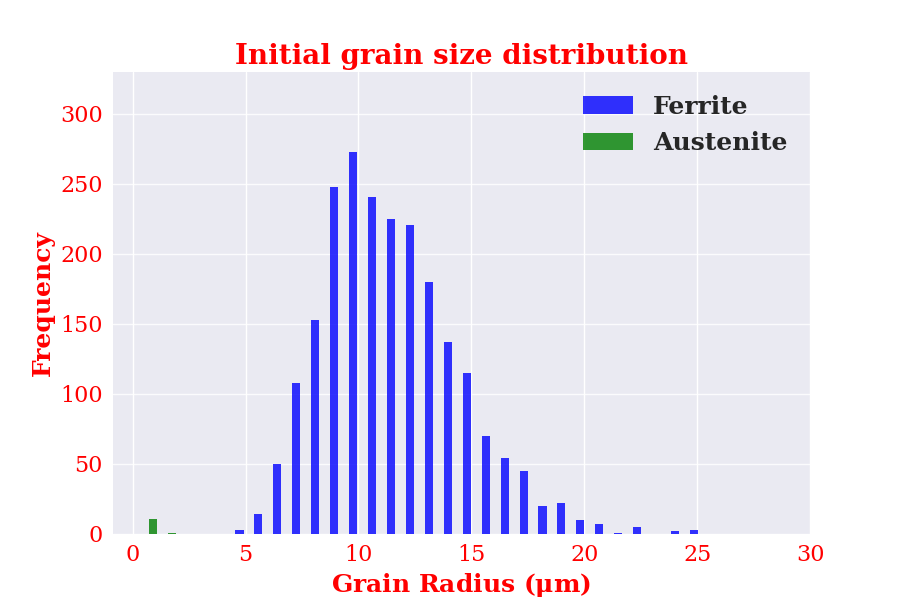}\label{LSC_GSD_GG_Initial}}
	\subfloat[At $t=\SI{1000}{\second}$]{\includegraphics[width=8cm]{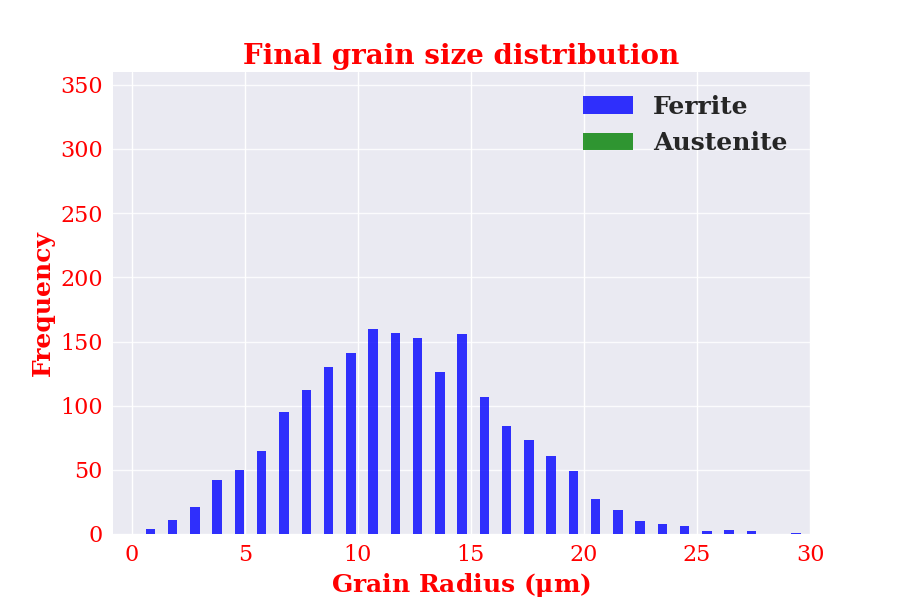}\label{LSC_GSD_GG_Final}}  
	\caption{Ferrite grain size distribution during the grain growth (GG) stage}
	\label{LSC_GSD_GG}
\end{figure}

\begin{figure}[htbp]
	\centering
	\captionsetup{justification=centering,margin=2cm}
        \subfloat[Initial microstructure]{\includegraphics[width=7cm]{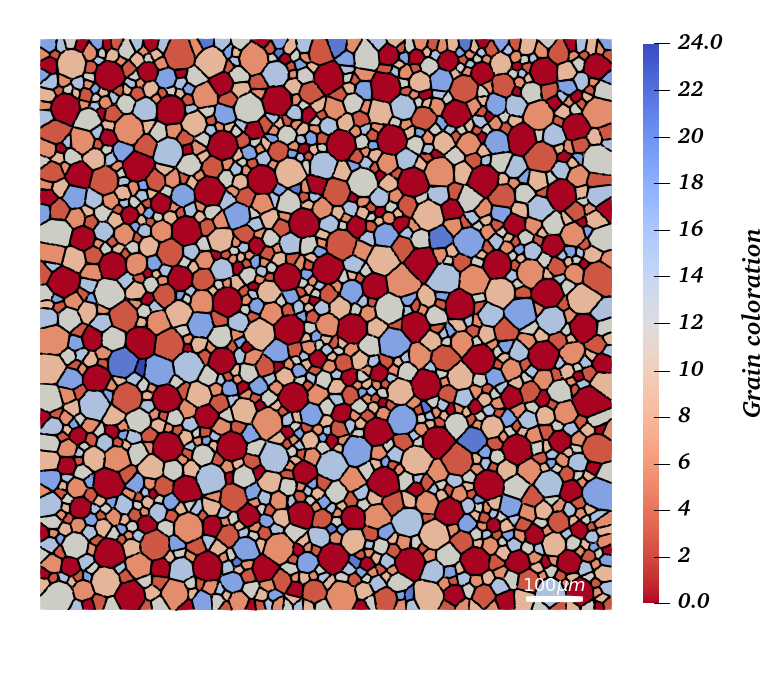}\label{PolyDSSPT_LSC_GC_Initial}}
	\subfloat[Final microstructure]{\includegraphics[width=7cm]{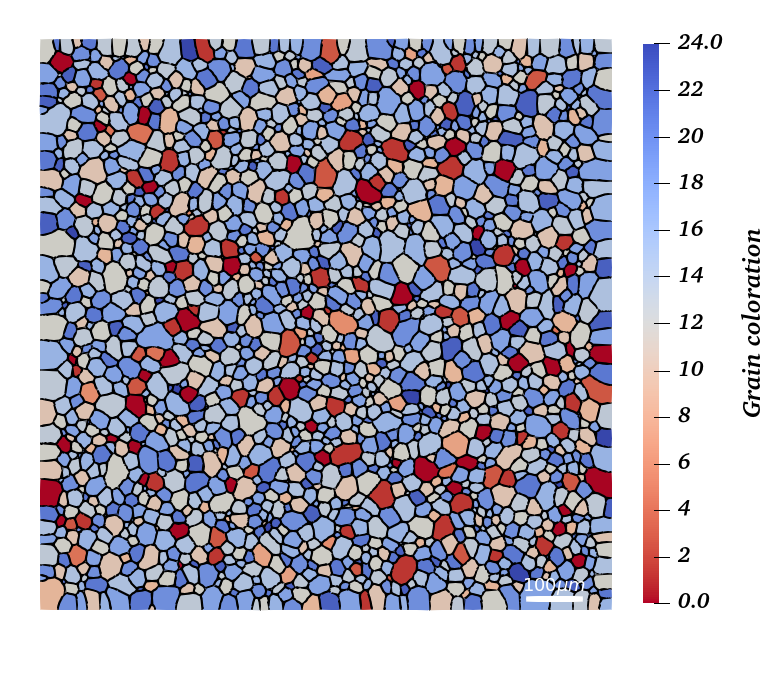}\label{PolyDSSPT_LSC_GC_Final}}
	\caption{Comparison of the microstructure morphologies before and after the complete thermal treatment: represented by grain coloration}
	\label{PolyDSSPT_LSC_GC}
\end{figure}

The influence of GG phenomenon on the resulting ferrite grain structure is portrayed in Figs.\ref{LSC_GG}. The growth of larger grains at the expense of smaller ones, aimed at minimizing the interfacial area, is clearly evident. Notably, the triple junctions approach the equilibrium angles (\SI{120}{\degree}) characteristic of isotropic GG. The differences in the ferrite grain size distributions obtained before the onset of GG regime, and after GG effects at $t=\SI{1000}{\second}$ are showcased in Figs.\ref{LSC_GSD_GG}. As the GG effects become prominent, the grain size distribution becomes more dispersed, with the average grain size gradually increasing.
The initial and the final microstructures resulting from this thermal treatment are depicted in Figs.\ref{PolyDSSPT_LSC_GC} using grain coloration. This scenario produces a finer microstructure after both the transformation and the subsequent grain growth. The current model does not account for the reverse phase transformation during the reheating stage. However, since the driving pressure description remains consistent, adapting the existing framework to control the direction of the phase transformation should be relatively straightforward. This could be a potential avenue for future research, involving additional complexities such as nucleation for the reverse transformation.
\subsubsection*{Ternary alloy: Influence of solute drags effects}
\begin{figure}[htbp]
	\centering
	\includegraphics[width=6cm]{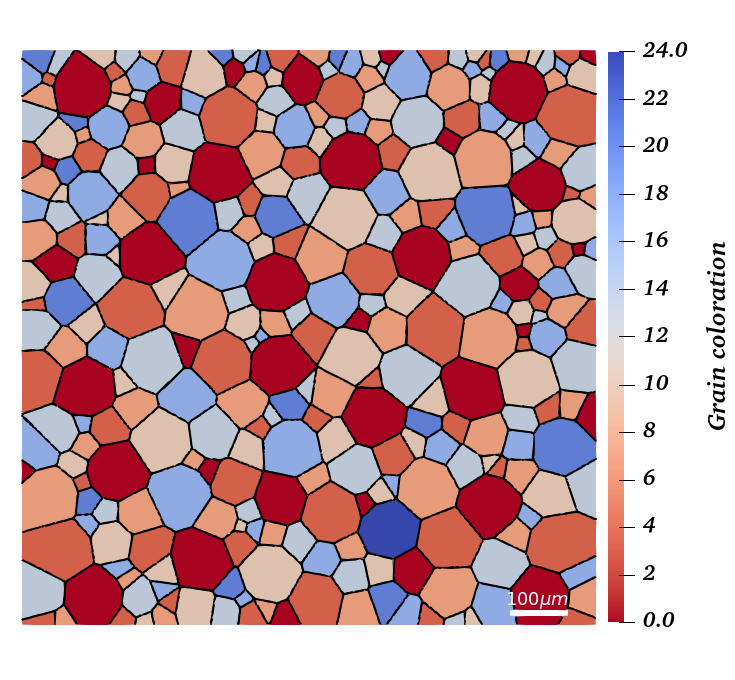}
	\captionsetup{justification=centering,margin=2cm}
	\caption{Grain coloration adopted for the initial austenitic grain morphology}
	\label{graincolinitCompNuc}
\end{figure}

We shall now examine diffusive phase transformation in a ternary alloy (Fe-\SI{0.1}{\wtpercent}C-\SI{0.5}{\wtpercent}Mn) in the presence of a substitutional Mn element. The influence of Mn on the transformation kinetics are incorporated through solute drag aspects under para-equilibrium hypothesis. For this case, a simple microstructure illustrated in Fig.\ref{graincolinitCompNuc} is considered with $317$ austenite grains.
For this alloy, the calculated austenitization temperature under the PE hypothesis is approximately \SI{1118}{\kelvin}. The initial temperature is set to this $T_{A3}^{PE}$ temperature. The microstructure is cooled to a final temperature of $T^f=\SI{910}{\kelvin}$ at a cooling rate of \SI{-10}{\kelvin\per\second}. At end of cooling, the microstructure is maintained at the final temperature for an additional \SI{30}{\second} to provide ample time for phase transformation and the associated solute redistribution. The interfacial energy, mobility, diffusivity data, and most numerical parameters are fixed the same as in the previous case. The time step is chosen as $\Delta t=\SI{0.02}{\second}$ for this case. The ferrite nucleation has been limited to grain corners. For simplicity, the number of ferrite nuclei stochastically introduced have been restricted to $300$ for this case. The nucleation start temperature is set to $T_{N_s}=\SI{1112}{\kelvin}$, and the range is assumed as $\delta T_N=\SI{30}{\kelvin}$.

\begin{figure}[htbp]
	\centering
	\captionsetup{justification=centering,margin=1cm}
        \subfloat[$\alpha_C(T)$]{\includegraphics[width=7cm]{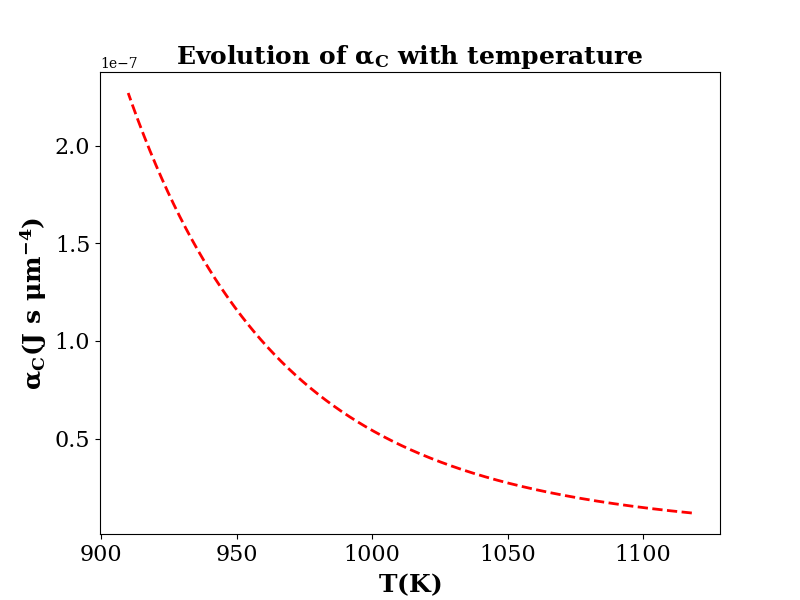}\label{CompSD_alphaC}}
	\subfloat[$\beta_C^2(T)$]{\includegraphics[width=7cm]{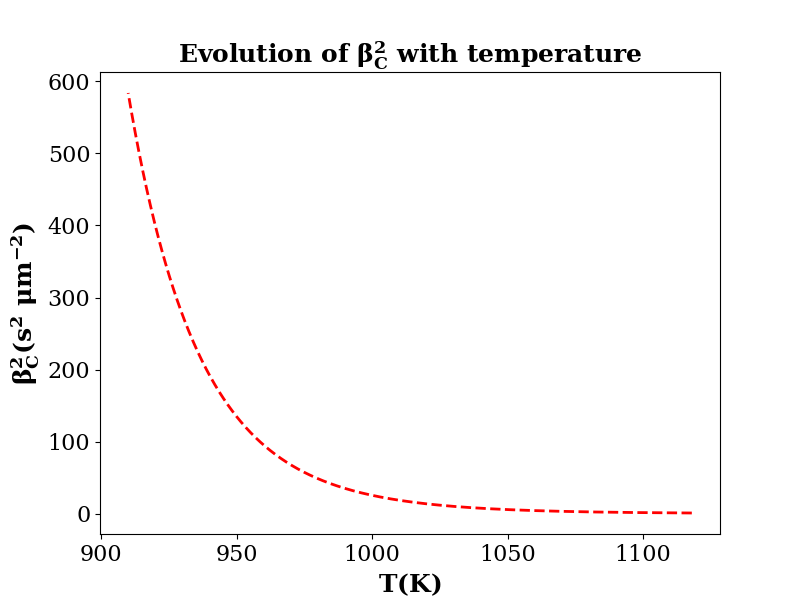}\label{CompSD_betaC}}   
	\caption{Evolution of Cahn's solute drag parameters with temperature for the chosen case}
	\label{SD_CahnPar_Evol}
\end{figure}

To model solute drag aspects, the simplified description provided by Cahn's model \cite{cahn1962impurity} is adopted as detailed in the second section. In practical scenarios, the objective is to utilize the two parameters ($\alpha, \beta$) introduced by Cahn's solute drag pressure as fitting model parameters to agree with the experimental results, rather than employing the analytical expressions provided by Cahn in Eqs.\eqref{alphaCahndef} and \eqref{betaCahndef}. However, for demonstration purposes in this context, we utilize the analytical expressions based on literature \cite{fazeli2005application} to establish some parameters in Eqs.\eqref{alphaCahndef} and \eqref{betaCahndef}, aiming for as realistic a representation as possible. Figs.\ref{SD_CahnPar_Evol} provide a visualization of the Cahn's parameters computed in this simulation, depicting their variation with temperature. The graph illustrating the variation of $\beta_C^2$ in Fig.\ref{CompSD_betaC} indicates a rise in solute drag pressure at elevated temperatures, gradually weakening as the temperature decreases.

\begin{figure}[htbp]
	\centering
	\includegraphics[width=7cm]{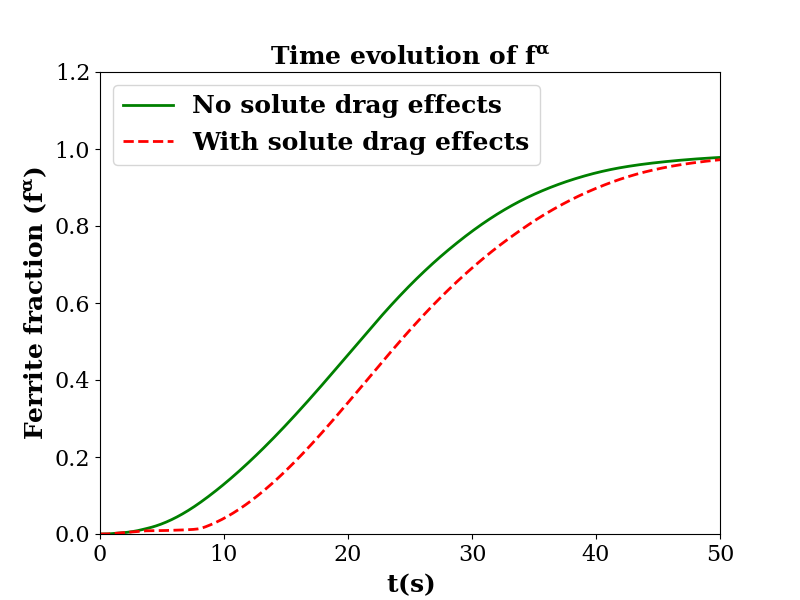}
	\captionsetup{justification=centering,margin=2cm}
	\caption{Evolution of ferrite fraction with time for the difference cases}
	\label{CompSD_phfracevolvtime}
\end{figure}

Fig.\ref{CompSD_phfracevolvtime} showcases the evolution of the ferrite phase fraction over time for two scenarios: one without any solute drag effects, and the other incorporating solute drag effects. It could be inferred that, enhanced drag effects at higher temperatures limit the phase fraction initially. However, with sufficient time, both the scenarios eventually converge to a similar ferrite fraction value as depicted in Fig.\ref{CompSD_phfracevolvtime}, since at lower temperatures the solute drag effects are significantly reduced.

\begin{figure}[!htbp]
	\centering
	\captionsetup{justification=centering,margin=2cm}
        \subfloat[Without solute drag effects]{\includegraphics[width=12cm]{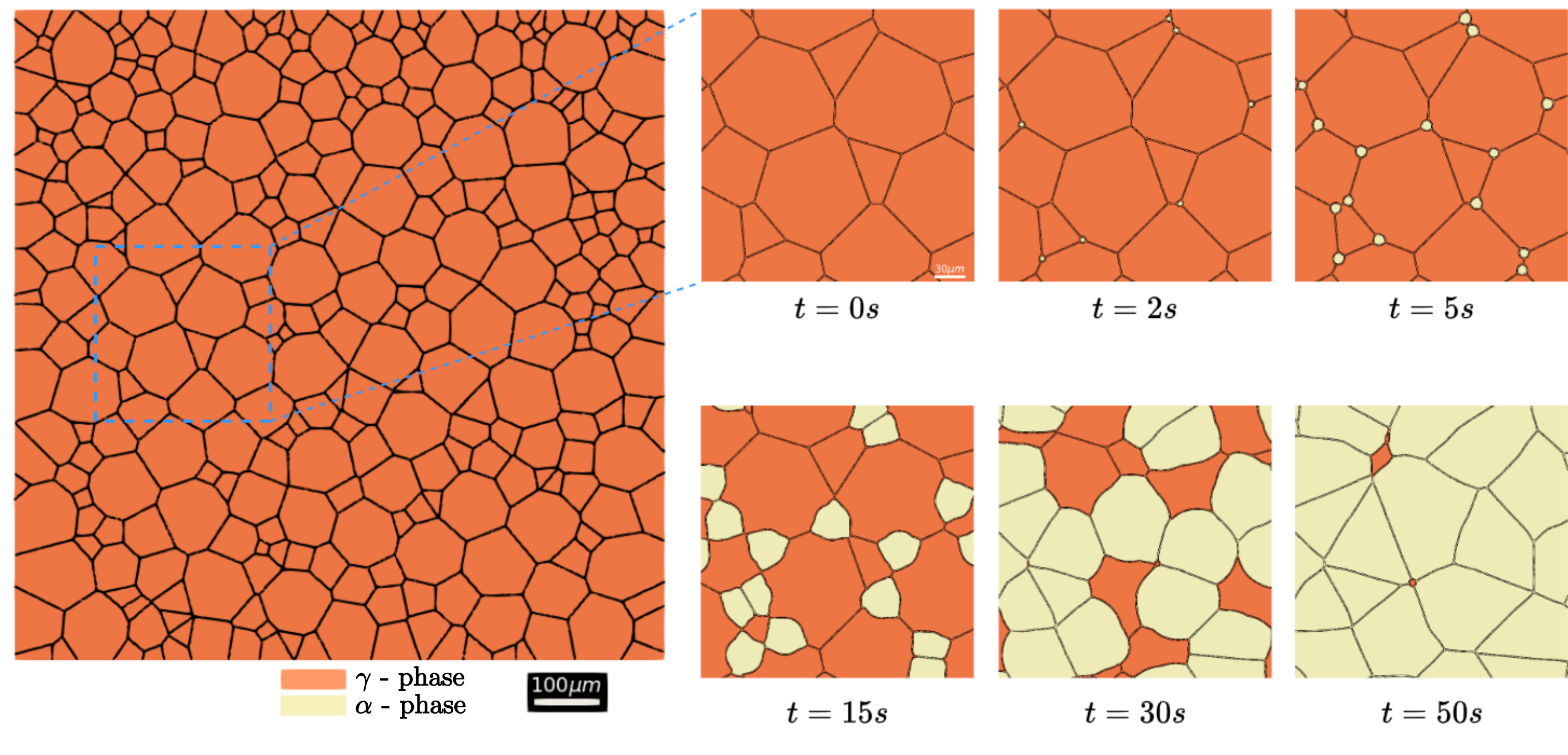}\label{PolyDSSPT_SSD_PD}}\\
	\subfloat[With solute drag effects]{\includegraphics[width=12cm]{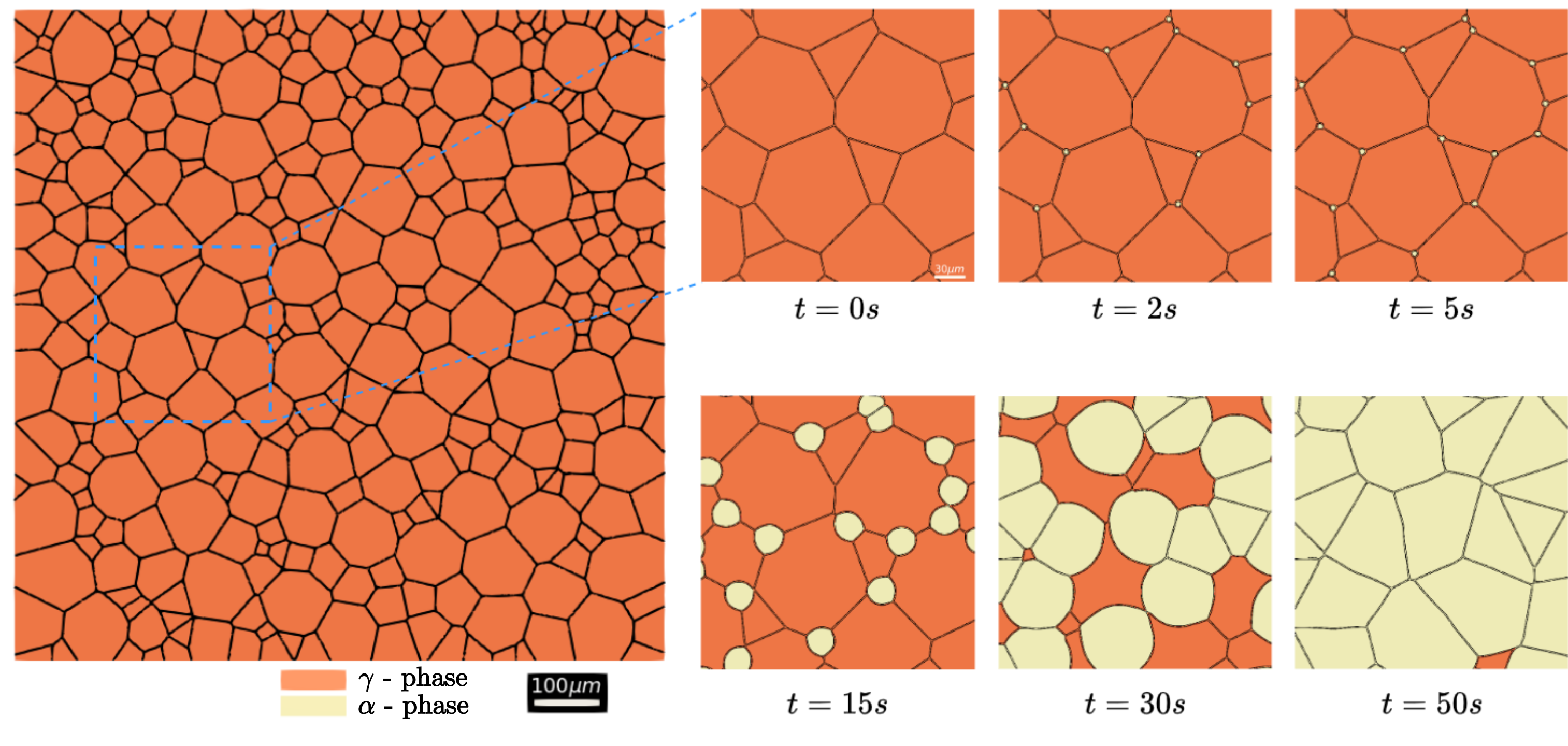}\label{PolyDSSPT_WSD_PD}}
	\caption{Snapshots of phase evolution with and without the consideration of solute drag effects}
	\label{PolyDSSPT_CompSD_PD}
\end{figure}

\begin{figure}[htbp]
	\centering
	\captionsetup{justification=centering,margin=2cm}
        \subfloat[Without solute drag effects]{\includegraphics[width=12cm]{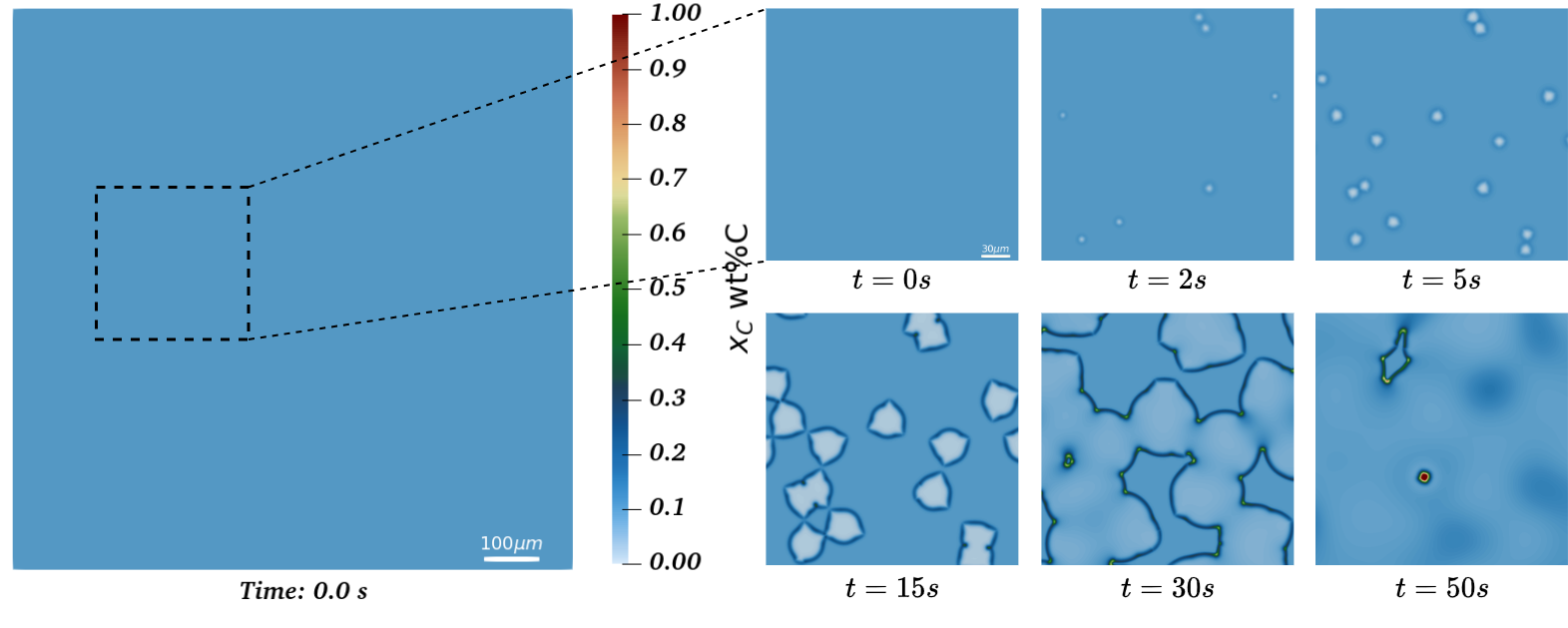}\label{PolyDSSPT_SSD_SD}}\\
	\subfloat[With solute drag effects]{\includegraphics[width=12cm]{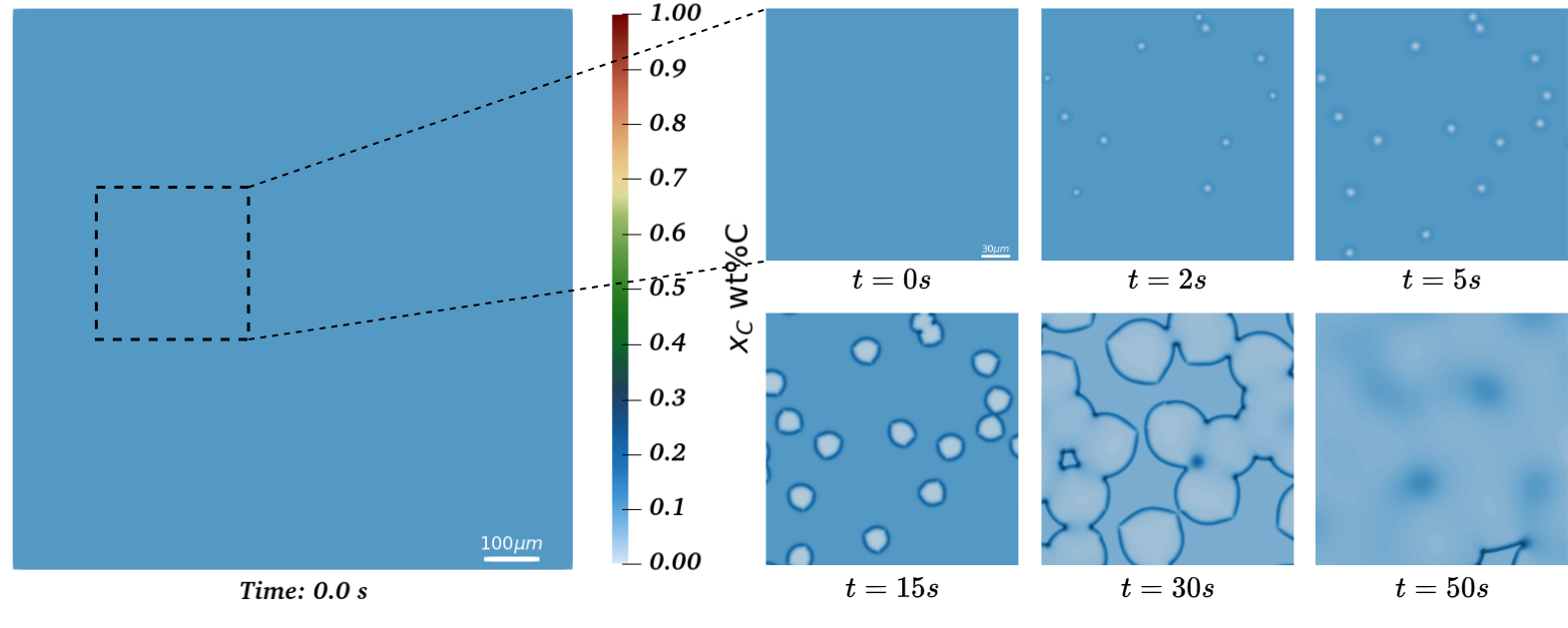}\label{PolyDSSPT_WSD_SD}}
	\caption{Snapshots of carbon distribution with and without the consideration of solute drag effects}
	\label{PolyDSSPT_CompSD_SD}
\end{figure}

\begin{figure}[htbp]
	\centering
	\captionsetup{justification=centering,margin=2cm}
        \subfloat[No solute drag effects]{\includegraphics[width=6cm]{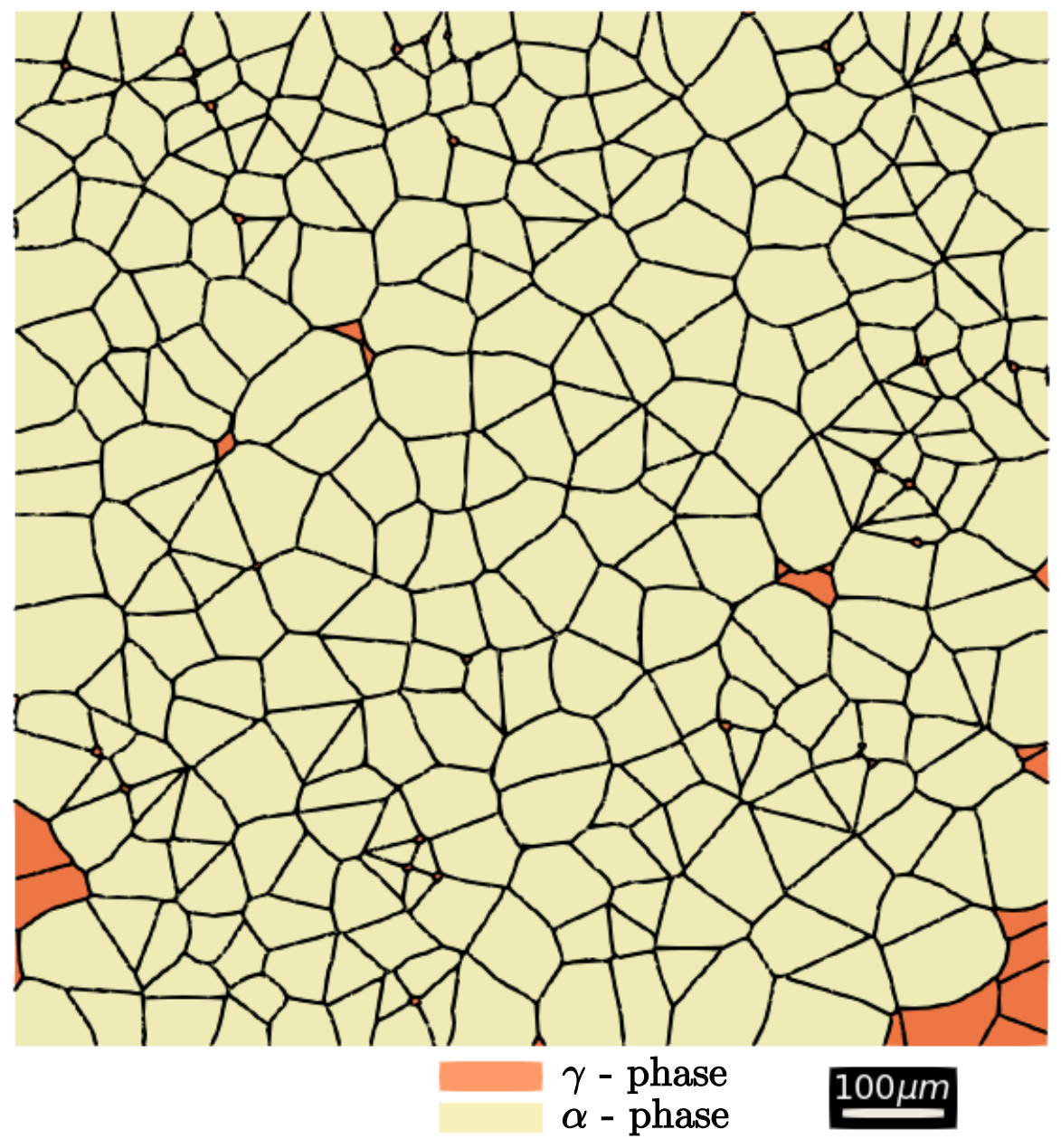}\label{PolyDSSPT_SSSD_PD_Final}}\hspace{1em}
	\subfloat[With solute drag effects]{\includegraphics[width=6cm]{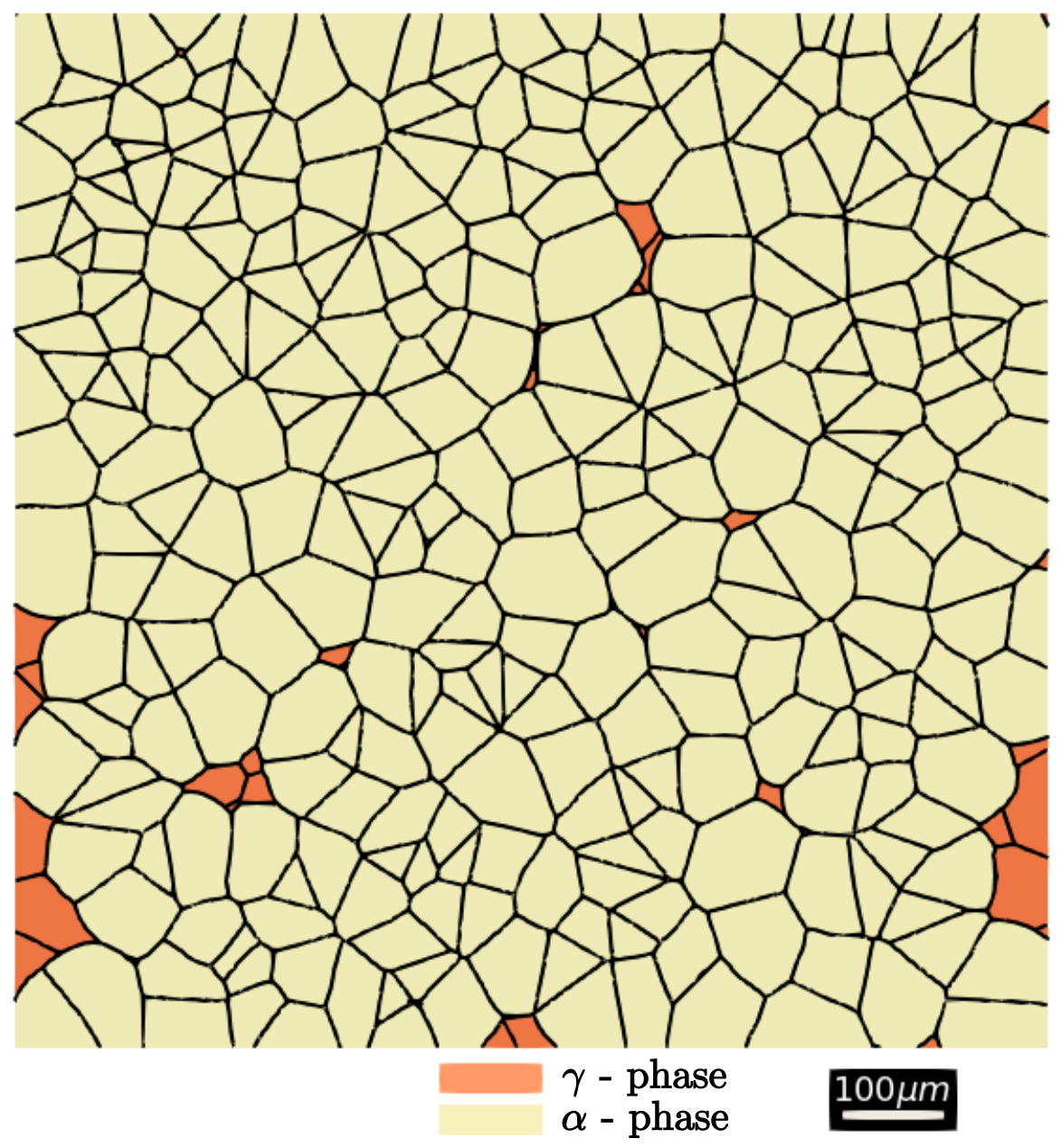}\label{PolyDSSPT_WSD_PD_Final}}
	\caption{Final phase distribution obtained with and without solute drag effects}
	\label{PolyDSSPT_PD_CompSD_Final}
\end{figure}

\begin{figure}[!htbp]
	\centering
	\captionsetup{justification=centering,margin=2cm}
        \subfloat[Without solute drag effects]{\includegraphics[width=7cm]{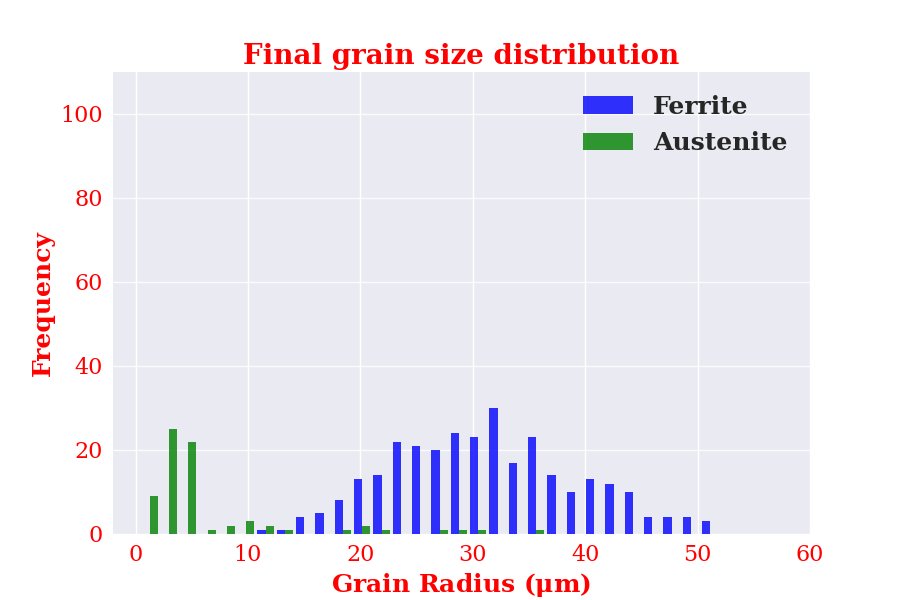}\label{PolyDSSPT_SSD_GSD_Final}}
	\subfloat[With solute drag effects]{\includegraphics[width=7cm]{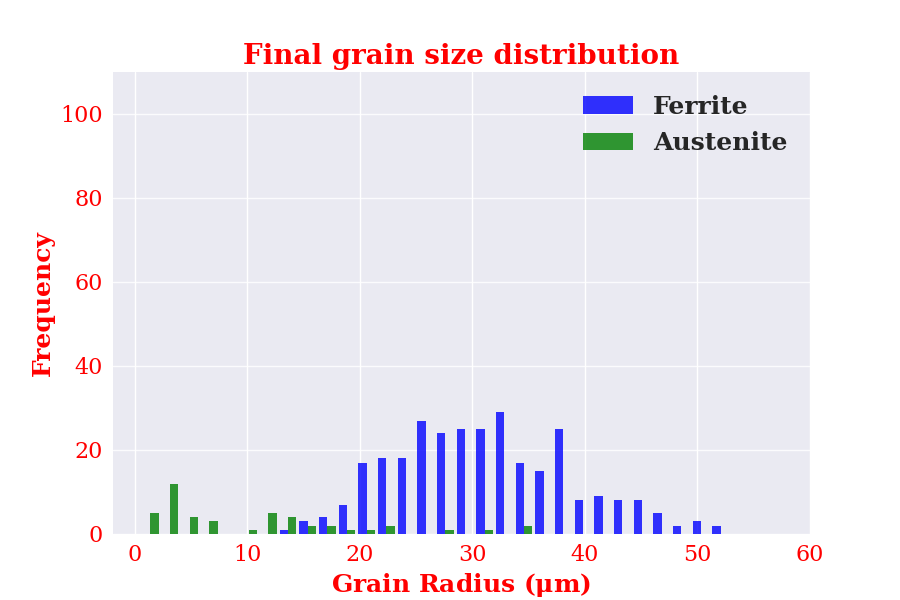}\label{PolyDSSPT_WSD_GSD_Final}}
	\caption{Final grain size distributions obtained for the two cases}
	\label{GSD_Final_CompSD}
\end{figure}

Figs.\ref{PolyDSSPT_CompSD_PD}, and \ref{PolyDSSPT_CompSD_SD} compare the phase distribution and the solute distribution during the transformation between the scenarios with and without the solute drag effects. Both the scenarios globally result in equiaxed grains. However, in the instance incorporating solute drag effects, the growth of ferrite nuclei in the beginning at higher temperatures is sluggish due to strong drag effects. Generally, this could limit the potential dispersion in grain size distribution caused by continuous nucleation, resulting in a relatively uniform distribution of grains. However, as evident in Figs.\ref{GSD_Final_CompSD}, in this case, the difference are minimal, possibly due to a smaller temperature range of nucleation for the considered cooling rate. The comparison between the final microstructure morphologies illustrated in Figs.\ref{PolyDSSPT_PD_CompSD_Final} demonstrates similar grain characteristics, including a comparable mean ferrite grain size. The noticeable distinctions primarily lie in the transformation kinetics between the two scenarios for the assumed solute drag parameters in this illustration.
\subsection{Simulation of other solid-state diffusive phenomena: Particle coarsening}

After a first-order diffusive phase transformation where second-phase particles (SPPs) separate within a matrix phase, local equilibrium might be achieved between the matrix and the SPPs. However, this state may not represent the system's minimum free energy configuration, particularly when the matrix phase consists of a dispersed or multi-modal distribution of SPPs. The excess free energy contributed by the interfacial energy of the of SPPs renders the system thermodynamically unfavorable. Consequently, the system attempts to liberate this excess free energy by dissolving smaller particles in favor of larger ones through diffusional mass transport. The curvature of SPPs alter the local equilibrium concentrations with the matrix phase due to the Gibbs-Thomson effect. The multi-modal distribution of SPPs thus creates a gradient in concentration which drives the diffusional mass transport between a region of higher concentration and lower concentration. This phenomenon is commonly referred to as particle coarsening or Ostwald ripening (OR) \cite{ostwald1901blocking, voorhees}. In the absence of elastic misfits between the two phases, the kinetics of particle coarsening are governed by the capillarity induced driving pressure and the driving pressure due to the phase transformation. It should thus be possible to simulate this phenomenon using the existing numerical framework.

\begin{figure}[htbp]
	\centering
	\includegraphics[width=7cm]{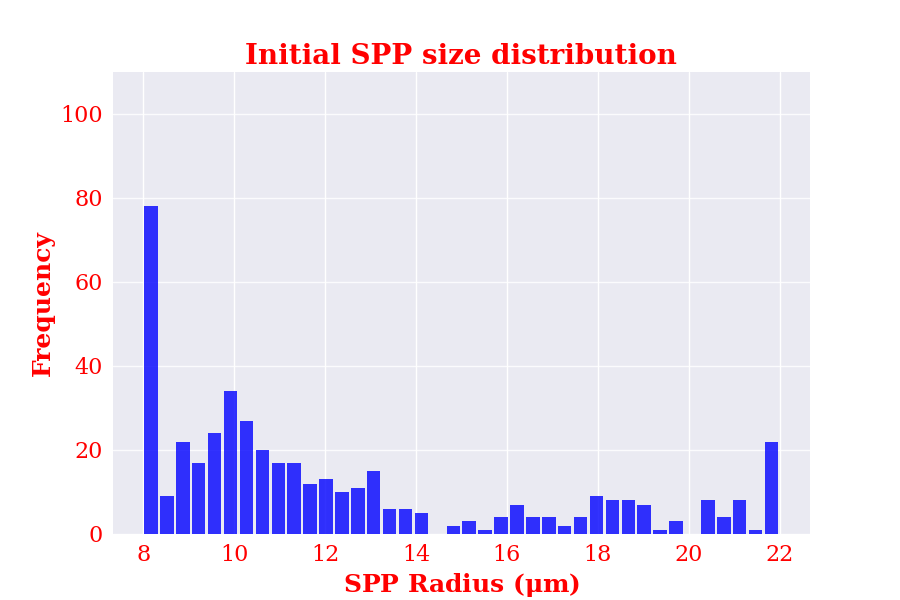}
	\captionsetup{justification=centering,margin=2cm}
	\caption{Initial size distribution of second-phase particles (SPPs)}
	\label{OR_gsd_init}
\end{figure}

\begin{figure}[htbp]
	\centering
	\includegraphics[width=7cm]{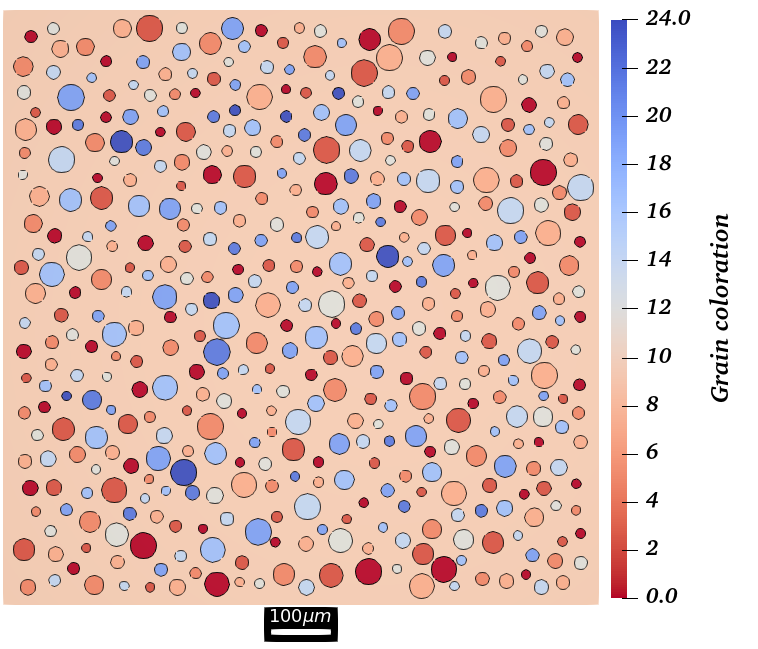}
	\captionsetup{justification=centering,margin=2cm}
	\caption{Initial grain coloration of the matrix-SPPs setup}
	\label{OR_graincol_init}
\end{figure}

To simulate this phenomenon, we've chosen to use a hypothetical scenario: representing the SPPs as the ferrite phase ($\alpha$) and the matrix as the austenite phase ($\gamma$). It's crucial to highlight that despite this hypothetical setup not mirroring the conventional appearance of Ostwald ripening, the inherent kinetics and dynamics remain identical. The primary objective is to showcase the numerical model's capability in simulating various diffusive solid-state phenomena, including particle coarsening, utilizing the same kinetic framework. Therefore, our simulation involves an austenite phase matrix within a \SI{1}{\milli\meter}-sized domain, wherein the second phase consists of ferrite, forming a bimodal distribution of particles, as illustrated in Fig.\ref{OR_gsd_init}. The initial configuration is represented in Fig.\ref{OR_graincol_init} through the grain coloration employed for this case.

The initial condition is slightly offset ($T^i=\SI{1162}{\kelvin}$) from the isothermal simulation temperature of $T=\SI{1160}{\kelvin}$ for the Fe-\SI{0.02}{\wtpercent}C alloy. This proximity ensures that the system is nearly at local equilibrium for phase transformation and can swiftly transition to the particle coarsening regime. The resolution time step is fixed at $\Delta t = \SI{0.5}{\second}$. The mobility and carbon diffusivity data are consistent with previous cases. However, considering the long time scale of this phenomenon, for illustrative purposes, the interfacial energy is set to a higher value ($\sigma_{\gamma\alpha}=\SI{2e-6}{\joule\per\milli\meter\squared}$) to accentuate the coarsening process. The other numerical parameters remain the same as the previous cases. The simulation is run for a duration of \SI{10000}{\second}.

\begin{figure}[htbp]
	\centering
	\includegraphics[width=10cm]{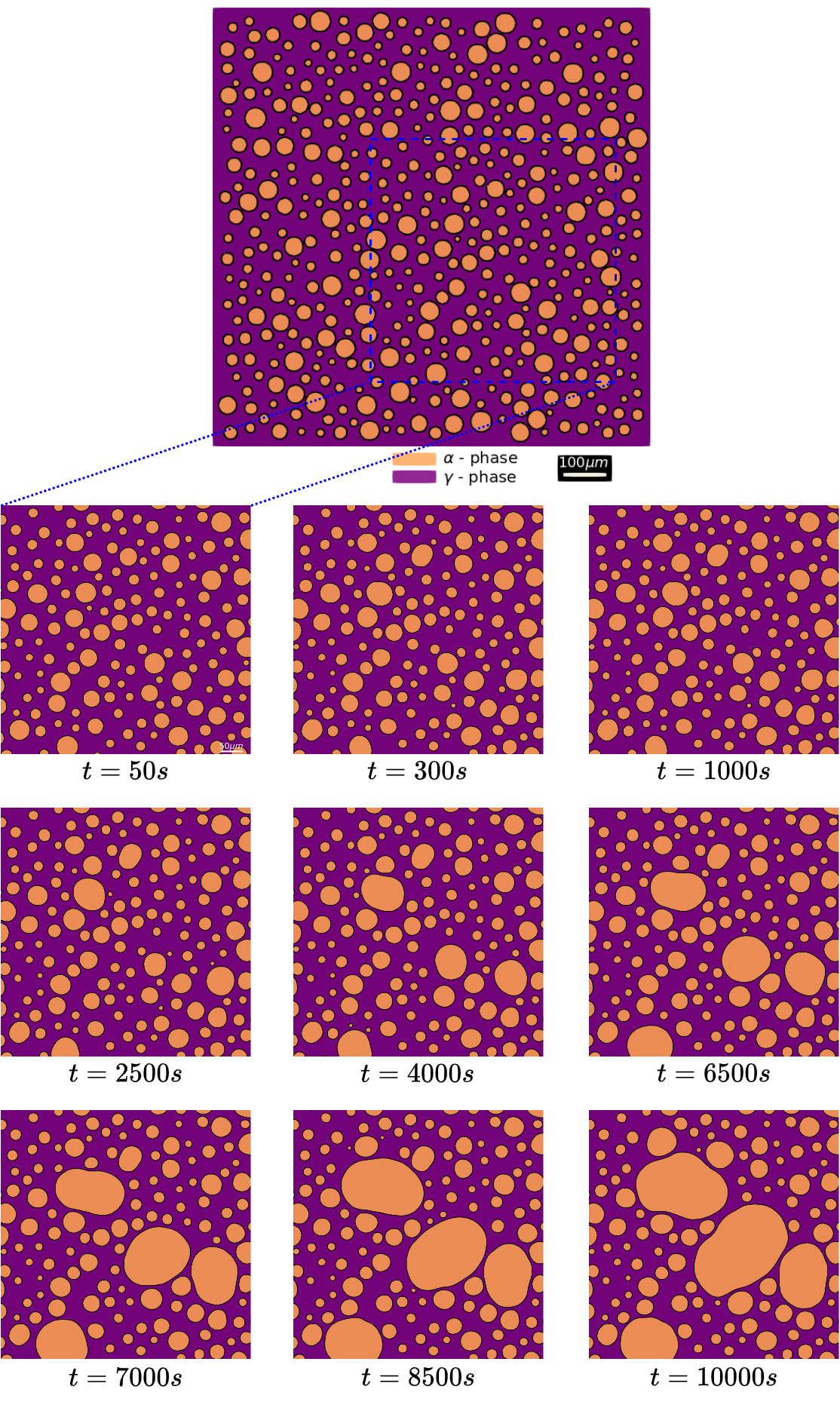}
	\captionsetup{justification=centering,margin=2cm}
	\caption{Snapshots of coarsening and dissolution of second phase particles (SPPs)}
	\label{OR_PD_Evol}
\end{figure}

Figs.\ref{OR_PD_Evol} present a series of snapshots illustrating the particle evolution at different instants. The system is anticipated to reach local equilibrium after the initial phase transformation at approximately \SI{50}{\second}. Thus, the particle coarsening regime is assumed to commence from $t=\SI{50}{\second}$. The expected coarsening of larger particles is visibly apparent, as the smaller particles dissolve over time. It's important to note that physical coalescence aspects are not accounted for in this simulation. Whenever two particles approach close proximity, the local diffusional flux and the interaction with capillarity pressure seem to flatten their interfaces.

\begin{figure}[htbp]
	\centering
	\includegraphics[width=7cm]{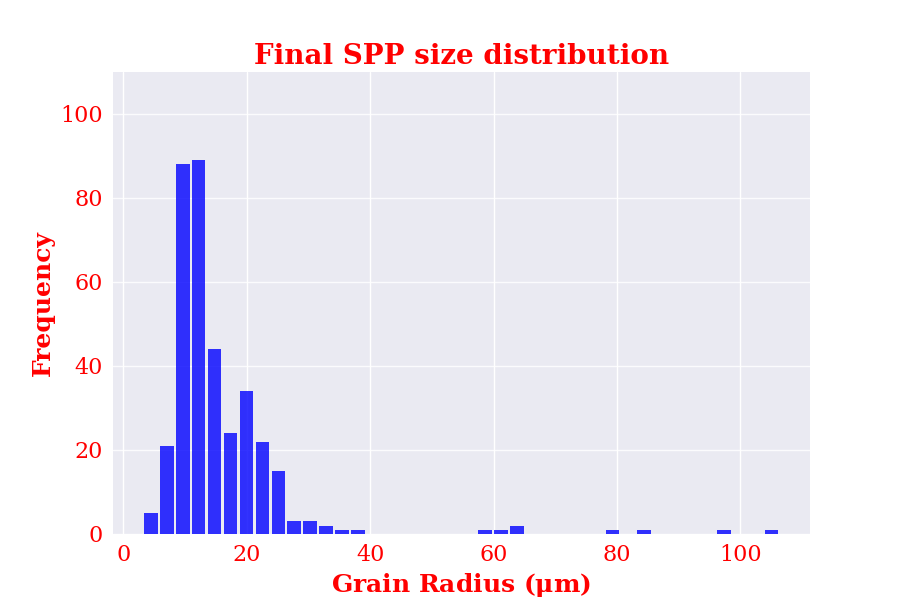}
	\captionsetup{justification=centering,margin=2cm}
	\caption{Final size distribution of second-phase particles (SPPs)}
	\label{OR_gsd_final}
\end{figure}

Fig.\ref{OR_gsd_final} demonstrates the particle size distribution obtained at the end of the simulation. It seems evident that, in comparison to Fig.\ref{OR_gsd_init}, there has been an increase in the frequency of larger particles over time as a result of coarsening.

\begin{figure}[htbp]
	\centering
	\includegraphics[width=8cm]{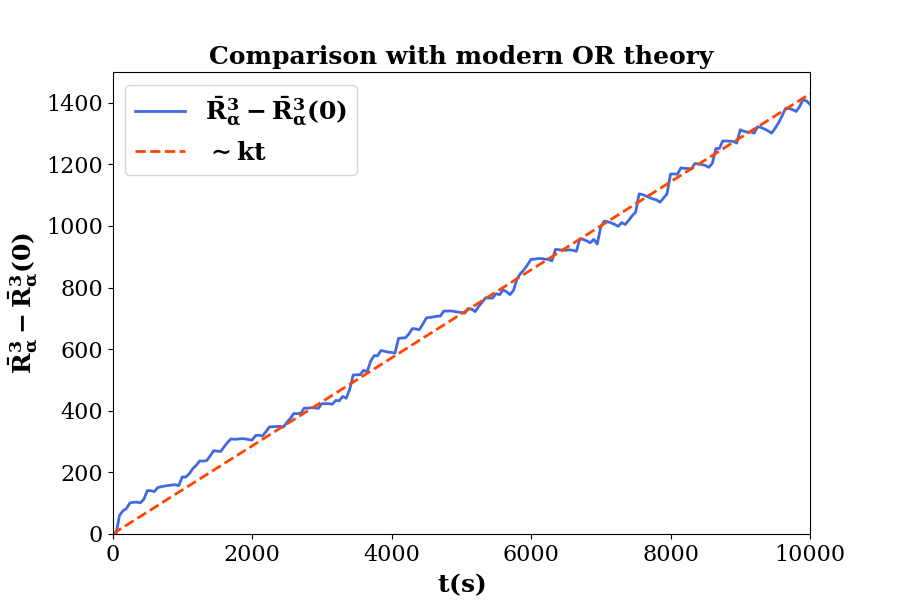}
	\captionsetup{justification=centering,margin=2cm}
	\caption{Comparison with modern Ostwald ripening theories}
	\label{OR_Revol_comp2}
\end{figure}

In the Fig.\ref{OR_Revol_comp2}, the kinetics of the particle size evolution are compared with the modern Ostwald ripening theories \cite{baldan2002review}. These theories are based on a generic expression:
\begin{equation}
    \bar{R}_\alpha^3 - \bar{R}_\alpha^3 (0) = kt, 
    \label{modernORtheory}
\end{equation} where $\bar{R}_\alpha$ is the mean particle size, $\bar{R}_\alpha^3(0)$ is the initial mean particle size, and $k$ is a parameter characterized based on the concerned theory. It appears that the simulated mean particle size can be fit into the aforementioned expression, resembling the quantitative description provided by the OR theories. Therefore, the hypothetical scenario we've adopted demonstrates kinetic characteristics akin to those of a typical Ostwald ripening phenomenon. This serves as evidence of the numerical model's seamless potential to simulate a range of diffusive solid-state phenomena.

\pagebreak
\section{Conclusions and perspectives}
\label{concl}
A generalization of the global level-set numerical framework used to primarily simulate diffusive solid-state phase transformation in metallic polycrystalline materials is presented. The generalized framework incorporates solute drag aspects into the transformation kinetics in the presence of any substitutional solutes. A coupling with a thermodynamic database was established to provide seamless extraction of necessary thermodynamic data for the transformation.
The level-set numerical model was benchmarked against a state-of-the-art sharp interface semi-analytical model, alongside comparisons with predictions from a corresponding phase-field numerical model. The potential of the proposed numerical framework to replicate the phase transformation behavior in complex large-scale microstructures with thousands of grains, relevant to industrial settings was showcased. An illustrative case considering solute drag aspects in a ternary alloy was presented, outlining the model's ability to reproduce sluggish transformation kinetics. The adjustable solute drag parameters offer additional flexibility to control the transformation kinetics, to be in better agreement with the experimental transformation curves. The numerical model’s versatility to simulate other diffusive solid-state phenomena such as particle coarsening, without any modifications to the existing formulation was demonstrated. The generalized nature of the kinetic framework alongside the adaptive meshing capabilities offer diverse scope to seamlessly integrate other complex evolution aspects into the model, including simulations in 3D.\\

The positive outcomes of this work are expected to guide future research, potentially expanding the range of modeling approaches available alongside the established phase-field method in the realm of diffusive solid-state phase transformation. A comprehensive experimental analysis to validate the proposed numerical model is planned. Such an analysis would enable improvements to the current nucleation model, potentially refining it based on experimental observations. In the proposed numerical framework, solute diffusion is assumed homogeneous within the bulk of a grain and across the grains of the same phase. Consequently, exploring the impact of short-circuit diffusion effects during phase transformation emerges as a potential avenue for future work. Integrating the effects of elastic/plastic accommodation due to volume misfits at the interphase boundaries into the current kinetic framework could further enhance our understanding of interface kinetics, providing a more comprehensive description of the physics involved. The global objective of this work was to establish a kinetic framework capable of encompassing various evolution aspects simultaneously, notably in a high plastic deformation context. The existing numerical tools and concepts devised for recrystallization could be integrated into the current framework, enabling the simulation of dynamic recrystallization in a multiphase context and potentially considering, at the mesoscopic scale, the modeling of realistic industrial thermomechanical paths for multiphase materials. Another perspective of these developments concerns the potential improvement/optimization of state-of-the-art equations used in macroscale code to describe type II residual-stresses due to phase transformations mainly during quenching and surface hardening treatments.

\section*{Acknowledgements}
The authors thank ArcelorMittal, Aperam, Ascometal, Aubert \& Duval, CEA, Constellium, Framatome, and Safran companies and the ANR for their financial support
through the DIGIMU consortium and RealIMotion ANR industrial Chair (Grant No. ANR-22-CHIN-0003).

\bibliography{main} 

\end{document}